\documentclass[useAMS,usenatbib]{mn2e}
\usepackage{times,epsfig,graphics,graphicx,multicol,amssymb,paralist,mn2e-breakabs}
\usepackage{natbib}
\usepackage{rotating}
\usepackage{longtable}

\usepackage[T1]{fontenc}
\usepackage[latin1]{inputenc}
\usepackage[font=small,labelfont=bf,tableposition=top]{caption}
\usepackage{booktabs}
\usepackage{rotating}

\usepackage[english]{babel}

\newcommand{\NII}{[N\,{\sc ii}]}
\newcommand{\OIII}{[O\,{\sc iii}]}

\newcommand{\Ha}{H$\alpha$}
\newcommand{\Hb}{H$\beta$}

\newcommand{\Msun}{~M$_{\odot}$}

\newcommand{\Luv}{~$L_{\rm UV}$}
\newcommand{\Lsun}{~L$_{\odot}$}
\newcommand{\LHa}{~$L_{H\alpha}$}

\newcommand{\LIR}{~$L_{\rm IR}$}

\def\Lc{~L_{100}}
\def\Lcs{~L_{160}}

\usepackage{savesym}
\usepackage{amsmath}
\savesymbol{iint}
\usepackage{txfonts}
\restoresymbol{TXF}{iint}

\bibpunct{(}{)}{;}{a}{}{,}

\title[\textit{Herschel} Far-IR counterparts of SDSS galaxies]{\textit{Herschel} Far-IR counterparts of SDSS galaxies: Analysis of commonly used Star Formation Rate estimates}
\author[H. Dom\'inguez S\'anchez et al.]{H. Dom\'inguez S\'anchez $^{1,2,3}$\thanks{E-mail:
helena@iac.es},  A. Bongiovanni$^{1,3}$, M. A. Lara-L\'opez$^4$, I. Oteo$^{1,3}$, 
\newauthor
J. Cepa$^{1,3}$, A. M. P\'erez Garc\'ia$^{1,3}$, M. S\'anchez-Portal$^5$, A. Ederoclite$^{6}$, D. Lutz$^7$,
\newauthor
 G. Cresci$^8$, I. Delvecchio$^9$, S. Berta$^7$, B. Magnelli$^7$, P. Popesso$^7$, F. Pozzi$^9$, L. Riguccini$^{10,11}$\\
$^{1}$Instituto de Astrof\'{\i}sica de Canarias, 38200 La Laguna, Spain\\
$^{2}$Centro de Astrobiolog\'{\i}a, Departamento de Astrof\'{\i}sica, CSIC-INTA, Ctra. de Ajalvir km. 4, 28850 Torrej\'on de Ardoz, Madrid, Spain\\
$^{3}$Departamento de Astrof\'{\i}sica, Universidad de la Laguna, Spain\\
$^{4}$Australian Astronomical Observatory, PO Box 915, North Ryde, NSW 1670, Australia\\
$^{5}$Herschel Science Center, INSA/ESAC, Madrid, Spain\\
$^{6}$Centro de Estudios de F\'{\i}sica del Cosmos de Arag\'on, Plaza de San Juan, 1, 44001 Teruel, Spain\\
$^{7}$MPE, Postfach 1312, 85741, Garching, Germany \\
$^{8}$INAF - Osservatorio Astrofisico di Arcetri, Largo E. Fermi 5, 50125 Firenze, Italy\\
$^{9}$Dipartimento di Fisica e Astronomia, Universita di Bologna, via Ranzani 1, I-40127 Bologna, Italy\\
$^{10}$NASA Ames Research Center, Moffett Field, CA, USA \\
$^{11}$BAER Institute, Sonoma, CA, USA\\}

\begin{document}

\date{Accepted 2014 March 11.  Received 2014 March 11; in original form 2013 September 30  }

\pagerange{\pageref{firstpage}--\pageref{lastpage}} \pubyear{2012}

\maketitle

\label{firstpage}

\begin{abstract}

We study a hundred of galaxies from the spectroscopic Sloan Digital Sky 
Survey with individual detections in the Far-Infrared Herschel PACS bands 
(100 or 160 $\mu$m) and in the GALEX Far-UltraViolet band up to z$\sim$0.4 
in the COSMOS and Lockman Hole fields. The galaxies are divided into 4 spectral 
and 4 morphological types. For the star forming and unclassifiable galaxies we 
calculate dust extinctions from the UV slope, the H$\alpha$/H$\beta$ ratio and 
the $L_{\rm IR}/L_{\rm UV}$  ratio. There is a tight correlation between the 
dust extinction and both $L_{\rm IR}$ and metallicity. We calculate SFR$_{total}$ 
and compare it with other SFR estimates (H$\alpha$, UV, SDSS) finding 
a very good agreement between them with smaller dispersions than typical SFR uncertainties. 
We study the effect of mass and metallicity, finding that it is only significant 
at high masses for SFR$_{H\alpha}$. For the AGN and composite galaxies we find 
a tight correlation between SFR and L$_{IR}$ ($\sigma\sim$0.29), while the 
dispersion in the SFR - L$_{UV}$ relation is larger ($\sigma\sim$0.57). The 
galaxies follow the prescriptions of the Fundamental Plane in the M-Z-SFR space.

\end{abstract}



\section{Introduction}

One of the main aspects to understand galaxy formation and evolution focuses on the mass assembly of galaxies at different epochs. There are three galaxy properties which are fundamental when studying these processes and that are strongly interrelated to each other: the galaxy stellar mass (M), the metallicity (Z) and the star formation rate (SFR). The existence of a main sequence, MS, in the M-SFR relation (and its evolution with redshift) has been widely demonstrated \citep{Brinchmann04, Noeske2007, Elbaz2007, Daddi2007, Rodighiero2010}, as well as the mass-metallicity M-Z relation \citep{Tremonti2004, Lara09b}. Moreover, in the last few years it has been shown that the M-SFR and M-Z relations for star-forming galaxies are particular cases of a more general relationship, defined as the Fundamental Plane (FP) by \citet{Laralopez2010b} or the Fundamental Metallicity Relation (FMR) by \citet{Mannucci2010}. In a more recent work, \citet{LaraLopez2013} refine the parameters of the original representation
of this plane, where the mass is a function of the metallicity and the SFR (M=f(Z,SFR)). This suggests that the stellar mass can be calculated as a linear combination of the rate at which a galaxy is currently forming stars (SFR) plus a measure of the star formation history, represented by the metallicity (corresponding to the amount of gas reprocessed by past stellar generations).

It is of special relevance to have accurate measurements of these three galaxy parameters. The stellar masses can be estimated by fitting the photometric spectral energy distributions (SEDs) or the spectral features, while for the metallicity derivation it is strictly necessary to have robust emission line measurements from spectroscopy. The SFR, however, can be measured through different indicators in a wide wavelength range. Among the many SFR indicators we have  X-rays, tracing X-ray binary emission (e.g. \citealt{Ranalli2003}); the UV, where the recently formed massive stars emit the bulk of their energy \citep{Schmitt2006,Rosa-Gonzalez2007}; optical wavelengths, from the recombination lines emission of the young massive population  \citep{Kewley2002};  the mid-IR and FIR, since a significant fraction of the UV light of a galaxy is absorbed by the interstellar dust and reemitted in the infrared  (e.g., \citealt{Kennicutt1998},  \citealt{Calzetti2005}, \citealt{Calzetti2007},  \citealt{Alonso-Herrero2006},  \citealt{Calzetti2009}, \citealt{Calzetti2010}); or the radio wavelengths, which traces supernova activity (e.g.  \citealt{Yun2001}). One of the major problems when using the SFR indicators in the UV or optical regimes is the absorption of a great part of the light emitted by the young population by the dust surrounding the star-forming regions. Previous works combine optical and infrared observations to derive attenuation-corrected  H${\alpha}$  and UV continuum luminosities of galaxies (e.g. \citealt{Gordon2000}, \citealt{Inoue2001}, \citealt{Hirashita2001}, \citealt{Bell2003}, \citealt{Hirashita2003}, \citealt{Iglesias-Paramo2006}, \citealt{Cortese2008}, \citealt{Kennicutt2009}, \citealt{Wuyts2011a}). The main advantage of the SFR from the FIR emission is that it is not affected by the dust extinction. However, in the pre-\textit{Herschel} era, the \LIR~estimates had to rely on the detection of IR emission at 24 or 70 $\mu$m from the \textit{Spitzer} data, meaning that the emission at longer wavelengths had to be extrapolated.

Due to the different physical mechanisms and assumptions made to estimate the SFR at different wavelengths it is of great importance to see how these SFRs indicators compare to each other and which galaxy properties have a more important impact on their agreement/disagreement. Testing the validity of the SFRs indicators with complete samples of galaxies at low-z is fundamental to extend their validity at higher redshifts where the available data is usually scarcer. In a recent work, \cite{DominguezSanchez2012} studied a \textit{Herschel} selected sample at z $\leq$ 0.46 with \Ha~emission from the zCOSMOS survey \citep{Lilly2007,Lilly2009} and  compared the SFR from the FIR with the SFR from the dust corrected \Ha~emission. We found an excellent agreement between the SFR indicators, except for very metal rich/poor galaxies.

In this work we extend this analysis and compare various SFR indicators  with each other (UV, \Ha, IR, SDSS, total). We combine for the first time the deep IR data from the latest PEP (PACS Evolutionary Probe, \citealt{Lutz2011}) \textit{Herschel} public data release  with the extensive and already processed ancillary data from  the Sloan Digital Sky Survey - Data release 7 (SDSS-DR7, \citealt{Abazajian2009}) in the COSMOS and Lockman Hole fields. The SDSS-DR7 data include masses, metallicities, emission line fluxes (e.g., \Ha~and \Hb) and SFRs for $\sim 10^6$ galaxies up to $z \sim 0.6$. We also use The Galaxy Evolution Explorer satellite (\textit{GALEX},  \citealt{Martin2005}) data in the far and near Ultra-Violet (FUV, NUV). The \textit{Herschel Space Telescope} has performed the  deepest surveys in the FIR bands, which sample the IR peak of the galaxy spectra, helping to derive accurate \LIR~values. Using \LIR~and \Luv~ we derive SFR$_{total}$=SFR$_{UV}$+SFR$_{FIR}$, i.e., the obscured (SFR from the FIR) 
plus the unobscured SFR (SFR from the UV uncorrected for dust extinction). We also derive SFRs from the FUV and \Ha~fluxes, using  two extinction correction approaches (from the observed \Ha/\Hb~ratio and  the UV slope). We compare the 
predictions of the different SFRs estimators and study how the galaxy stellar mass and metallicity affect the comparison.  We also investigate the relation between SFR and \LIR~and \Luv~  for a sample of AGN and composite galaxies. We locate the  FIR counterparts of the SDSS galaxies in the M-Z-SFR space. 

This paper is organised as follows. In Section 2 we explain the sample selection and the data from the SDSS-DR7, \textit{Herschel} and \textit{GALEX} surveys. In section 3 we study the differences between the whole SDSS sample and the FIR detected sample. In Section 4 we derive luminosities at different wavelengths (\Luv, \LHa, \LIR, $L_{100}$ and $L_{160}$). Then we separately study the SF and unclassifiable galaxies in Section 5 (where we calculate dust extinctions and SFRs and compare the results), while in Section 6 we study the AGN and composite galaxies and the correlation of their \LIR~and \Luv~with SFR. In Section 7 we study the location of our sample of galaxies in the M-Z-SFR space. Finally in Section 8 we summarise our results and highlight the most important conclusions.

	Throughout this paper we use a standard cosmology ($\Omega_{m}=0.3,\Omega_{\Lambda}=0.7$), with $H_{0}=70$  km s$^{-1}$ Mpc$ ^{-1}$. The stellar masses are given in units of solar masses (M$_{\odot}$), and both the SFRs and the stellar masses assume a \cite{Kroupa2001} IMF. The SFR estimates from the SDSS-DR7 are derived with the  Kroupa IMF. We convert the SFR derived with the \cite{Kennicutt1998} (K98 herafter) recipes, which assume a \citet{Salpeter1955} IMF into Kroupa IMF, by dividing the Salpeter SFR values by 1.5 \citep{Brinchmann04}.


\section{Data processing and sample selection}
\label{sect:data}

In this paper we present results  for galaxies from the SDSS-DR7,  with a counterpart from the  PACS Evolutionary Probe (PEP, \citealt{Lutz2011}) \textit{Herschel} survey in two different fields: the Cosmic Evolution Survey (COSMOS) field  \citep{Scoville07} and the Lockman Hole (LH, hereafter; \citealt{Lockman1986}). The COSMOS survey is designed to probe the evolution of galaxies and active galactic nuclei (AGNs) up to z $\sim$ 6 over a large enough sky region, $\sim$2 deg$^2$, to address the role of environment and large scale structures and star formation rate (SFR) without problems related to cosmic variance that smaller surveys can suffer. It is based on deep multi-wavelength observations from X-ray to radio wavelengths \citep[e.g.][]{Zamojski2007, Hasinger07, Taniguchi07, Capak2007, Sanders07, Bertoldi07, Schinnerer07, Elvis2009, McCracken2010, Lutz2011}. This multiwavelength coverage is of the outmost importance to derive fundamental properties of galaxies, such as photometric 
redshifts, galaxy stellar masses or SFRs. On the other hand, the LH is an area of $\sim$ 0.5 deg$^{2}$ on the sky  with the lowest galactic hydrogen column density along the line of sight (NH $\approx$ 5.7 $\times$ 10$^{19}$ cm$^{2}$, \citealt{Lockman1986}; \citealt{Schlegel1998}).
This physical characteristic provides the opportunity to perform extra-galactic observations without significant absorption of the
radiation in the soft X-rays and the ultraviolet and with minimal galactic cirrus emission in the infrared. For this reason the field
has been observed in all energy bands from the radio to the X-rays, including the FIR with \textit{Herschel} (an area  $\sim$ 0.2 deg$^{2}$ has been observed with the PACS instrument). A detailed overview of the various observations is given in  \citet{Rovilos2009, Fotopoulou2012}. 

We also make use of the photometric information from the \textit{GALEX} satellite. \textit{GALEX} was a NASA mission led by the California Institute of Technology to investigate how star formation in galaxies evolved from the early Universe up to the present. \textit{GALEX} used microchannel plate detectors to obtain direct images in the NUV and FUV and a grism to disperse light for low resolution spectroscopy. We therefore study two fundamental wavelength ranges for the SFR, the UV and the FIR, which provide complementary aspects of the emission of newly born stars. Besides, thanks to the ancillary data analysis from the SDSS survey, we are able to compare the results with main galaxy properties, such us stellar masses or metallicities.

\subsection{Optical data}

Our study was carried out using optical data for galaxies from SDSS-DR7. We used the emission-line analysis of SDSS-DR7 galaxy spectra performed by the MPA-JHU group\footnote{http://www.mpa-garching.mpg.de/SDSS/DR7/}. It includes metallicities, stellar masses, and SFRs from the analysis of a total of 927552 galaxy spectra derived  following the methods described below.  In the following sections we briefly describe how the main galaxy parameters from the SDSS-DR7 have been derived; please refer to the corresponding papers for a more detailed description on each quantity.

\subsubsection{Spectral Classification}

 As explained in \cite{Brinchmann04} (B04 hereafter), galaxies were classified into different spectral types on the basis of their S/N ratio and their location in the BPT diagram (\citealt{Baldwin1981}, \citealt{Veilleux1987}):

\begin{description}
\item[\textbf{All}] The set of all galaxies in the sample regardless
  of the S/N of their emission lines.
\item[\textbf{SF}] The star forming galaxies. These are the galaxies
  with S/N$>3$ in all four BPT lines that lie below the most
  conservative AGN rejection criterion (see Fig. 1 from B04):

  log[{O\,\textsc{iii}}] $\lambda$5007/{H$\beta$} $\le$ 0.61/$\{$log([{N\,\textsc{ii}}] /{H$\alpha$})-0.05$\}$ + 1.3

 \item[\textbf{C}] We define as Composite galaxies the objects with S/N$>3$ in all four BPT lines that
  are between the upper and lower lines in the BPT diagram. Up to 40$\%$ of their \Ha~-luminosity might come
  from an AGN. 
\item[\textbf{AGN}] The AGN population consists of the galaxies above
  the upper line in Fig. 1 from B04, following the \cite{Kewley06} criteria. This line
  corresponds to the theoretical upper limit for pure starburst models
  so that a substantial AGN contribution to the line fluxes is
  required to move a galaxy above this line. 
\item[\textbf{Low S/N AGNs}] A minimum classification for AGN galaxies
  is that they have \NII/\Ha~ $>0.6$ (and S/N$>3$ in both lines). It is therefore possible to
  classify these even if \OIII~ and/or \Hb~ have too low S/N to be
  useful.
\item[\textbf{Low S/N SF}] The remaining galaxies with S/N$>2$
  in \Ha~ which are not classified as AGN, composites or low S/N AGNs are considered  low S/N
  star forming.
\item[\textbf{Unclassifiable}] Those remaining galaxies that are
  impossible to classify using the BPT diagram. This class is mostly
  made up of galaxies with no or very weak emission lines.
\end{description}

In this work we  consider 4 main spectral types: SF (including both SF and low S/N SF galaxies), composites, AGNs (including both AGNs and low S/N AGNs) and unclassifiable galaxies. The number of galaxies of each galaxy type is summarised in Table \ref{class}.

\subsubsection{Morphological classification}

In addition to a spectral classification, we  also divide our galaxy sample into different morphological types based on the morphological catalog from \citet{Huertas-Company2011A}. The catalog is an automated morphological classification in 2 broad classes (early or late type, i.e., E/S0 or S) and in 4 detailed types (E, S0, Sab, Scd) of 698420 galaxies at z $<$ 0.25 from the SDSS-DR7 spectroscopic sample with good photometric data and clean spectra. The main new property of the classification is that they associate a probability
to each galaxy of being in the four morphological classes instead of assigning a single class. The classification is therefore better
adapted to nature where we expect a continuous transition between different morphological types. The algorithm is trained with a
visual classification and then compared to several independent visual classifications including the Galaxy Zoo first-release catalog. In Table \ref{morph} we summarise the number of galaxies of each morphological type. Note that the number of galaxies with a morphological classification is not the same as the sum of the galaxies of each morphological type. This is because each galaxy has a given probability of belonging to each morphological class. We consider that a galaxy belongs to the E type when the probability of being elliptical is larger than the probability of being of any other morphological type. However, there are galaxies for which the probabilities of belonging to two or more morphological types are equivalent (i.e. 50-50 $\%$). We do not include these galaxies in our analysis when the morphology classification is used, as their morphological type is not well defined (these only represent $\sim$ $1\%$ of the morphological sample).

\subsubsection{Stellar Masses}

 The masses from the SDSS-DR7 were estimated using fits to the  the broad-band u, g, r, i, z photometry. These magnitudes are corrected for emission lines by assuming that the relative contribution of emission lines to the broad-band magnitudes is the same inside the fibre as outside. The fits are made to a large grid of models from \cite{BC2003} spanning a large range in star formation histories. For each model a likelihood is calculated from $\chi^2$. The likelihood of 
all 
models is then marginalised onto the mass axis and a likelihood distribution for the mass is obtained.

\subsubsection{Star Formation Rates}
\label{sect:SFR-SDSS}

The SDSS-DR7 total SFRs for SF galaxies were inferred directly from the emission lines, based on the careful modelling discussed in B04, who modelled the emission lines in the galaxies following the  \citet{Charlot01} prescription, achieving a robust dust correction. The metallicity dependence of the case B recombination {H$\alpha$}/{H$\beta$} ratio is also taken into account. The B04 method offers a more robust SFR estimate than using, for example, a fixed conversion factor between {H$\alpha$} luminosity and SFR \citep[e.g.][]{Kennicutt1998}.

 For other classes of galaxies (AGN, composite and unclassifiable) there is a slight modification in the procedure. In B04, the authors constructed the likelihood distribution of the specific SFR (sSFR=SFR/M), as a function of the 4000 $\AA{}$ break, D4000, using the star-forming sample. Then, the value of D4000 was used to estimate the sSFR for AGNs, composite or unclassifiable galaxies. Likewise, for the low-S/N SF galaxies they constructed the average conversion factor from observed \Ha~ luminosity to SFR and used this to estimate SFRs for those galaxies. However, this method implicitly applies a dust correction similar to that of the average in the SDSS SF sample, about A$_V$ $\sim$ 1.

In B04 the SDSS-DR2 was used and the sample size did not allow to  properly take into account the effect of different dust attenuations. In this new data release, when the galaxy has \Ha~ and \Hb~ with S/N$>$3, average probability distribution functions are constructed using galaxies with similar \Ha/\Hb~. This removes all trends with dust attenuation. The SFRs estimated for AGN and composite galaxies are typically  larger than with the previous technique.

\subsubsection{Metallicities}

The metallicities from the SDSS-DR7  were estimated statistically using Bayesian techniques by \citet{Tremonti04}, based on simultaneous fits of all the most prominent emission lines ([{O\,\textsc{ii}}], {H$\beta$}, [{O\,\textsc{iii}}], {H$\alpha$}, [{N\,\textsc{ii}}], [{S\,\textsc{ii}}]) using a model designed for the interpretation of integrated galaxy spectra \citep{Charlot01}. Since the metallicities derived with this technique are discretely sampled, they exhibit small random offsets \citep[see for details][]{Tremonti04}. Any dependence of SFR on the estimated metallicity would be minor \citep[][]{Tremonti04,Brinchmann08}. All of the galaxies analysed in this work have 12+log(O/H) $>$ 8.4, corresponding to the upper branch of the R$_{23}$ (R$_{23}$=([OII]$\lambda$3727+ [OIII]$\lambda \lambda$4959, 5007)/H$\beta$).

\subsubsection{Aperture corrections}
\label{sect:aperture}

The SDSS-DR7 aperture corrections were derived following \cite{Salim2007}, by calculating the light outside the fiber for each galaxy, and then fitting stochastic models  to the photometry. This removes most of the bias found for galaxies with low level  star formation when using the empirical aperture corrections from B04, which were based on the distribution of SFR/M at a given (g-r, r-i) colour. We will use the aperture corrections to convert the \Ha~ fluxes from the fiber into total   \Ha~ fluxes (to derive SFR$_{H\alpha}$, see Section \ref{sect:SFR}).

\subsection{\textit{Herschel} Far-IR counterparts}

Taking the advantage of the recently released public FIR data from the \textit{Herschel Space Observatory} from the PEP survey, we look for FIR counterparts of the SDSS galaxies in the COSMOS and LH fields.  We use data from the Herschel PEP survey, which provides fluxes at 100 and 160 $\mu$m from the Photodetector Array Camera and Spectrometer \citep[PACS, ][]{Poglitsch2010}. This wavelength range covered by \textit{Herschel} samples the FIR peak of the galaxy spectra, which is fundamental to study the relation between SFR and \LIR.

 We use PACS catalogs extracted with 24 $\mu$m priors from the data release 1 (DR1) for the COSMOS and LH fields. The COSMOS 24 $\mu$m prior catalog also includes observations with the  24 $\mu$m bandpass of the Multi-band Imaging and Photometer for \textit{Spitzer}, (MIPS, \citealt{Rieke04}, \citealt{LeFloch09}). The 3$\sigma$ \textit{Herschel} depths reached for the final sample in the 100 and 160 $\mu$m bands are 4.10 and 9.21 mJy in the Cosmos field and 10.65 and 11.37 mJy in the LH.

\begin{figure}
 \centering
\includegraphics[scale=0.3]{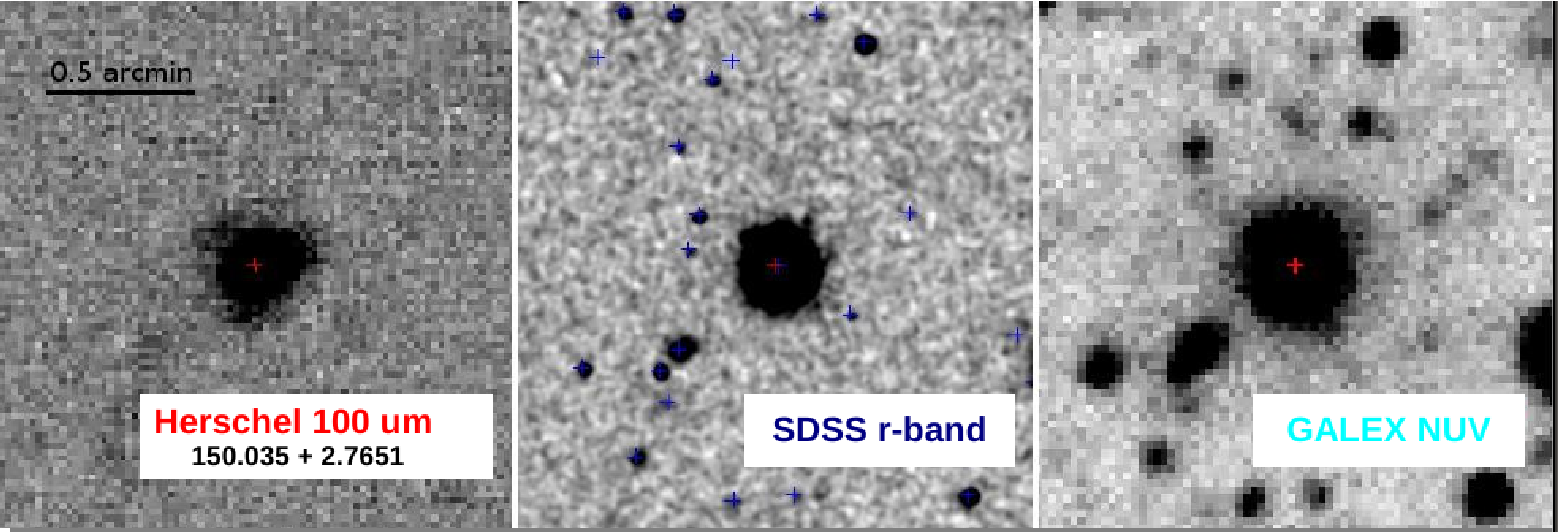}
\includegraphics[scale=0.3]{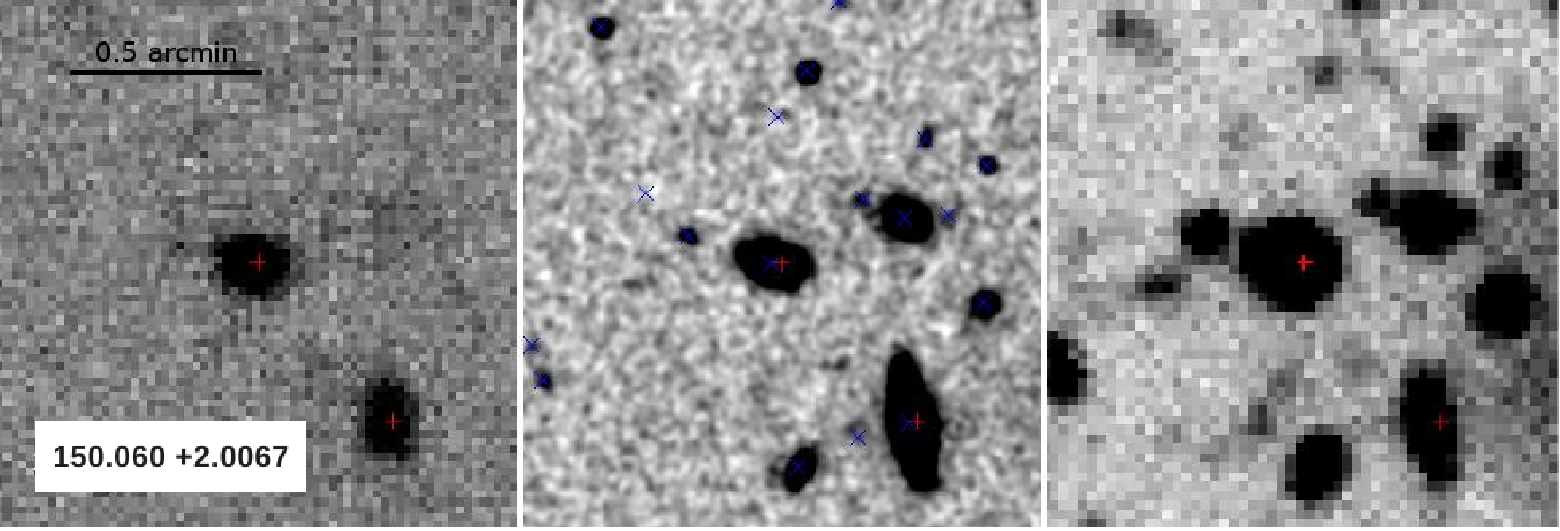}
\includegraphics[scale=0.3]{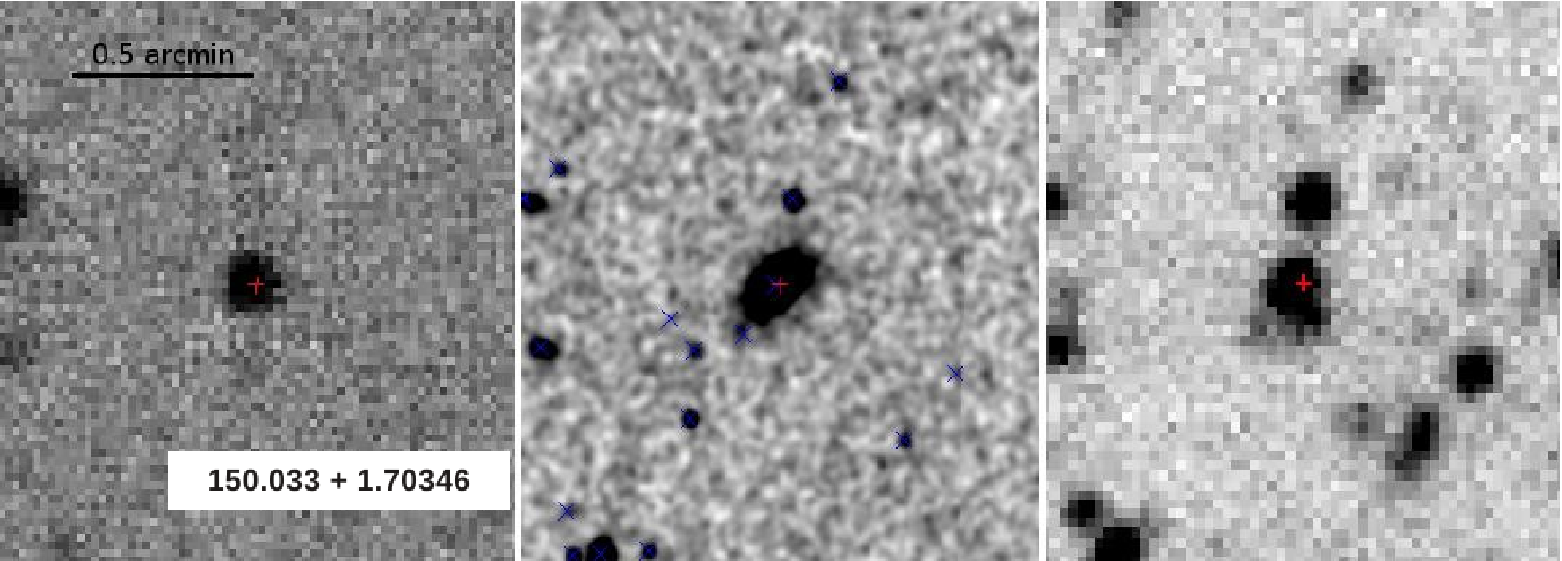}
\caption{Cutouts of three typical galaxies of our sample in the 100 $\mu$m \textit{Herschel} band (\textit{left panel}), the SDSS optical r-band (\textit{central panel}) and the NUV \textit{GALEX} band (\textit{right panel}). Red crosses are the \textit{Herschel} FIR detections, while blue crosses are the SDSS detections. Due to the low redshift considered ($<$ 0.5) and the brightness of the galaxies, the majority of them are not affected by confusion in the \textit{GALEX} or \textit{Herschel} bands.}
\label{cutouts}
\end{figure}

 We looked for the \textit{Herschel} counterparts of the SDSS galaxies in the COSMOS and LH fields. After considering only  detections with S/N $>$ 3, we found a FIR counterpart with detection in at least one band and a separation $<$ 3 arcsec for 123 galaxies (114 from COSMOS, 9 from the Lockman Hole). We visually inspected all the galaxies in the FIR, optical (r-band from SDSS) and \textit{GALEX} (see Section \ref{Sect:GALEX}) bands and eliminated the problematic sources (due to confusion in the \textit{Herschel} bands or wrong photometry).  In Fig. \ref{cutouts} we show typical cutouts of our galaxies in the \textit{Herschel}, SDSS-r and \textit{GALEX} bands. It can be seen that the counterparts  are well identified and they are not affected by confusion, despite the larger PSF of the \textit{Herschel} and \textit{GALEX} data with respect to the SDSS data.

 \subsection{GALEX counterparts}
 \label{Sect:GALEX}
 
  We also make use of the \textit{GALEX} satellite data (see \citealt{Morrissey2005}, \citealt{Morrissey2007} for technical details). \textit{GALEX} simultaneously observes  two broadband filters in the far-UV (FUV) and near-UV (NUV), with effective wavelengths of 1528 and 2271 $\AA{}$, respectively and limiting magnitudes of FUV=NUV=22.7 mag (see \citealt{Salim2007}). In this work we have used the data from the DIS (GALEX Deep Imaging Survey) DR6/DR7, released in February 2013.
 
 We cross-correlated our sample of 123 SDSS galaxies with \textit{Herschel} detection with the combination of the \textit{GALEX} tiles in the two considered fields (COSMOS and LH). We found a \textit{GALEX} counterpart for 105 galaxies ($\sim 85\%$) with detected FUV flux (we will use the FUV flux to derive the SFR$_{UV}$, see Section \ref{sect:SFR}). We imposed the separation between the SDSS and  \textit{GALEX} positions to be $<$ 2 arcsec. In this case we constrain the best match to only 2 arcsec (instead of 3 arcsec as for the FIR counterparts) due to the better angular resolution for  the \textit{GALEX} data and the more similar wavelengths considered. 
 
 A table with the main parameters of the 123 galaxies which compose the final sample is available on-line.
 
 \section{Comparison of the whole SDSS sample and the FIR detected galaxies}

\begin{table*}
  \centering
  \begin{tabular}{*{4}{c}}\hline
    \multicolumn{1}{c}{Subsample}  &
    \multicolumn{1}{c}{SDSS (whole sample)} &
     \multicolumn{1}{c}{\textit{Herschel} detected (COSMOS \& LH)}&
       \multicolumn{1}{c}{\textit{Herschel} $\&$ FUV detected (COSMOS \& LH)} \\
      \\ \hline
     All           &   927552  (100.0$\%$)  &  123  (100.0$\%$)&  105  (100.0$\%$) \\
     SF            &   384599  (41.5$\%$)   &   67  (54.5$\%$) &   58  (55.2$\%$) \\
     AGN           &   91477  (9.9$\%$)     &   26  (21.1$\%$) &   23  (21.9$\%$) \\
     C             &   47704   (5.1$\%$)    &   13  (10.6$\%$) &   12  (11.4$\%$) \\
     UnClass       &   403772  (43.5$\%$)   &   17  (13.8$\%$) &   12  (11.4$\%$) \\
 \hline
  \end{tabular}
  \caption{Number of galaxies of the different spectral types described in the text for the whole SDSS-DR7 sample, for the sample of galaxies considered in this work (the 123 SDSS \textit{Herschel} detected galaxies in the COSMOS and LH fields)  and for the subsample of galaxies with FUV detection (105). }
  \label{class}
\end{table*}

\begin{table*}
  \centering
  \begin{tabular}{*{4}{c}}\hline
    \multicolumn{1}{c}{Subsample} &
    \multicolumn{1}{c}{SDSS (whole sample)} &
     \multicolumn{1}{c}{\textit{Herschel} + morphology (COSMOS \& LH)}& 
       \multicolumn{1}{c}{\textit{Herschel} $\&$ FUV + morphology (COSMOS \& LH)}\\
      \\ \hline
     All  &   698420  (100.0$\%$) &  118  (100$\%$)  &  101  (100$\%$) \\
     E &   139635  (20.2$\%$)  &   11  (9.3$\%$)  &    7  (6.9$\%$) \\
     S0   &   141305   (20.5$\%$) &   25  (21.2$\%$) &   22  (21.8$\%$) \\
     Sab  &   254316  (36.8$\%$)  &   47  (39.8$\%$)&    40  (39.6$\%$) \\
     Scd  &   155839  (22.5$\%$)  &   35  (29.7$\%$) &   32  (31.7$\%$) \\
 \hline
  \end{tabular}
  \caption{Number of galaxies of the different morphologies from the \citet{Huertas-Company2011A} catalog for the whole SDSS sample, for the sample of 118 FIR detected galaxies with morphological classification  in the COSMOS and LH fields and for the subsample of galaxies with FUV detection in the COSMOS and LH fields (101). }
  \label{morph}
\end{table*}

\begin{figure}
 \centering
\includegraphics[scale=0.4]{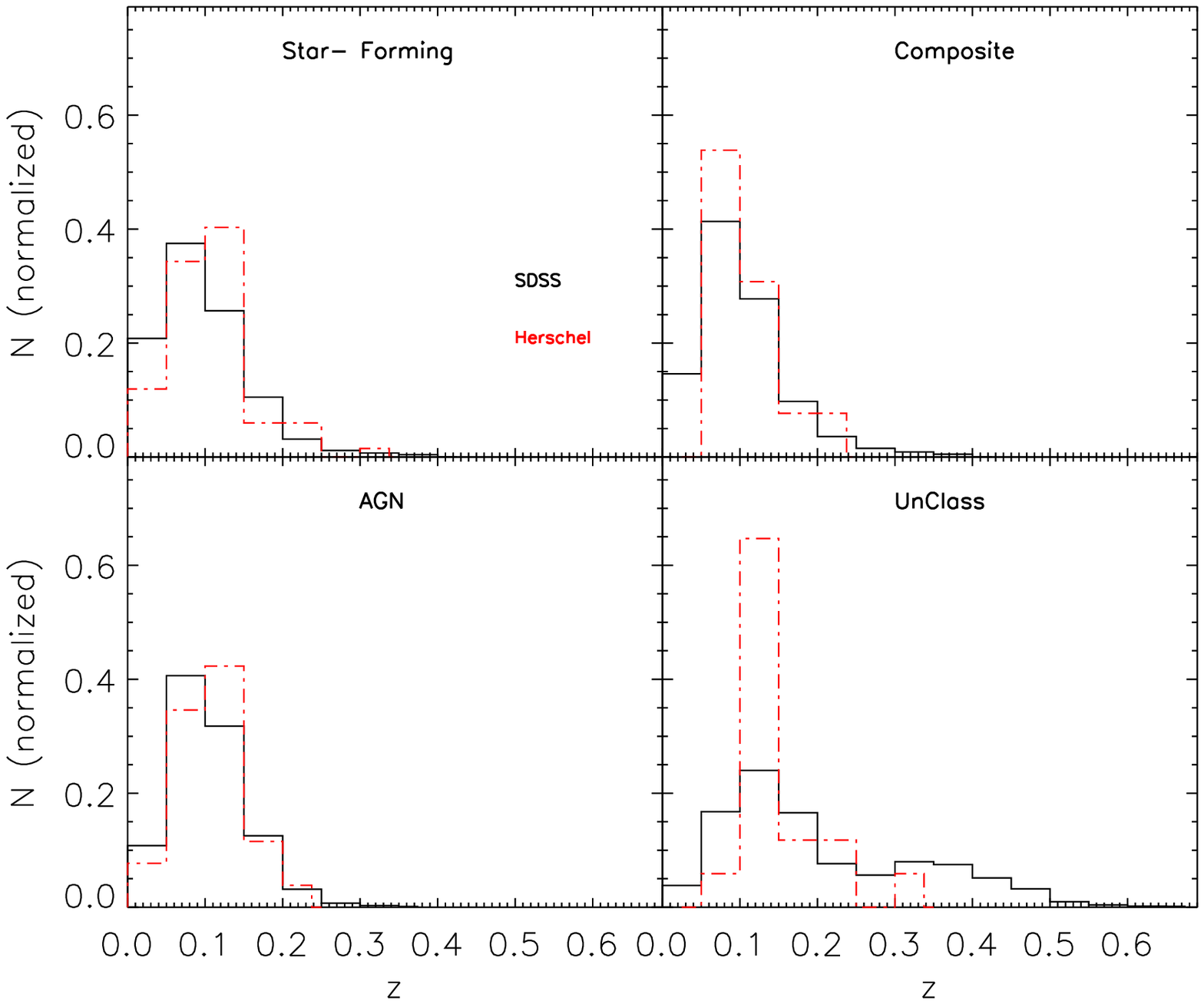}
\includegraphics[scale=0.4]{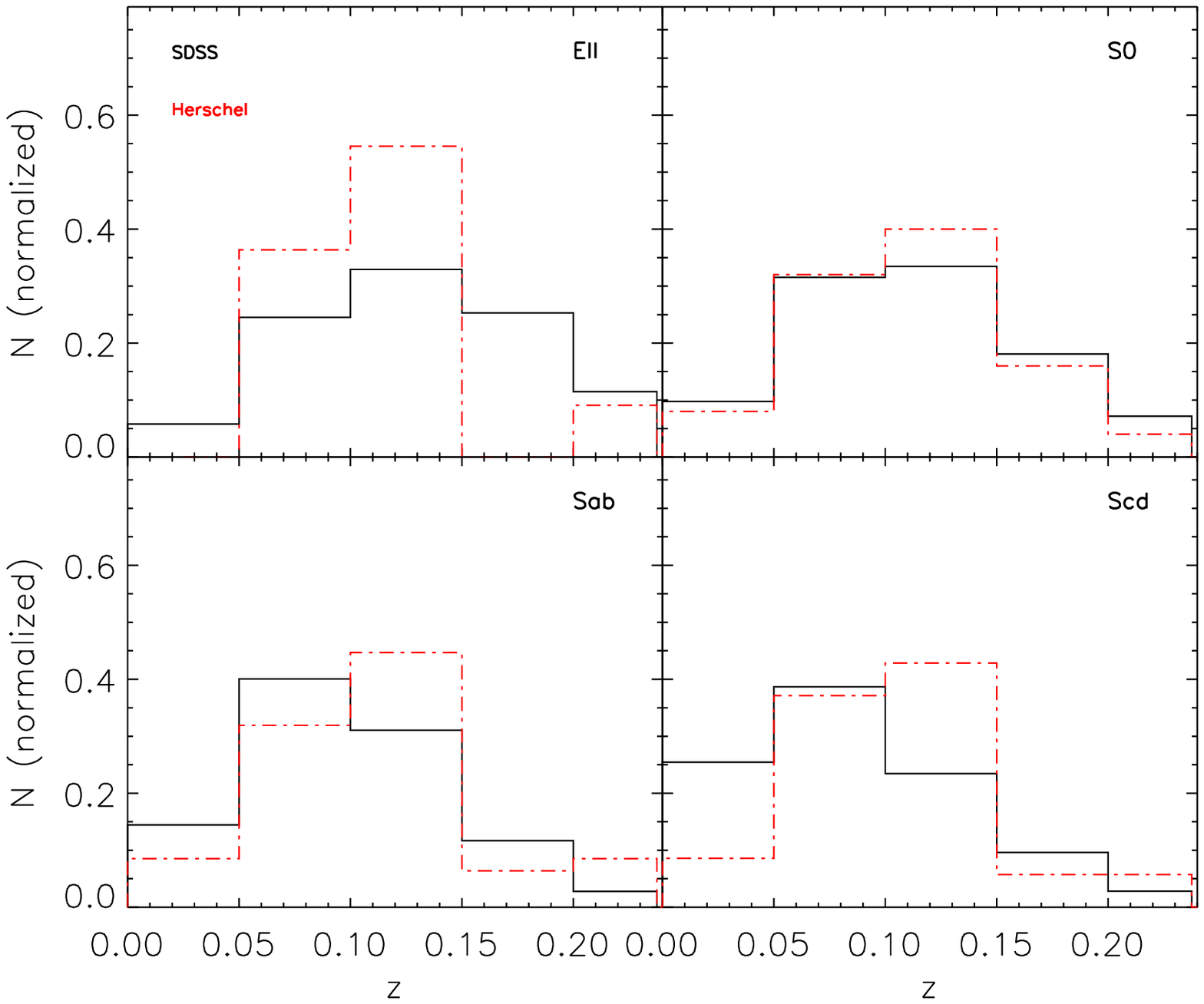}
\caption{Normalised redshift distribution for the whole SDSS-DR7 sample (thick black line) and for the sample of SDSS galaxies with  \textit{Herschel} detection (red dotted-dashed line) divided in different spectral (\textit{upper panel}) and morphological types (\textit{bottom panel}).}
\label{z-distribution}
\end{figure}

In this section we compare the properties of the whole spectroscopic sample of SDSS galaxies with the galaxies with FIR and FUV  detection that we study in detail in this article. In Fig. \ref{z-distribution} we show the normalised redshift distribution for the whole SDSS sample  and for the galaxies with \textit{Herschel} detection  for different spectral  and morphological classifications. The redshift distributions are very similar for the two subsamples, even when divided into different classes (note that only galaxies with z $<$ 0.25 have a morphological classification), except for a very pronounced peak in the redshift distribution of unclassified galaxies at z $\sim$ 0.15, which may be explained by the small number of this galaxy type (17) and which, not so pronounced,  but also exists for the whole SDSS sample. 

 In Table  \ref{class} and \ref{morph} we summarise the spectral and morphological classifications for the whole SDSS sample and for  our FIR  and FUV detected samples in the studied fields (COSMOS and LH), which explains the large difference in the number of galaxies considered.  The percentage of FIR detections of the SDSS-DR7 spectroscopic sample is  46 and 53$\%$ for the COSMOS and LH fields, respectively.  If we focus on the differences between the whole SDSS sample and our studied sample we can see that the percentage of unclassified galaxies substantially decreases, from 43.5$\%$ for the whole SDSS sample to  13.8$\%$ for galaxies with a FIR counterpart. The percentage of the other spectral types (SF, AGN and composites) increases for the FIR SDSS counterparts. The decrease in unclassifiable galaxies (with the corresponding increase in SF, AGNs or composite galaxies) is understandable 
because we are considering galaxies with FIR emission. FIR emission is associated with star-forming events or nuclear activity and  both of the process produce an increase in the emission lines intensity. As already mentioned, the unclassifiable galaxies are those for which the emission lines are too weak to be able to  locate them in the BPT diagram. When we include \textit{GALEX} detection in the FUV band, more than half of the sample (55.2$\%$) becomes SF and less than 12$\%$ of the galaxies are unclassifiable. 

If we focus on the morphological classification (Table \ref{morph}), we can see that the percentage of elliptical galaxies (E) decreases from 20.2$\%$ for the whole SDSS sample to 9.3$\%$ for the FIR detected galaxies, while the Scd galaxies  increase from  22.5$\%$ to 29.7$\%$. Clearly, when selecting galaxies detected in the FIR we have a bias towards later spectral types. This as a consequence of the FIR emission being related to star-formation events. The percentage decreases/increases even more for the E/Scd types when including the \textit{GALEX} detection (6.9 $\%$ for the E and 31.7 $\%$ for the Scd galaxies).  

\begin{figure}
 \centering
\includegraphics[scale=0.6]{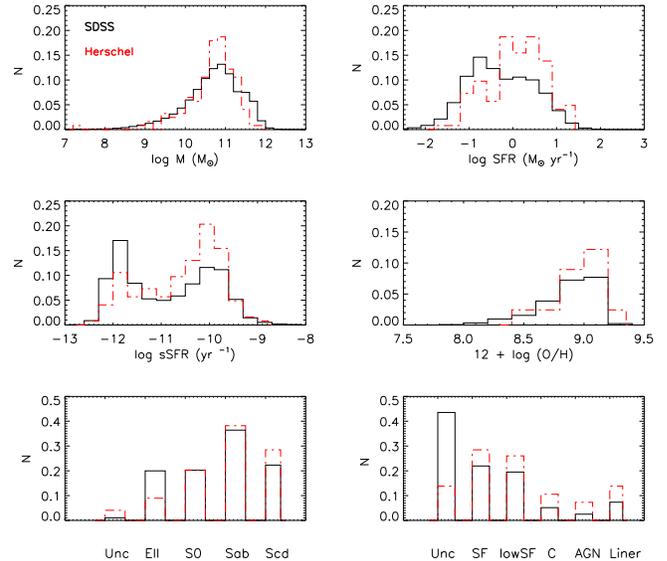}
\caption{Normalised distribution of different properties for the whole SDSS-DR7 sample (thick black line) and for the sample of SDSS galaxies with FIR  \textit{Herschel} detection (red dotted-dashed line). }
\label{properties}
\end{figure}

Besides the different spectral types, we show in Fig. \ref{properties} the distribution of  various properties for the whole  and the FIR detected SDSS sample: stellar mass, SFR, sSFR and metallicity. It can be seen that the stellar mass distributions are very similar for the two samples, although somewhat narrower for the FIR detected sample (we may be probably missing very massive passive E or very low mass galaxies with low SFRs). The FIR detected SDSS galaxies show larger SFR and sSFR values than the whole SDSS galaxies and also slightly larger metallicities. 

To summarise, the FIR detection of the SDSS sample does not significantly  affect the redshift, mass or metallicity distributions, but introduces a bias towards larger SFRs (and therefore sSFRs) and later morphological types. The number of SF or late type galaxies increases for the FIR detected galaxies and the unclassifiable galaxies are reduced. This selection effects are a consequence of  the FIR emission arising from the re-emission by dust of the light emitted by the young stellar populations.

\section{Luminosities}

There are many methods to derive the SFR of a galaxy. Most of them rely on simple conversions from luminosities to SFRs (see Section \ref{sect:SFR}). We have derived the luminosities from the FUV, the FIR and the \Ha~emission line fluxes to calculate the SFR in different ways.

\begin{description}

\item{\textbf{Ultra-Violet and  \Ha~ Luminosities}} \hfill \\

The luminosity in one band can be obtained when the flux and luminosity distance are known. We have used the FUV fluxes and redshifts of the 105 galaxies with FUV detection  to derive \Luv~  using Eq. \ref{eq:L}:

\begin{equation}
 L_{UV}=4 \pi d_{L}^2\times F_{FUV  }\times \nu_{RF}
\label{eq:L} 
 \end{equation}

 where  $F_{FUV}$ is the observed flux at $1500 ~ \AA{}$, d$_{L}$ is the luminosity distance, and $\nu_{RF}=\nu/(1+z)=c/\lambda_{RF}$ is the rest-frame frequency.

The derived log \Luv~ values range from 7.6 to 10.1 \Lsun, as it can be seen in Fig. \ref{LUV-distr}, where we plot the obtained distribution of \Luv. In the upper panel we separate our galaxies with respect to the field they belong to, in the central panel we plot the \Luv~  distribution for the different spectral types of galaxies and in the bottom panel we show the same by dividing our sample into different morphological types. 
 There is a clear bimodality in the \Luv~ distribution: the SF galaxies present the largest \Luv~ values and the unclassifiable galaxies the lowest \Luv~ values, while the composite and AGNs are almost uniformly distributed. As the \Luv~ is directly related with the SFR it is reasonable that the SF galaxies show the largest \Luv~and that the unclassifiable galaxies (with weaker emission lines) have fainter \Luv.  If we separate our galaxies with respect to their morphology, the Scd galaxies clearly take the largest \Luv~ values, while the E galaxies are distributed at the faint end of the \Luv~ distribution. This is related to  the late type galaxies being  more efficient star-forming galaxies.

\hfill \\

\begin{figure}
 \centering
\includegraphics[scale=0.45]{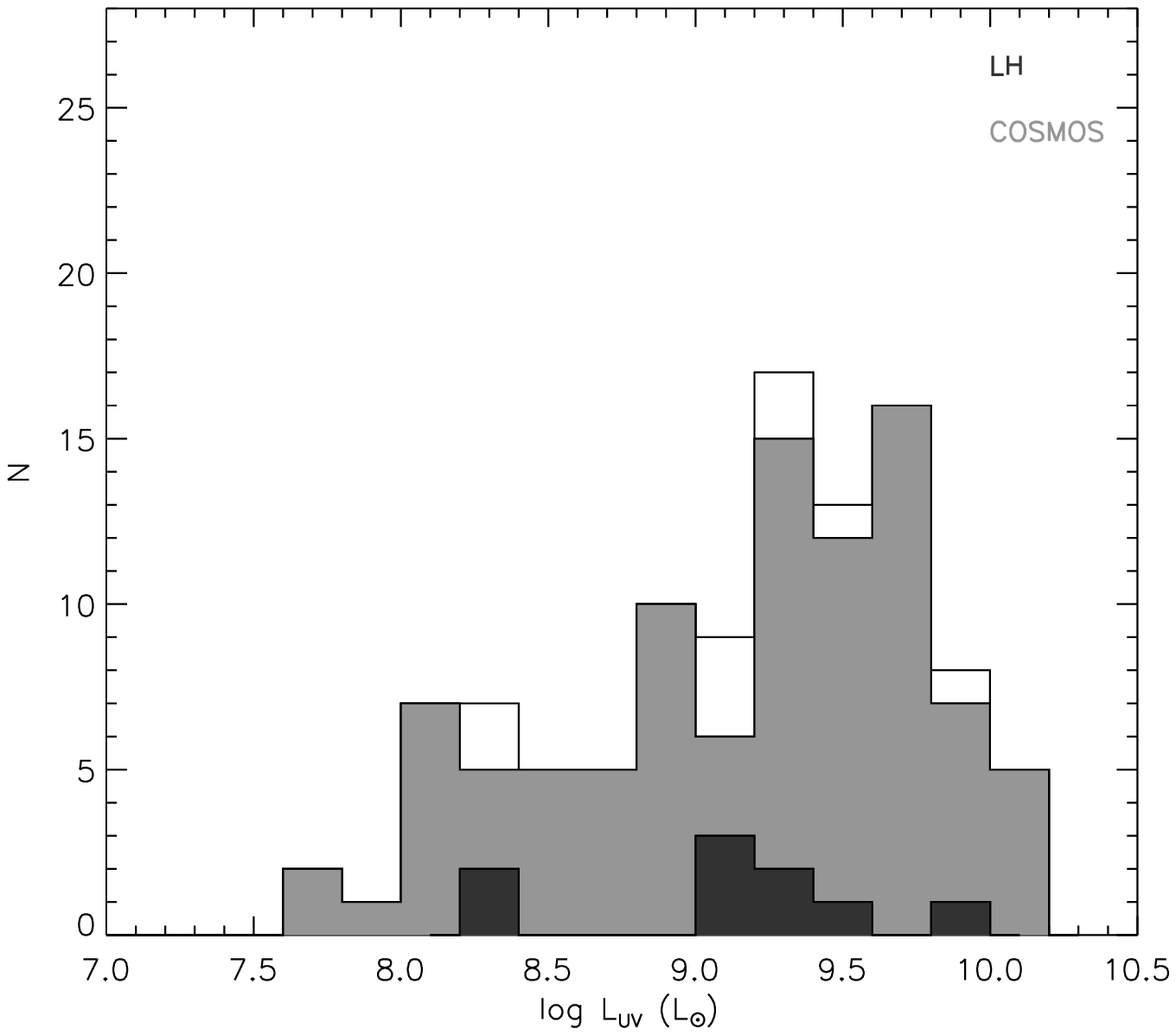}
\includegraphics[scale=0.55]{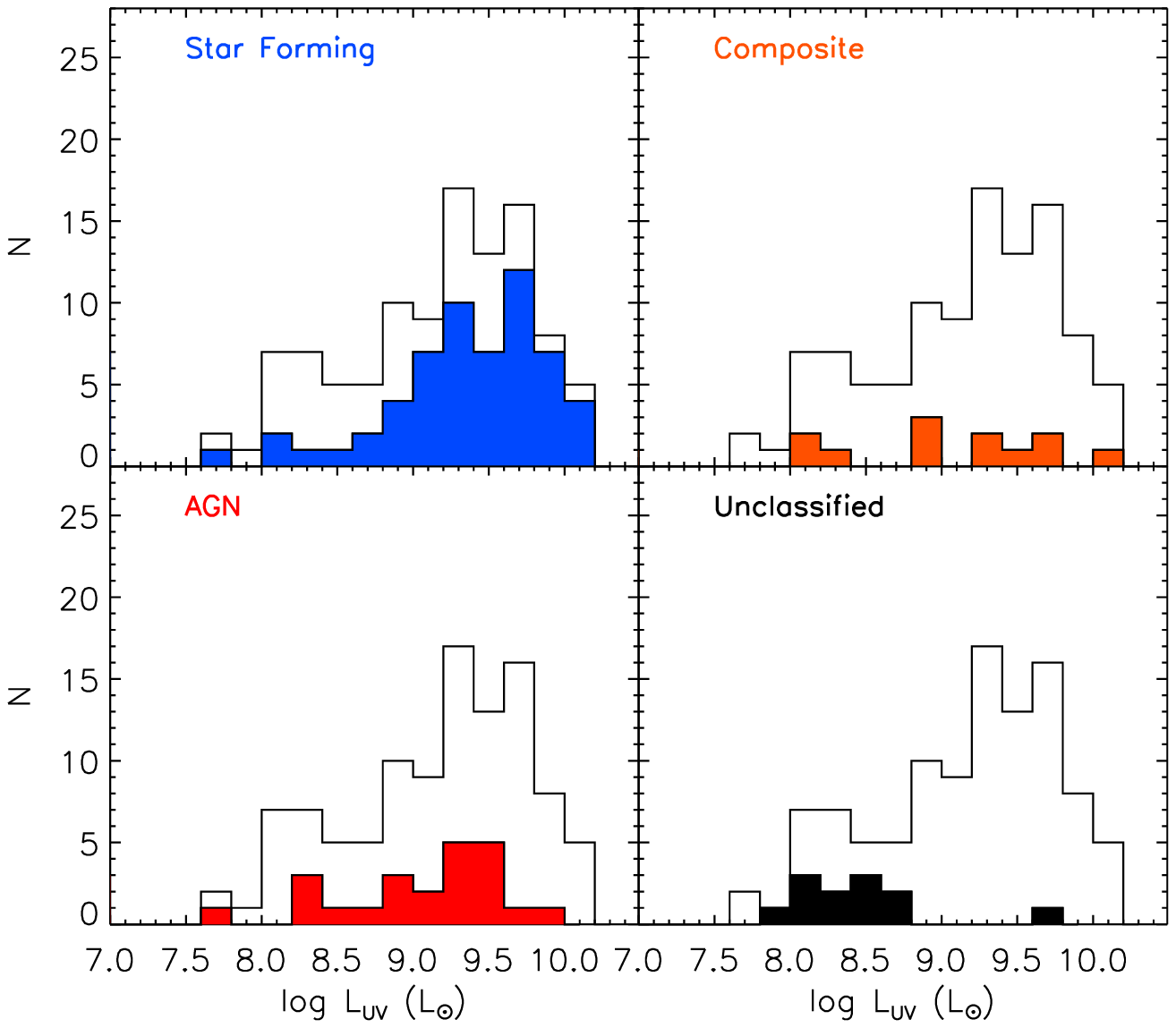}
\includegraphics[scale=0.55]{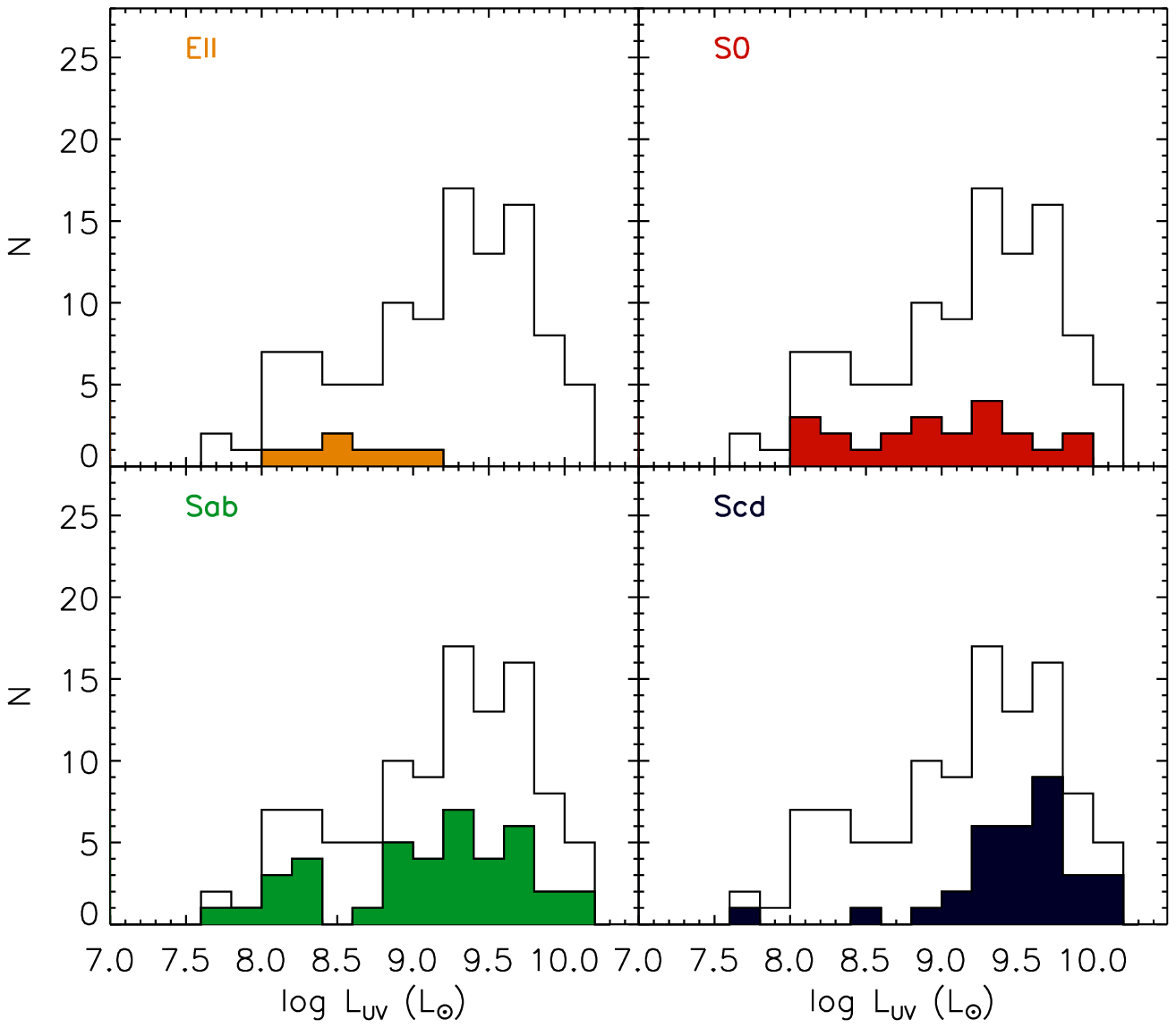}
\caption{\textit{Upper panel}: \Luv~ distribution for the whole sample of galaxies (empty histogram) and for the galaxies belonging to the different (light grey for COSMOS, dark grey for LH). \textit{Central panel}: \Luv~ distribution for the galaxies separated into different spectral types: SF (blue), composites (orange), AGNs (red)  and unclassifiable (black). \textit{Lower panel}: \Luv~ distribution for the galaxies separated into different morphological types: E (yellow), S0 (red), Sab (green) and Scd (dark blue).}
\label{LUV-distr}
\end{figure}

We have derived the \Ha~ luminosities, \LHa, from the observed \Ha~emission line flux following equation \ref{eq:L}, by substituting the FUV fluxes by the \Ha~fluxes. Another significant difference is that the \Ha~fluxes are measured inside the fiber, with a given aperture width, and they need to be corrected by the aperture effect (not all the \Ha~ flux is detected inside the fiber) using the aperture corrections explained in Sect. \ref{sect:aperture}.  The resulting \LHa~distributions, for the different fields, galaxy types and morphologies can be seen in Fig. \ref{LHa-distr}. As for the \Luv~ distributions, SF and late type galaxies show the largest \LHa~values, while the unclassified and E galaxies show lower \LHa.

\begin{figure}
 \centering
\includegraphics[scale=0.45]{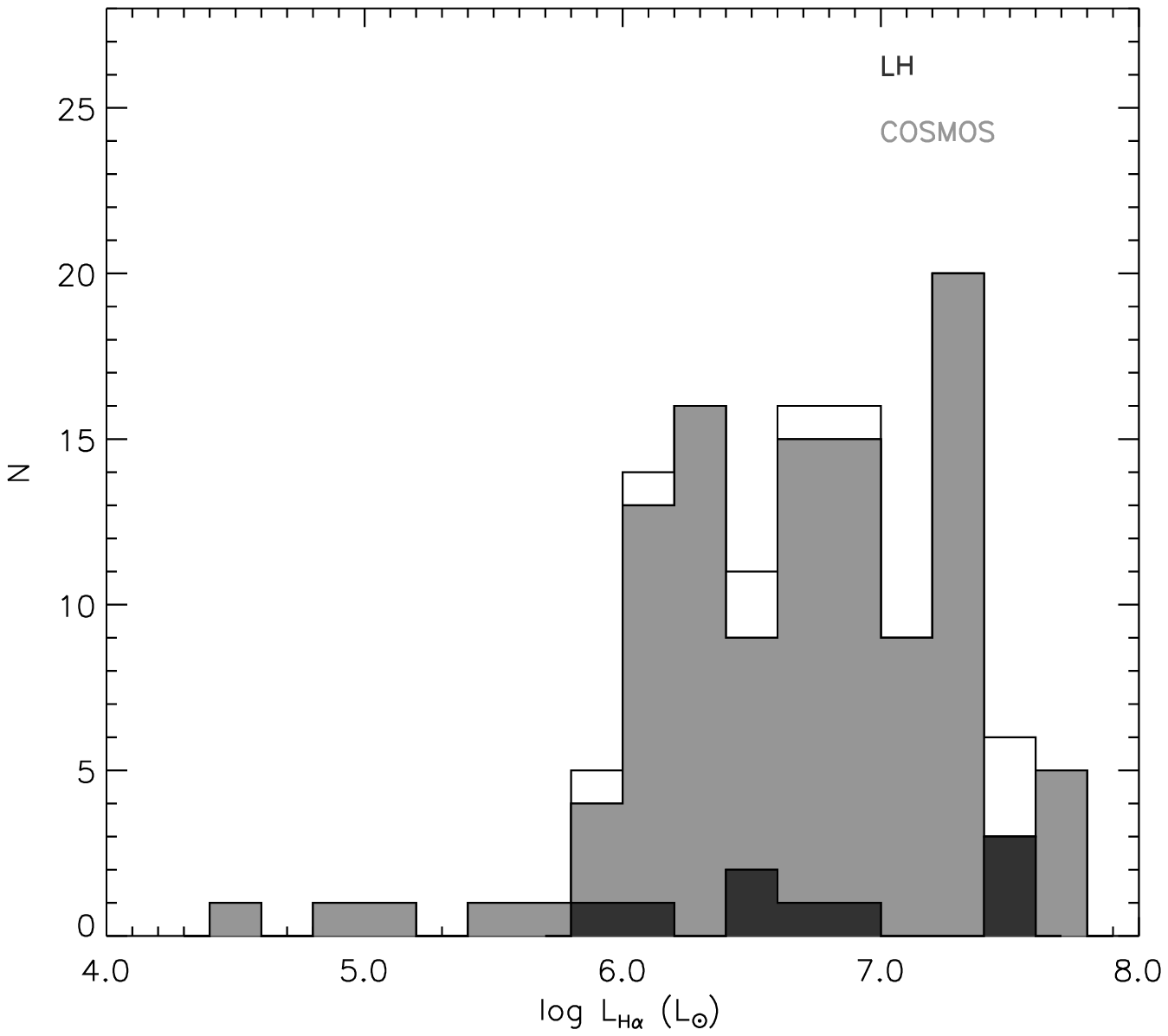}
\includegraphics[scale=0.55]{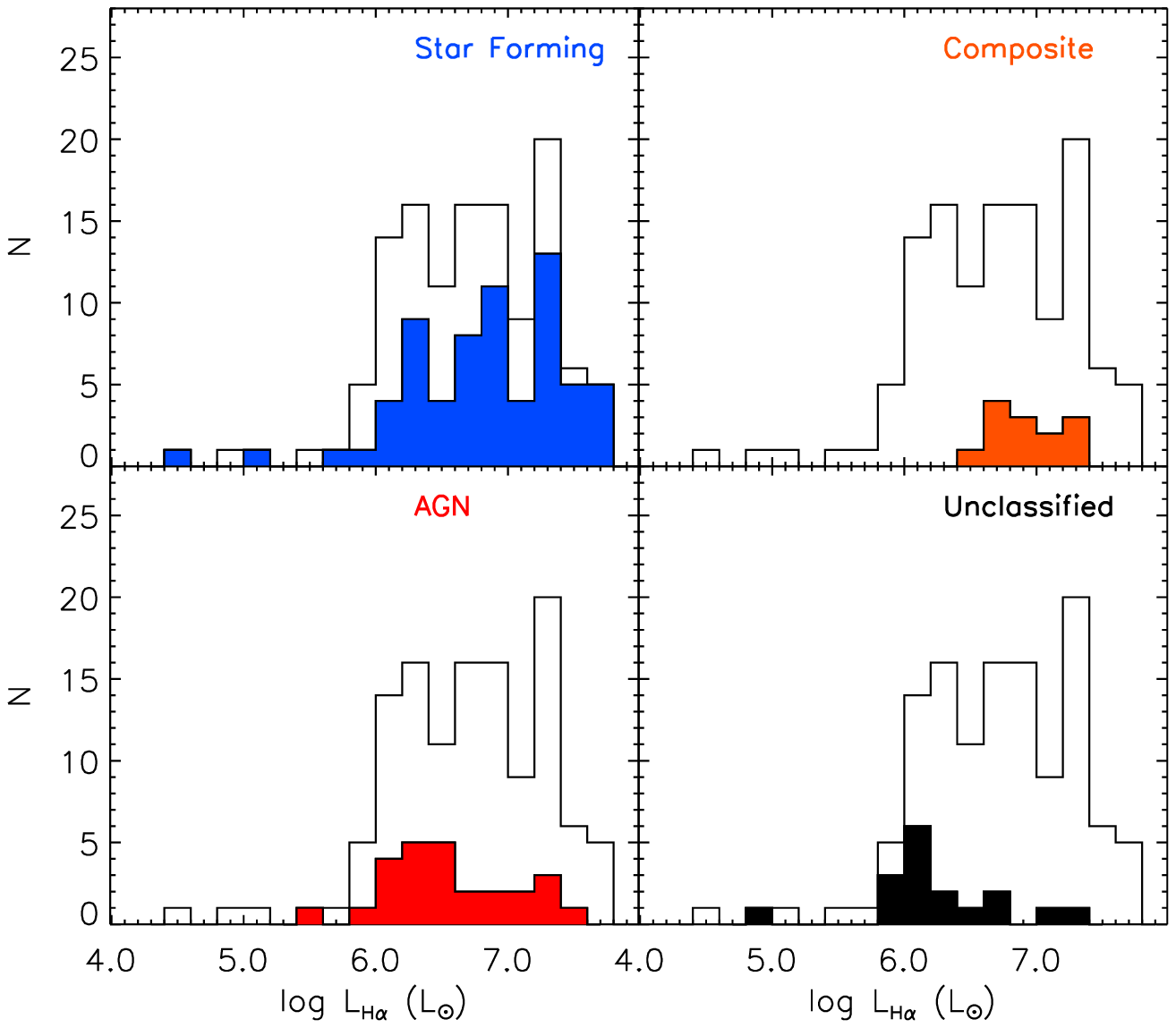}
\includegraphics[scale=0.55]{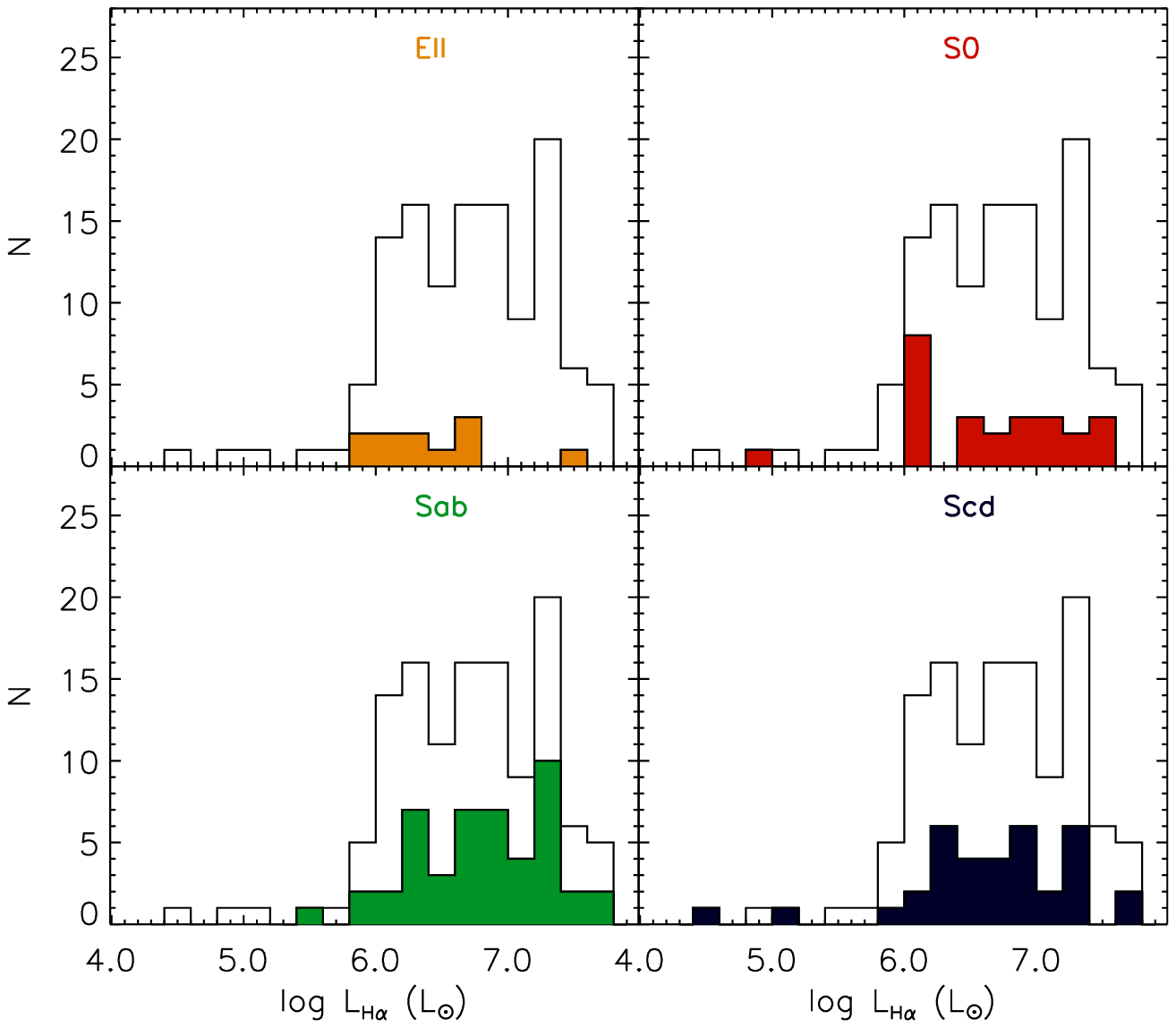}
\caption{\LHa~ distributions, colors as in Fig.  \ref{LUV-distr}.}
\label{LHa-distr}
\end{figure}

\hfill \\

\item{\textbf{Infrared Luminosities}} \hfill \\

\begin{figure}
 \centering
\includegraphics[scale=0.4]{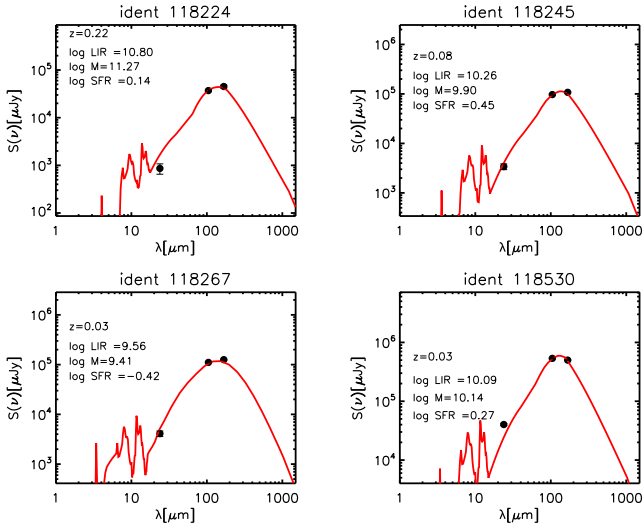}
\caption{Example of FIR SED-fitting to derive the \LIR. The observed data are plotted as black dots, while the best-fitting template is the red thick line. Also shown the redshift, the \LIR~(derived in this work), the masses and SFR (from the SDSS-DR7). }
\label{SEDs}
\end{figure}

\end{description}

\begin{figure}
 \centering
\includegraphics[scale=0.45]{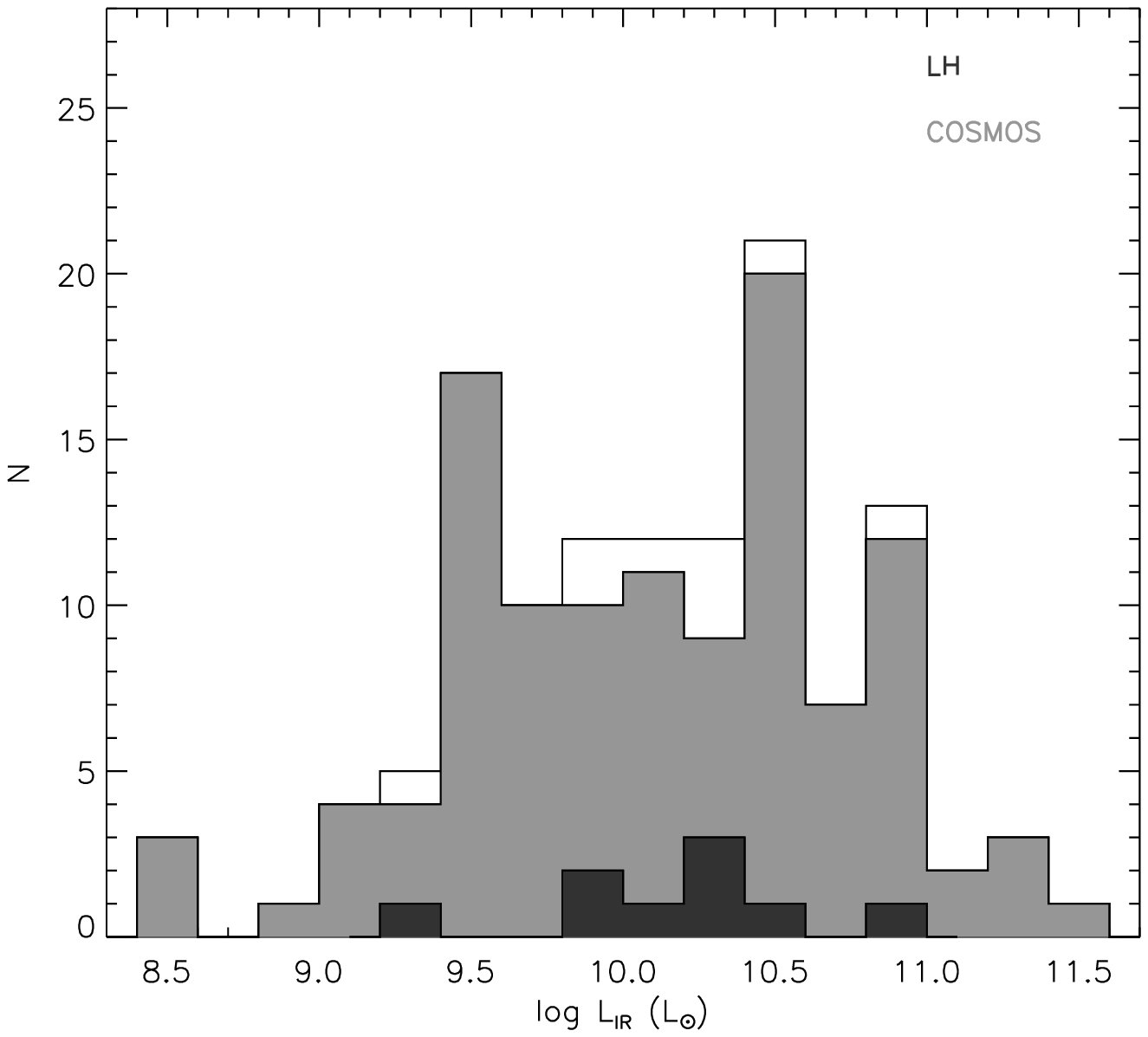}
\includegraphics[scale=0.55]{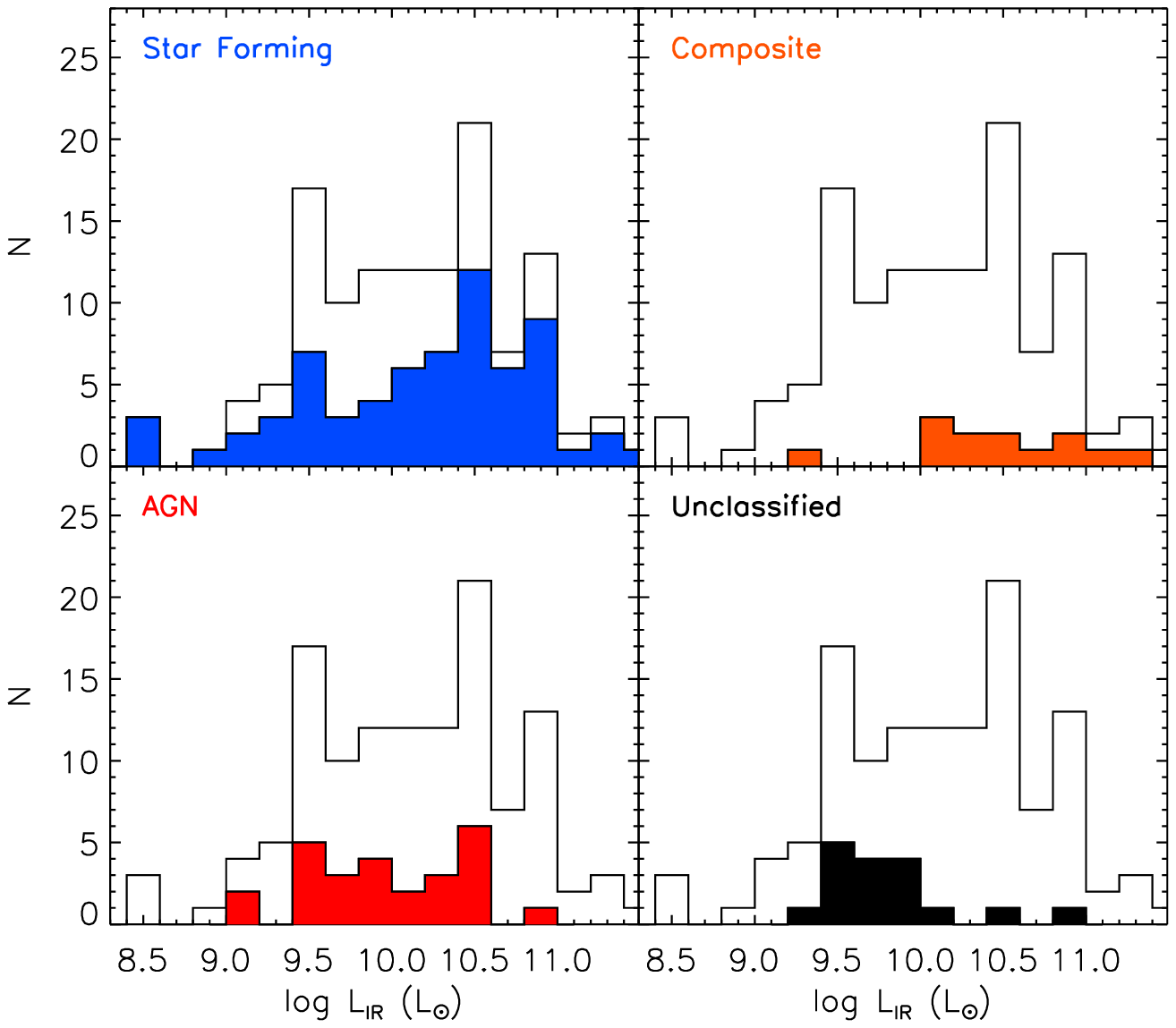}
\includegraphics[scale=0.55]{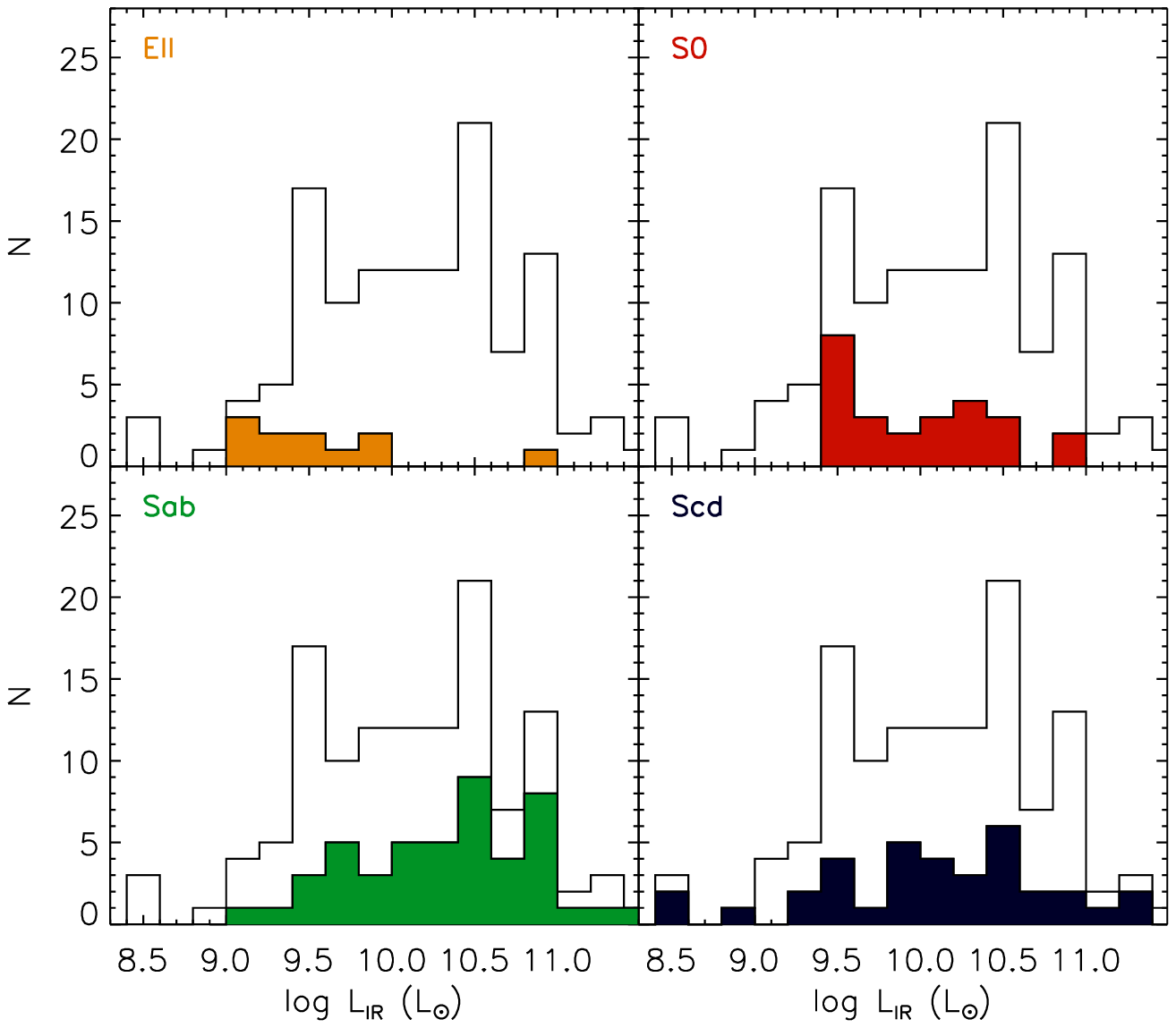}
\caption{\LIR~distribution for the whole sample of galaxies (empty histogram) and for the galaxies belonging to the different fields considered (\textit{upper panel}), spectral types (\textit{central panel}) and morphological types (\textit{bottom panel}). Colours as in Fig. \ref{LUV-distr}.}
\label{LIR-distr}
\end{figure}

The infrared luminosity, \LIR,  is defined as the integral of the luminosity in the 8-1000 $\mu$m range.  To estimate the \LIR~of our galaxies  we have performed SED-fitting from 24 to 160 $\mu$m using the \textit{LePhare} \footnote{http://www.cfht.hawaii.edu/~arnouts/lephare.ht} code (\citealt{Arnouts1999}, \citealt{Ilbert2006}). The code calculates the \LIR~by integrating each best-fit template between 8 and 1000 $\mu$m. The code directly gives as an output the \LIR. We fit our data with the \citet{CE2001} templates (CE01), which have been widely used in the literature to estimate \LIR~(\citealt{Nordon2010, Nordon2012, Berta2013, Oteob2013}, \citealt{Oteoa2013}). The combination of CE01 templates with \textit{Herschel} FIR data provide reliable determinations of the total IR luminosity over a wide range of redshifts \citep{Elbaz2010,Elbaz2011}.

Four  examples of SED fitting can be seen in Fig. \ref{SEDs}, with the derived \LIR~and the redshift of each galaxy. The log \LIR~ values range from 8.4 to 11.4 \Lsun. There are 6 LIRGs (Luminous Infrared Galaxies, log \LIR~$>$ 11.0 \Lsun). This translates into a number density  of LIRGs, $\Phi = 10^{-3.67}$ Mpc$^{-3}$, which, given the scarce statistics,  is in agreement with the expected number density of  LIRGs from the latest IR Luminosity Function estimates \citep{Gruppioni2013}, $\Phi=10^{-3.46}$ Mpc$^{-3}$. In Fig. \ref{LIR-distr}  we show the \LIR~distribution for galaxies belonging to the COSMOS or LH field. The reduced statistics of the LH (only 9 galaxies) does not allow to confirm that the  \LIR~and \Luv~ distributions are the same for 
the two fields. However, it does not seem to be a preference in the luminosity distribution for the LH galaxies with respect to the COSMOS ones, but the values rather span over the whole luminosities range. Therefore, in the next sections we will not differentiate between galaxies from COSMOS or LH, but we will consider the two fields together.

In Fig. \ref{LIR-distr} we also show  the \LIR~distribution divided into the different spectral types  and morphological types. It can be observed that the SF and composite galaxies present the largest values of \LIR, while the lowest \LIR~region is mostly occupied by the unclassifiable galaxies (as it happened for the \Luv~ and \LHa). This is consistent with the SFR-\LIR~relation, where the most SF galaxies show larger values of \LIR~as these two quantities are very strongly related. On the other hand, the unclassifiable galaxies are related to weak emission lines, which is also associated with lower SF values and, therefore, \LIR. If we focus on the morphological types, the  \LIR~distribution is not so different for the different morphological types. Still, the lowest  \LIR~values belong to E galaxies, which usually present non or almost quenched star-formation, and therefore low FIR emission, while the largest \LIR~values belong to the Sab and Scd galaxies, i.e., late-type galaxies, where the SFR is supposed to be 
more efficient. 

\begin{figure*}
 \centering
\includegraphics[scale=0.55]{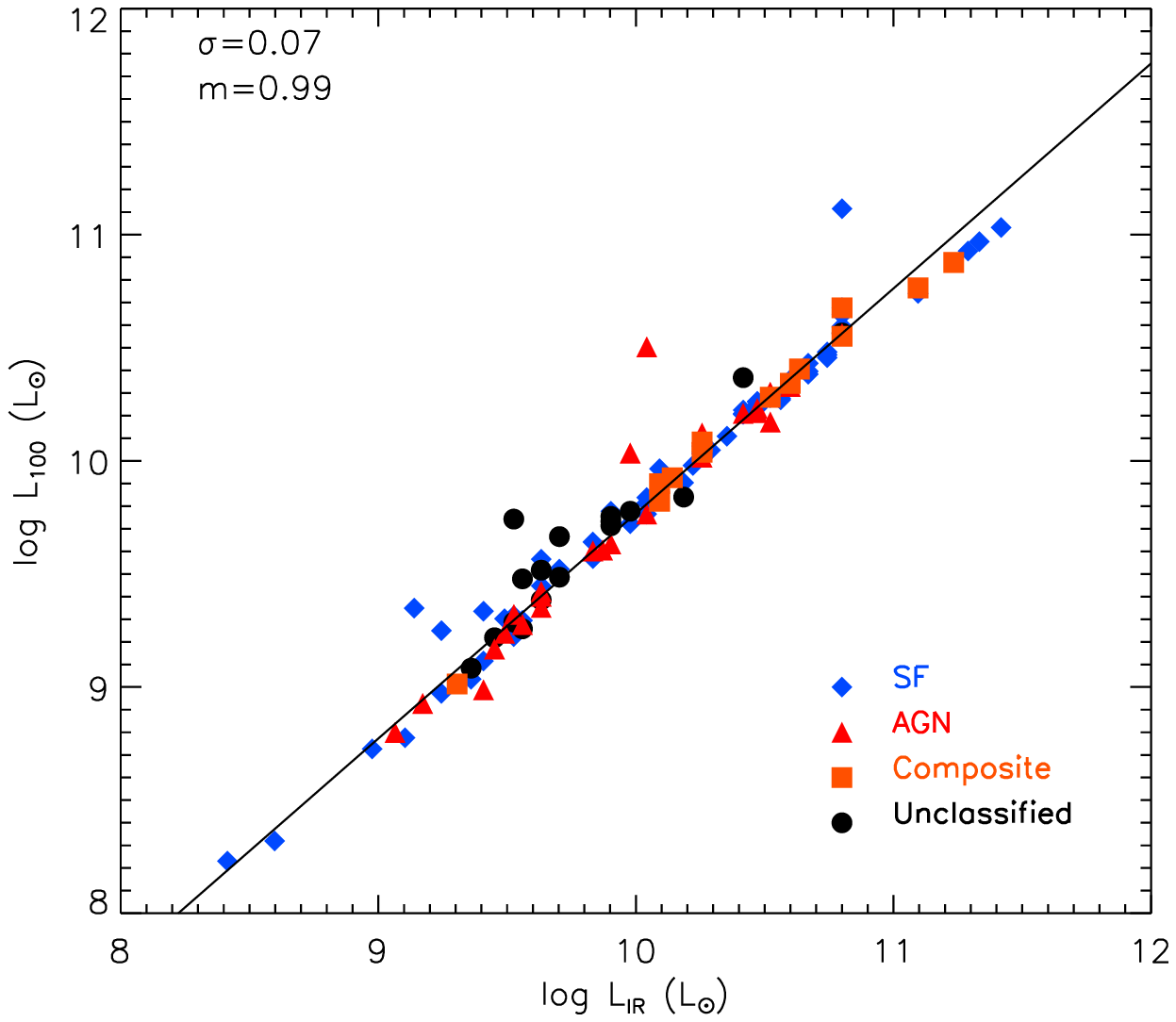}
\includegraphics[scale=0.55]{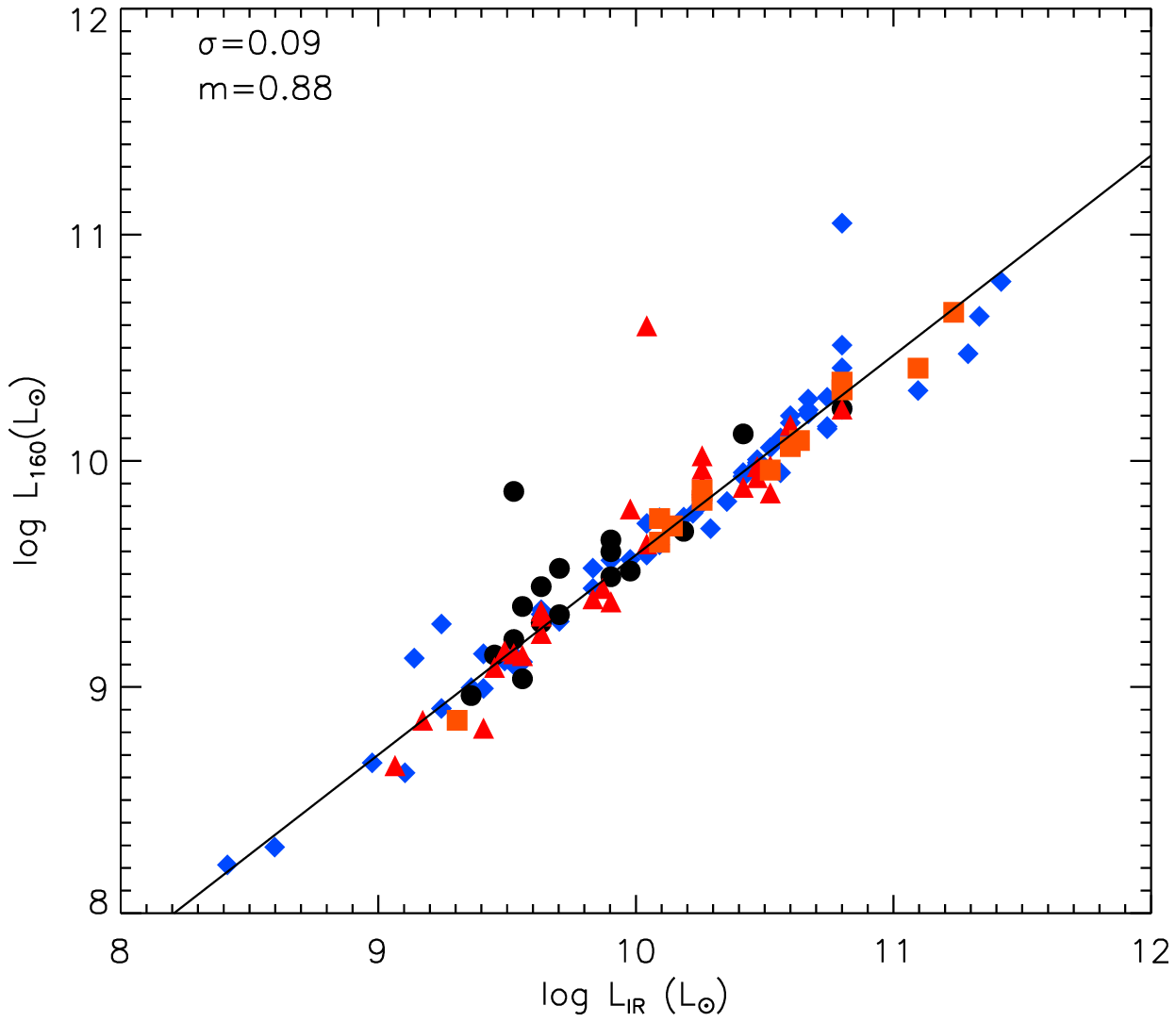}

\caption{Correlation between the total \LIR~and $\Lc$ and $\Lcs$. The solid line represents the best-fit to the data. Colours and symbols explained in the legend. Also shown the best-fit slope (m) and the dispersion ($\sigma$) values. }
\label{L_LIR}
\end{figure*}

We also derive the $\Lc$ and $\Lcs$ from the 100 and 160 $\mu$m fluxes (as in Eq. \ref{eq:L}). There is a very tight correlation between the total \LIR~and both $\Lc$ and $\Lcs$ for all the galaxy types, as shown in Fig. \ref{L_LIR}. We obtain slopes of $m_{100}=0.99$ and $m_{160}=0.88$ and dispersion values of $\sigma_{100}=0.07$ and $\sigma_{160}=0.09$. This implies that, at the redshifts considered,  both $\Lc$ and $\Lcs$ are a very good representation of the total \LIR, since these wavelengths sample the FIR maximum. The \LIR-$\Lc$ relation is a better \LIR~proxy at this z, presenting only minor deviation from the straight line and a small scatter.

\hfill \\

\section{Star Forming and unclassifiable galaxies}
\label{sect:SF}

One of the main purposes of this work is to compare different SFR indicators such as UV and \Ha~ luminosities with the total SFR derived from the obscured plus unobscured SFR (UV+IR). The general prescriptions to convert luminosities into SFR, as well as the dust extinction estimates have been derived for SF galaxies. The contribution of the AGN component to the UV, \Ha~ and IR fluxes is difficult to be taken into account, which complicates the SFR derivation for galaxies with nuclear activity. Therefore, in the next two sections we divided our sample into SF/unclassifiable galaxies and AGN/composites and we treat them separately. For the unclassifiable galaxies it was not possible to assign them a spectral classification based on their emission lines, but they have been morphologically classified.  Around 65 $\%$  (11/17) of the unclassifiable galaxies present early-type 
morphologies (E/S0), while only  13 $\%$  of the SF galaxies are morphologically classified as ETGs (3 E and 6 S0). The main problem when studying ETGs is that their UV and FIR emissions are not obviously related to young stars. The FIR in these galaxies  may be associated to the heating of the older stellar population (e.g. Calzetti 2012) or old metal-rich stars can contribute in the NUV band \citep{Donas2007}. This means that the SFRs based on FIR or UV luminosities, as well as dust extinction corrections, may not be appropiate and should be taken with care. However, we include the unclassified galaxies together with the SF to study how much these galaxies deviate from  the general SF galaxies'  trends. We have visually inspected the unclassifiable galaxies images, and they do not show particularly problematic photometry or strange shapes.  Anyway, all of the relations shown in the following sections have been derived for the SF sample only, so they should not be strongly  affected by problems related to ETGs.

\subsection{Dust extinction derivation}
\label{sect:dust}

One of the main problems that one has to overcome when deriving SFRs from luminosities in the UV or optical wavelengths is the effect of dust extinction. It is well known that the dust molecules and grains present in the galaxies absorb part of the light emitted by the stars. There are many empirical extinction laws derived for the Milky Way, the Large or the Small Magellanic Clouds  (e.g., \citealt{Allen1976,Prevot1984,Fitzpatrick1985}, respectively). In this work we will use the \citet{Calzetti2000} extinction law derived for starburst galaxies, which is widely used in literature and will allow to compare our results more easily. The dust extinction is more effective at lower wavelengths, meaning that when using \LHa~ or \Luv~ to calculate SFRs we will have to apply the dust correction.

The corrected flux at wavelength $\lambda$ can be written as:

\begin{equation}
 F_{corr}(\lambda)=F_{obs}(\lambda) \times 10^{[0.4 \times A(\lambda)]}=F_{obs} \times 10^{[0.4 \times K(\lambda)E(B-V)]}
\end{equation}

where $k(\lambda)\equiv A_{\lambda}/E(B-V)\equiv R_V A_{\lambda}/A_V$, $A_{\lambda}$ is the change in magnitude at wavelength $\lambda$ due to the extinction, E(B-V)=$A_B-A_V$ is the color excess between the B and V bands and $R_V$ is defined as $A_V/E(B-V)$. 
\\

\begin{itemize}
 \item Dust extinction from \Ha~/\Hb~
 
A method to measure the extinction is to compare the observed ratio of the H${\alpha}$ and H$\beta$ emission lines with the theoretical value ($R_{th}$=H$\alpha$/H$\beta$=2.87, for SF galaxies, assuming case B recombination and a electronic temperature T$_e$=1000 K; \citealt{Osterbrock&Ferland2006}). The reddening towards the nebular regions E(B-V) can be estimated following the \cite{Calzetti2000} extinction law (see  e.g., \citealt{DominguezSanchez2012} for details, DS2012 hereafter). We will refer to this dust extinction as E(B-V)$_R$.\\

\item Dust extinction from \LIR/\Luv

Another dust attenuation estimator is the ratio of IR to UV emission. This method is built on an energy budget consideration (e.g., \citealt{Meurer1999}, \citealt{Buat1999}). According to the energy balance argument, all the starlight absorbed at UV and optical wavebands by the interstellar dust is re-emitted in the IR, so the combination of the \LIR~and the observed \Luv~ should be able to probe the dust-free UV luminosity \citep{WH1996}. Following \cite{Hao2011}:

\begin{eqnarray}
A_{FUV}=2.5 log (1+ a_{FUV}\times10^{IRX})\\\nonumber
IRX=log(L_{IR}/L_{UV})\\\nonumber
\end{eqnarray}

where, $a_{FUV}=0.46$ for star-forming galaxies. This is the dust extinction towards the continuum, and it is usually lower than the extinction derived towards the nebular emission lines. We will refer to this dust extinction as E(B-V)$_{IRX}$.\\

\item Dust attenuation from the UV slope

The dust attenuation is more effective at shorter wavelengths than at larger ones, which causes the reddening of the galaxy spectra. Therefore, the strength of the reddening and the absorption fraction of the UV are related. There are different proposed formulas to convert the UV slope into dust extinctions (e.g., \citealt{Meurer1999}, \citealt{Boissier2007}, \citealt{Overzier2011},\citealt{Wilkins2012}). We use the revised version of the Meurer et al. 1999 relation by \citet{Takeuchi2012}, which includes a better treatment of the UV aperture corrections.

\begin{equation}
 A_{UV}=3.06 + 1.58\times\beta
\end{equation}

where $\beta$ is the UV slope defined as the ratio between the FUV and NUV fluxes

\begin{equation}
 \beta=\frac{\log{f_{FUV}}-\log{f_{NUV}}}{\log{\lambda_{FUV}}-\log{\lambda_{NUV}}}
\end{equation}

with effective wavelengths $\lambda_{FUV}$=1520 $\AA{}$ and $\lambda_{NUV}$=2310 $\AA{}$. This is the dust extinction towards the stellar continuum and we will refer to this dust extinction as E(B-V)$_\beta$. 

\end{itemize}

\subsection{Dust extinction comparison}
\label{sect:dust-comparison}

\begin{figure}
 \centering
\includegraphics[scale=0.55]{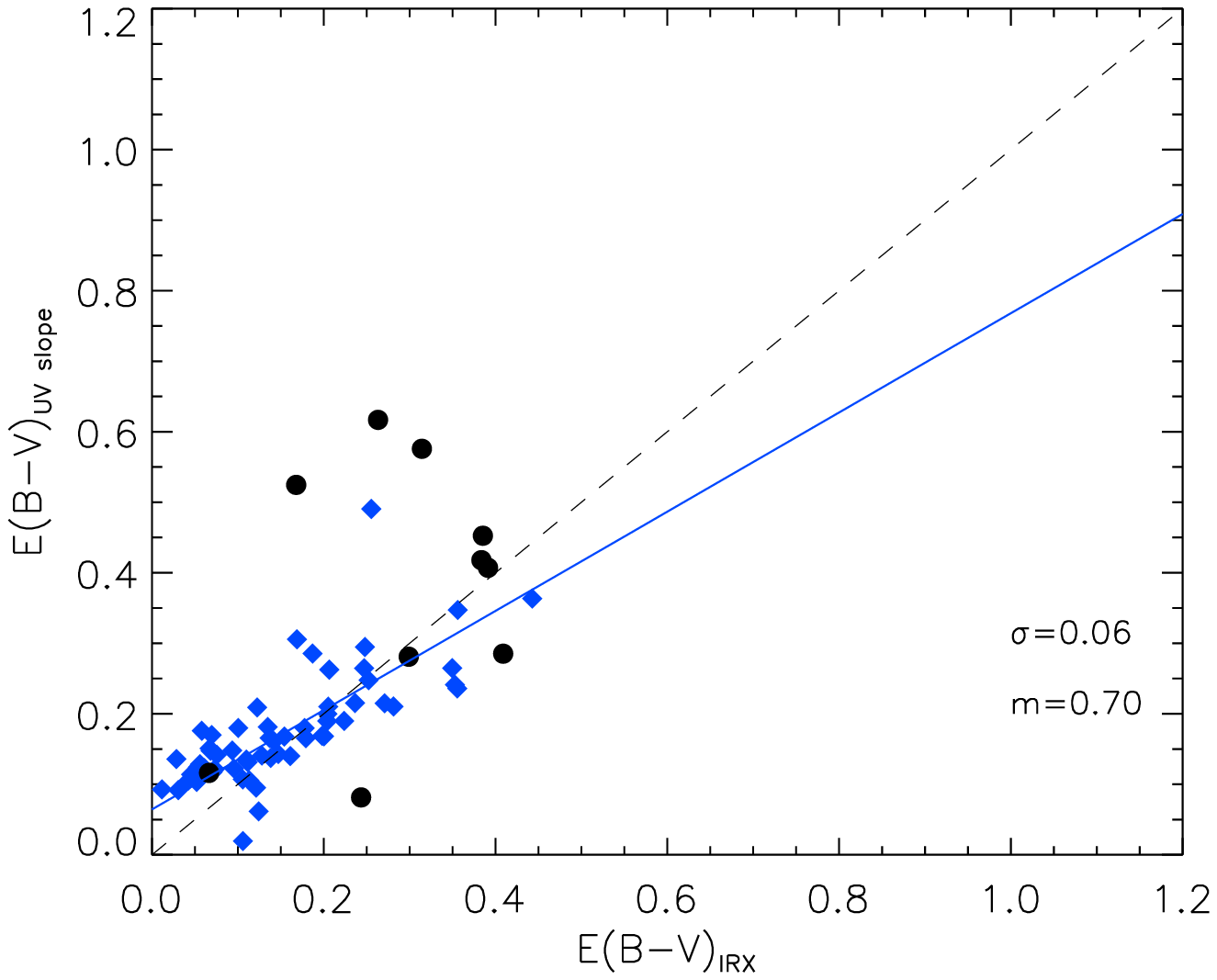}
\includegraphics[scale=0.55]{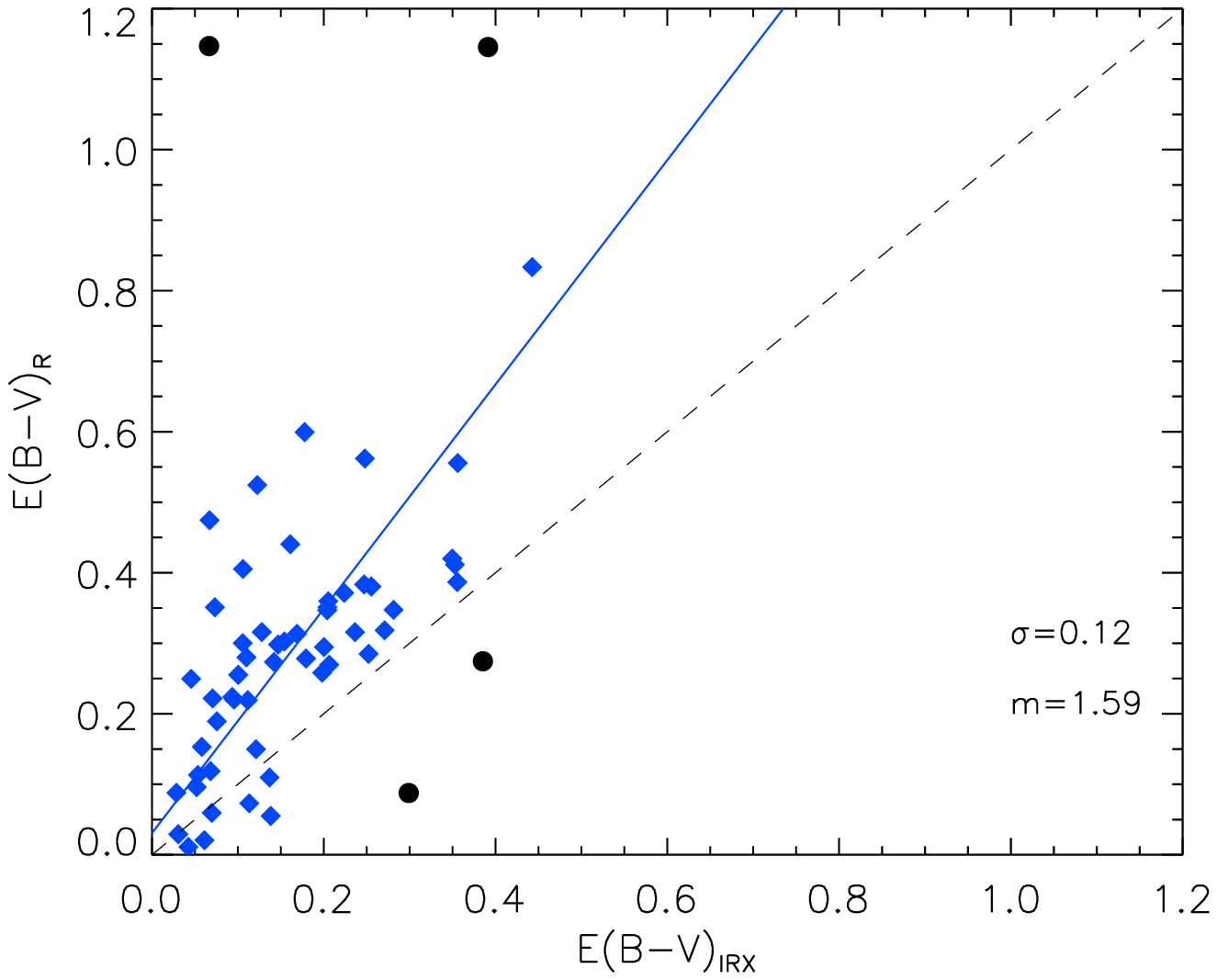}
\caption{\textit{Upper panel:} E(B-V)$_{IRX}$ versus  E(B-V)$_{\beta}$ for the SF (blue diamonds) and unclassifiable galaxies (black circles). The dashed line is the one to one relation and the thick blue line is the best-fit to the SF sample. Also shown the best-fit slope  (m) and dispersion ($\sigma$) for the SF galaxies. \textit{Lower panel:} Comparison of  E(B-V)$_{IRX}$ and E(B-V)$_{R}$; colours and symbols as in the upper panel.}
\label{EBV_IRX}
\end{figure}

In Fig. \ref{EBV_IRX} we plot the comparison of the dust extinction estimates for the SF and unclassifiable galaxies. In the upper panel we plot E(B-V)$_{IRX}$ versus  E(B-V)$_{\beta}$. For the SF sample we obtain a slope of m=0.70 and a dispersion $\sigma$=0.06. The value of the slope is mainly driven by  the larger E(B-V)$_{\beta}$ than E(B-V)$_{IRX}$ at low extinctions. In fact, $\sim$ 70 $\%$ of the galaxies have E(B-V)$_{IRX}$ $<$ E(B-V)$_{\beta}$, with the largest differences for the lowest E(B-V) values. The mean difference is 0.09, i.e., $\sim$ 40 $\%$  of the mean E(B-V) value. Other extinction recipes based on the UV slope such as \citealt{Wilkins2012, Meurer1999} would yield even higher dust extinction values than Takeuchi et al. 2012. The unclassifiable galaxies clearly show a stronger deviation from the one to one relation than the SF. 

The deviation is even larger when we compare 
E(B-V)$_{IRX}$ with E(B-V)$_{R}$, this is related to the fact that the unclassifiable galaxies have the lowest S/N in the \Ha~ and \Hb~ emission lines. In fact, the galaxies with E(B-V)$_{R}$ $\sim$ 1.2 have S/N$_{H\alpha}$=2.0 and 0.8 and S/N$_{H\beta}$=0.3 and 21, respectively. Unfortunately, there are no unclassifiable galaxies with S/N $>$ 3 in both \Ha~ and \Hb~lines and with valid extinction values from IRX . This means that E(B-V)$_{R}$ for the unclassifiable galaxies are not reliable. We think it is interesting to include them in the plot to show the different trend with respect to the SF galaxies (but recall that the relations derived hereafter are for the SF and late type galaxies only). In this case, the slope for the SF galaxies is m=1.59 and the dispersion is larger than for the comparison with E(B-V)$_{\beta}$ , $\sigma$=0.12. It is reasonable that E(B-V)$_{R}$ is larger than E(B-V)$_{IRX}$ because the extinction towards the emission lines is higher than the extinction towards the continuum. \citet{Calzetti2000} find a conversion factor of 0.44 when converting the extinction from the UV to extinction towards the nebular lines from the  \Ha/\Hb~ratio. This is in excellent agreement with what  we obtain for our sample of SF galaxies (0.70/1.59=0.44).

It is always useful to find galaxy properties that correlate with their intrinsic dust extinction, as it may help to derive fundamental relations that could provide extinction measurement approximations when there are not available quantities to calculate the extinction values (e.g. no emission lines or IR information are available). In DS2012 a tight correlation was obtained between the \LIR~ and the E(B-V) values from the measurement of the \Ha/\Hb~ratio from stacked spectra divided into different \LIR~bins. The DS2012 sample consisted on 474 galaxies with 0.06 $<$ z $<$ 0.46, mean log M=10.46 \Msun~ and mean log \LIR=9.95 \Lsun, similar to the properties of the sample in this work.

\begin{figure}
 \centering
\includegraphics[scale=0.55]{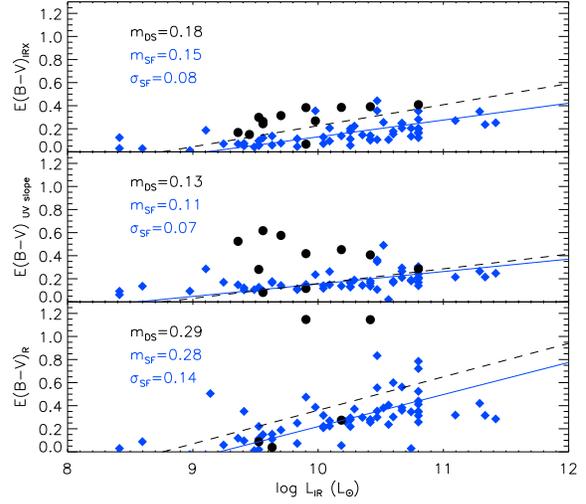}
\caption{\textit{Upper panel:} E(B-V)$_{IRX}$ versus \LIR~for the SF (blue diamonds) and the unclassifiable samples (black circles). The blue thick line is the best-fitting slope to the SF sample, while the black dashed line is the relation obtained by DS2012. We show the same for the  E(B-V)$_{\beta}$ (\textit{middle panel})  and E(B-V)$_{R}$ (\textit{bottom panel}). m$_{DS}$ are the slopes obtained in DS2012 for the E(B-V)-\LIR~relation (translated into the different E(B-V) using the relations from Fig. \ref{EBV_IRX}), while m$_{SF}$ and $\sigma_{SF}$ are the best-fitting slopes and dispersions for the SF sample.}
\label{ext-lir}
\end{figure}

In Fig. \ref{ext-lir} we show  our derived E(B-V)$_{IRX}$, E(B-V)$_{\beta}$ and E(B-V)$_{R}$  versus the \LIR~of each galaxy. We also show the best-fitting slopes for the SF sample, as well as the relation obtained by DS2012. In the upper and middle panel  we have translated the relation obtained by DS2012 into extinction towards the continuum using  the conversion factors from the relations obtained in Fig. \ref{EBV_IRX}. There is a very good agreement between the relation derived by DS2012 and the  E(B-V)-\LIR~relation for the  SF galaxies with the three dust extinctions considered. In particular, the agreement between the relation E(B-V)$_{\beta}$-\LIR~from DS2012 and that obtained in this work is excellent and it is also the one which shows a smaller scatter ($\sigma=0.07$), while the E(B-V)$_{R}$-\LIR~shows the largest dispersion ($\sigma=0.14$).

\begin{figure}
 \centering
\includegraphics[scale=0.55]{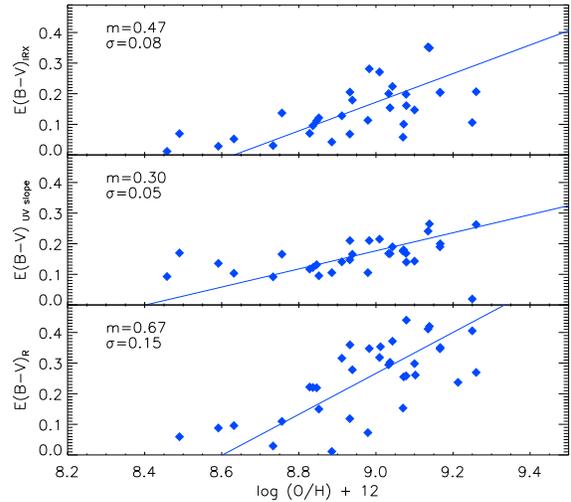}
\caption{\textit{Upper panel:} E(B-V)$_{IRX}$ versus metallicity for the SF (blue diamonds). The blue thick line is the best-fitting slope to the SF sample (m) and $\sigma$ is the dispersion. We show the same for the  E(B-V)$_{\beta}$ (\textit{middle panel})  and E(B-V)$_{R}$ (\textit{bottom panel}).}
\label{ext-met}
\end{figure}

	In Fig. \ref{ext-met} we show the extinction values as a function of the metallicity. In this case, we can only consider the high S/N SF sample (32 galaxies), as there are only accurate metallicity measurements for this galaxy type. There is a clear trend of increasing dust extinction for the more metal-rich galaxies, as already reported for nearby galaxies \citep{Heckman1998} and at z$\sim$1 and 2 \citep{Roseboom2012, Reddy2010}. We show in the plots the best-fitting slopes and dispersions. In this case, the E(B-V)$_{\beta}$-metallicity relation shows the smallest scatter value ($\sigma=0.05$), while the E(B-V)$_{R}$-\LIR~shows the largest dispersion ($\sigma=0.15$), as it happened for the E(B-V)-\LIR~relation.

\begin{figure}
 \centering
\includegraphics[scale=0.55]{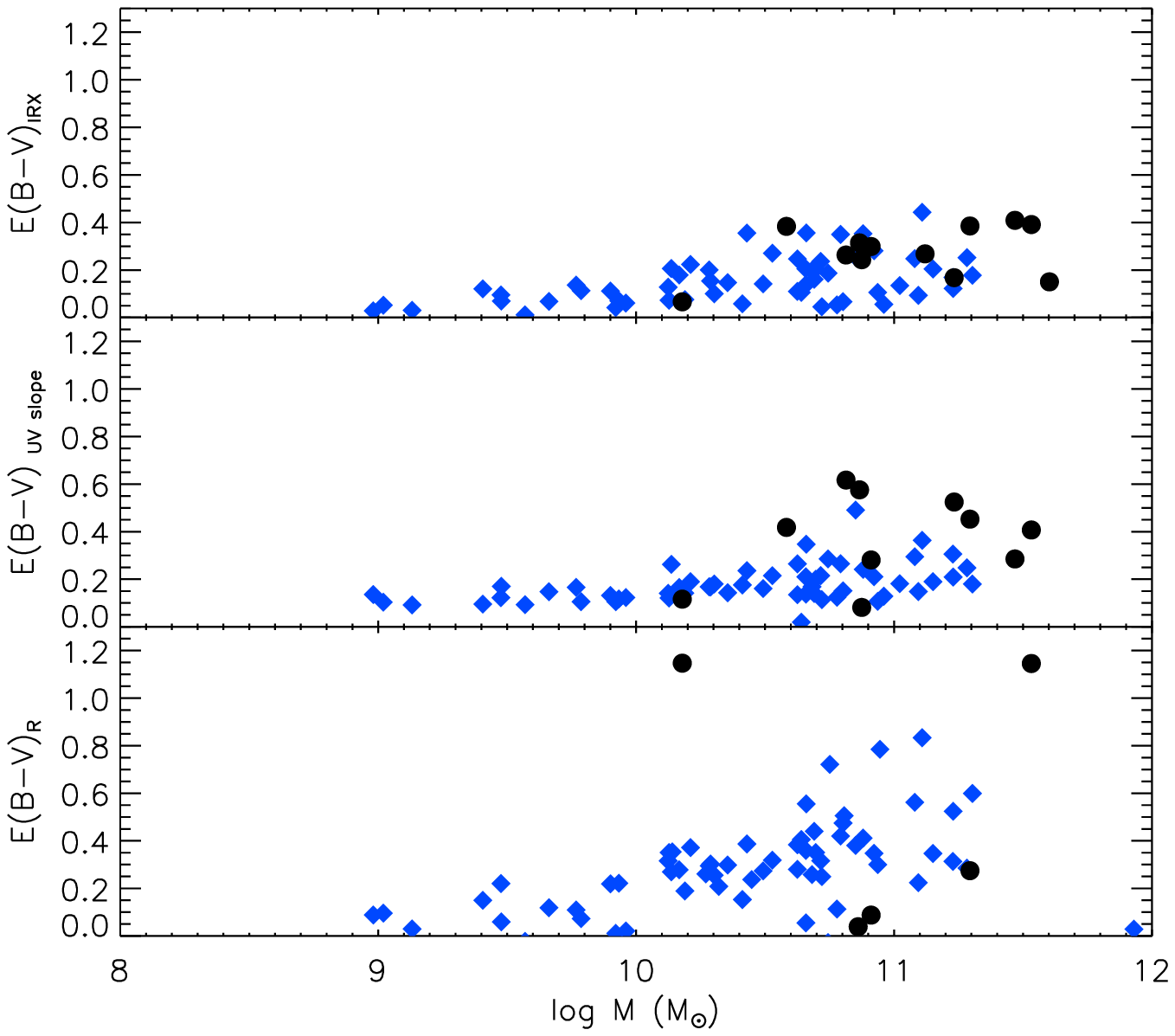}
\caption{Dust extinction values  as a function of the galaxy stellar mass. We plot E(B-V)$_{IRX}$ (\textit{upper panel}), E(B-V)$_{\beta}$ (\textit{middle panel})  and E(B-V)$_{R}$ (\textit{bottom panel}). Colour code as in Fig.\ref{ext-lir}.}
\label{ext-mass}
\end{figure}

Another quantity that may be related to the dust extinction is the galaxy stellar mass, as the most massive galaxies contain more dust mass and therefore present larger extinction values. In Fig. \ref{ext-mass} we show the three extinction values E(B-V) derived in this work versus the stellar masses. Although there is a trend between extinction and stellar mass (it can be observed that the lowest mass galaxies are also the ones that present lower extinctions), there is a very large dispersion at large mass values (log M $>$ 10 M$_\odot$), meaning that the dust extinction is not directly related to the stellar mass. This is an indication of the complexity of the galactic systems, that can be very dusty or not depending on their chemical composition, evolutionary stage or even the angle of sight.

\subsection{Star Formation Rate derivation}
\label{sect:SFR}

As already mentioned in the introduction, there are many methods to calculate the current SFR of a galaxy. In this work we have derived SFRs from the \Luv,\LHa~ and \LIR. The advantage of the \LIR~is that it is not affected by the dust extinction (as it is in fact, the dust emission itself), while when using luminosities in the UV or optical wavelengths an extinction correction must be applied (see Section \ref{sect:dust}). In this section we explain how we estimate the SFRs that will be used throughout the paper.

\begin{itemize}

\item \textbf{SFR from the \Luv~}\\

The SFR-\Luv~ calibration is based on the fact that the most massive stars, which emit most of their energy in the UV continuum, have lifetimes much shorter than the typical age of the galaxy. The UV part of the spectrum is dominated by young stars, so the SFR scales linearly with \Luv. The most widely used conversion comes from K98 (see reference for details on the synthesis model used to obtain this calibration):

\begin{equation}
SFR_{UV}(M_\odot ~ yr^{-1})=1.4\times10 ^{-28}L_{UV} (erg~ s^{-1}~Hz^{-1})
\end{equation}
\label{eq:SFR-UV}

In this case, the \Luv~ must be corrected for the dust extinction if one wants to obtain the total SFR, otherwise the SFR will only be representative of the unobscured star formation. We use the dust extinction derived from the UV slope, E(B-V)$_{\beta}$, and we will refer to it as $SFR_{UV}$.\\

 \item \textbf{SFR from the \LHa~}\\

The nebular lines are also a direct, sensitive probe of the young and massive stellar population and they are often used to estimate the SFR. The \Ha~ emission line is the best candidate due to its higher intensity in comparison with other lines (\Hb~, $P\beta$, $Br\alpha$ or $Br\gamma$). Besides, it is less affected by dust extinction than the UV and  presents lower stellar absorption than other  emission lines such as \Hb. The conversion from K98 is:

\begin{equation}
SFR_{H\alpha}(M_\odot yr^{-1})=7.9\times10 ^{-42}L(H\alpha) (erg~ s^{-1})
\end{equation}
\label{eq:SFR-Ha}

We use the aperture corrections described in Sect. \ref{sect:aperture} to recover the total SFR from the \LHa. Again, the \LHa~ must be corrected for the dust attenuation. We use the dust extinction derived from the \Ha/\Hb~ ratio, E(B-V)$_{R}$ and we will refer to it as SFR$_{H\alpha}$.\\

 \item \textbf{SFR from the SDSS analysis}\\

We will include in our analysis the SFR derived by the MPA-JHU group for the SDSS galaxies as explained in Section \ref{sect:data}. We will refer to it as SFR$_{SDSS}$. \\

 \item \textbf{SFR from the \LIR}\\

The \LIR~is a direct probe of the current SFR as the dust absorbs and re-emits the light of very bright and young OB stars. It is well known that the SFRs and \LIR~of galaxies are correlated and calibrations have been made to obtain SFRs from \LIR, with the most widely used being the \cite{Kennicutt1998} relation:

\begin{equation}
SFR_{IR}(M_\odot~ yr^{-1})=4.5\times10^{-44}L_{IR} (erg~s^{-1})
\label{eq:SFR-IR}
\end{equation}

As the \LIR~does not need to be corrected for dust extinction, we directly derive the SFR from the \LIR~by multiplying the \LIR~values by the K98 constant. We will refer to it as SFR$_{IR}$. Note that, as mentioned before,  the FIR emission for the ETGs is not obviously related to the heating young stars, so the SFR$_{IR}$ results for these galaxies should be taken with caution.\\

 \item \textbf{Total SFR}\\

 \begin{figure}
 \centering
\includegraphics[scale=0.55]{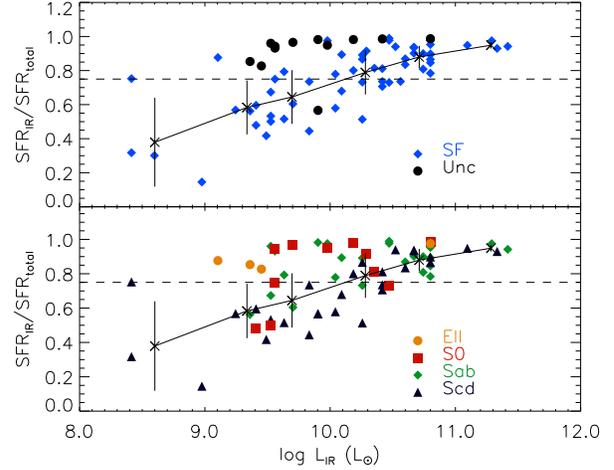}
\caption{\textit{Upper panel:} SFR$_{IR}$/SFR$_{total}$ versus the log \LIR~for the SF (blue diamonds) and the unclassifiable (black circles) galaxies.  Also shown the mean SFR$_{IR}$/SFR$_{total}$ for 6 \LIR~bins and their dispersions (error bars).  \textit{Bottom panel:} Same as in the upper panel, but galaxies have been color coded according to their morphological classification: E (yellow circles), S0 (red squares), Sab (green diamonds), Scd (dark blue triangles). }
\label{SFRIR/SFRtotal}
\end{figure}

Although the SFR$_{IR}$ is commonly used alone as a SFR indicator, it actually only accounts for the obscured SFR. One can derive the total SFR (obscured + unobscured) by adding the SFR derived from the UV, without applying any extinction correction, SFR$_{UV, unc}$. Then,  the total SFR becomes:

\end{itemize}

\begin{equation}
SFR_{total}=SFR_{UV,unc}+SFR_{IR}
\end{equation}
\label{eq:SFR-total}

Note that this is, by definition, equivalent to SFR$_{UV}$, if the FUV flux is corrected for dust extinction using the \Luv/\LIR~ ratio.

In Fig. \ref{SFRIR/SFRtotal} we show the ratio SFR$_{IR}$/SFR$_{total}$ versus the log \LIR~for the SF and unclassifiable galaxies. We also show the mean values of SFR$_{IR}$/SFR$_{total}$ in 6 \LIR~bins and their mean dispersion (error bars). The first thing that we observe is that the SFR$_{IR}$ accounts for more than half of SFR$_{total}$ for the majority of the galaxies (there are only 6 galaxies with SFR$_{IR}$/SFR$_{total}$ $<$ 0.5). There is also an increase of the SFR$_{IR}$ contribution to  SFR$_{total}$ with the \LIR~of the galaxies (e.g. \citealt{Oteo2012}). Only for galaxies  with log \LIR~$<$ 10 \Lsun~ does the SFR$_{IR}$ accounts for less than 75 $\%$ of the total SFR (dashed line), while for galaxies with log \LIR~$>$ 10.7 \Lsun~  the  SFR$_{IR}$ is $\gtrsim$ 90 $\%$. However, given the incompleteness of observed IR galaxies at log \LIR~$\lesssim$ 9.5 \Lsun~, we can only have firm conclusions at high \LIR.  A similar result was obtained by \
\citet{Buat2010} when analysing a sample of FIR and UV detected galaxies. The authors found that the SFR$_{IR}$ measures $\sim$ 94 $\%$ of the total SFR for 
galaxies with log \LIR~$>$ 11.5 \Lsun~ and $\sim$ 71 $\%$ for galaxies with log \LIR~$<$ 10 \Lsun. The unclassifiable galaxies show almost always the largest SFR$_{IR}$/SFR$_{total}$ ratios. For this galaxies, the unobscured SFR is almost negligible and the ratio of the two SFRs is independent of \LIR. In the lower panel of Fig. \ref{SFRIR/SFRtotal} we plot the same (i.e., the SF and unclassifiable galaxies), but  colour coded by their morphological classification. In general, the early type galaxies (E and S0) show the highest SFR$_{IR}$/SFR$_{total}$ ratios, while for the Scd galaxies the unobscured SFR seems to be more important and more than half of the Scd galaxies  (16/24) have SFR$_{IR}$/SFR$_{total}$ $<$ 0.75. This may imply that the later type galaxies have an important UV contribution that escapes from the 
dust attenuation. We suggest that caution may be taken when deriving SFR based only in the \LIR~for galaxies with low \LIR~values (log \LIR~$<$ 9.5 \Lsun~) or late type galaxies, as more than half of the SFR coming from the unextincted UV light may be missing.

\subsection{Star Formation Rate comparison}
\label{SFR-comparison}

\begin{table*}
  \centering
  \begin{tabular}{|cccc|}\hline
    \multicolumn{1}{c}{SFR$_{total}$ vs} &
     \multicolumn{1}{c}{SFR$_{SDSS}$} &
      \multicolumn{1}{c}{SFR$_{UV}$} &
       \multicolumn{1}{c}{SFR$_{H\alpha}$} 
     \\ \hline
     SF   & m=1.05, a=0.04 &   m=1.16 a=-0.01 &    m=1.11, a=0.28 \\
     Late & m=0.99, a=0.09 &   m=1.04, a=0.03  &    m=1.14, a=0.29  \\
    
 \hline
  \end{tabular}
  \caption{ Best-fitting slopes (m) and zero points (a) for the SF and late type galaxies when comparing the SFR$_{total}$ with the other studied SFR estimates. }
  \label{slopes}
\end{table*}

\begin{figure*}
 \centering
\includegraphics[scale=.8]{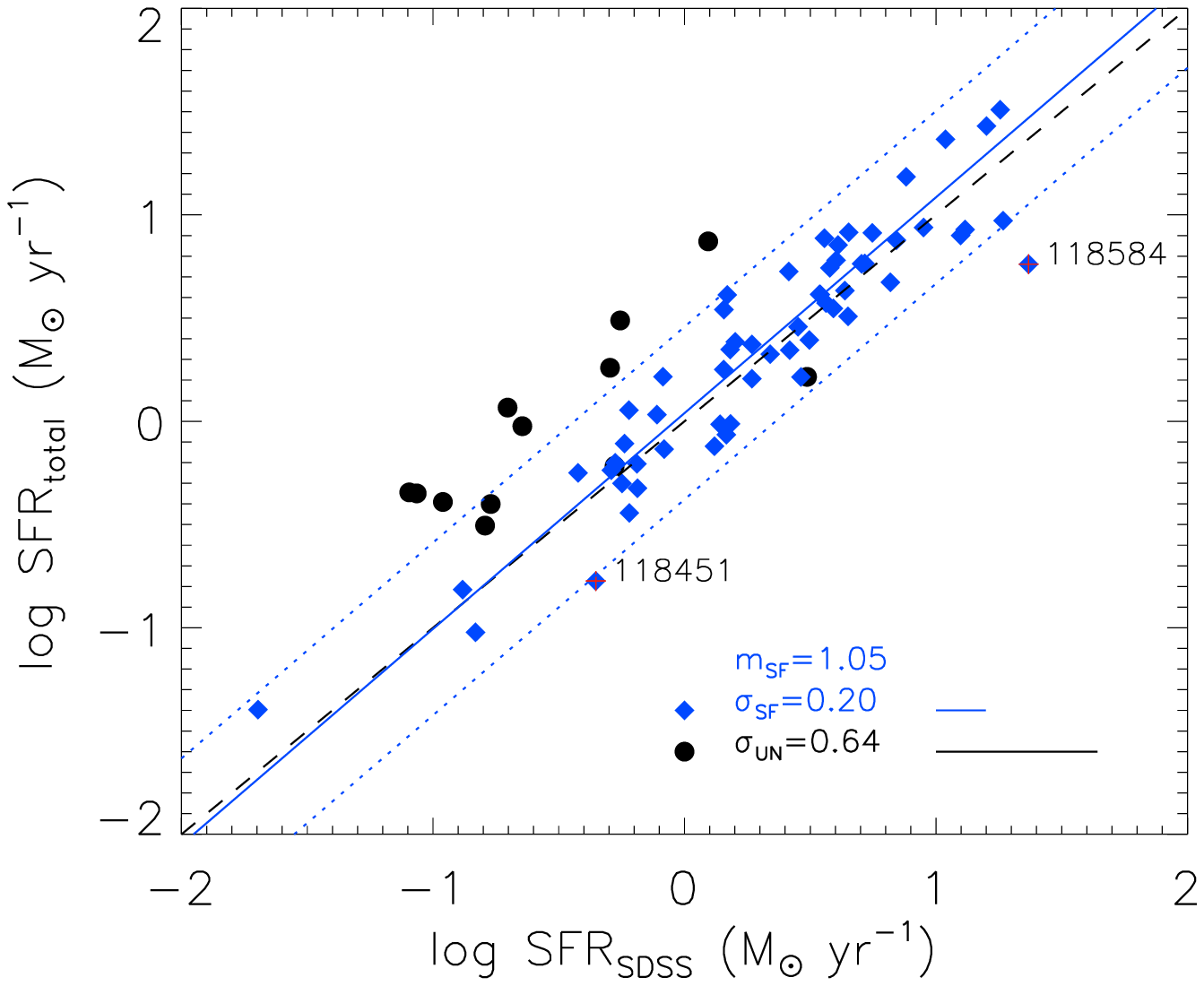}
\includegraphics[scale=.8]{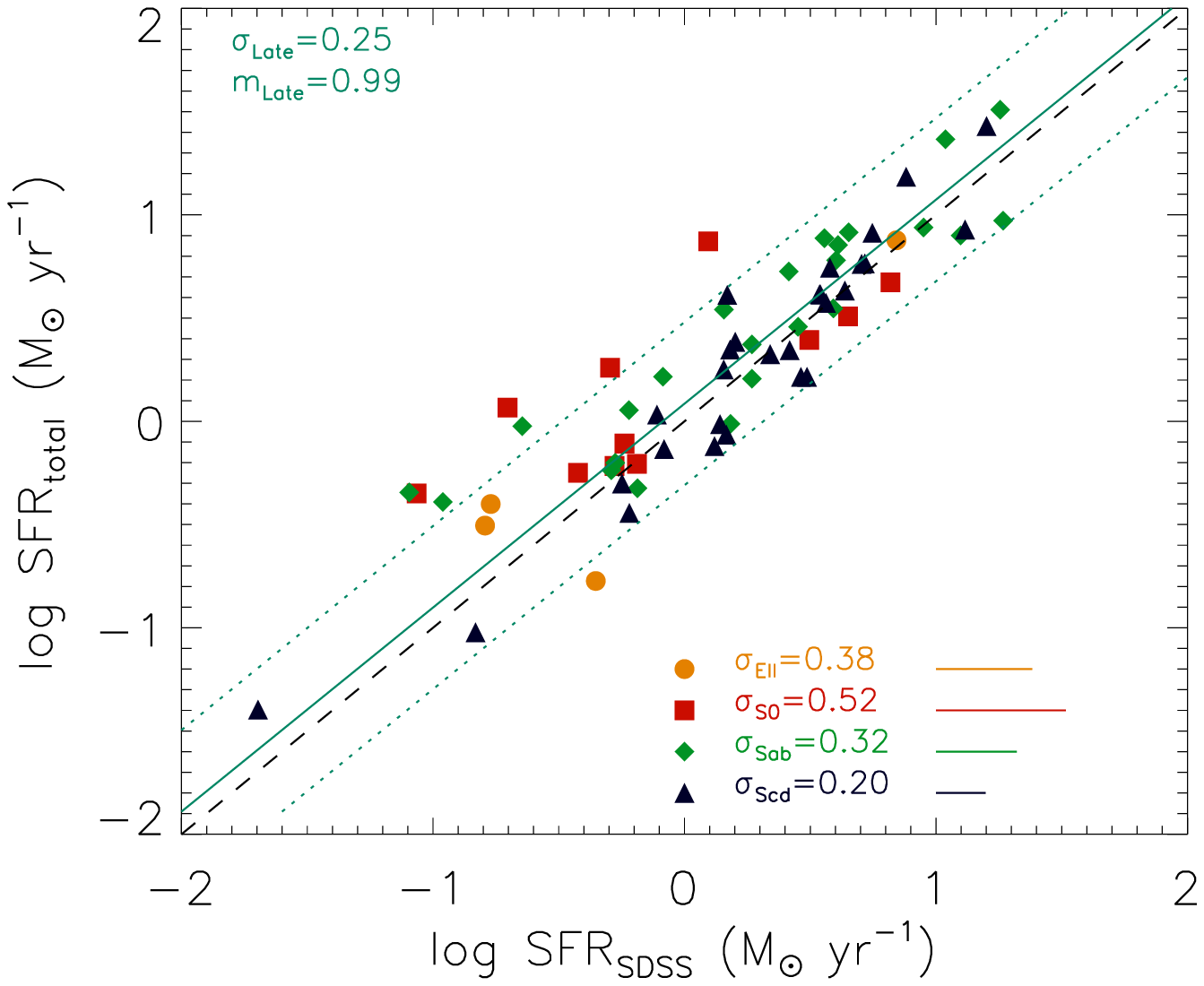}
\caption{\textit{Upper panel}: SFR$_{total}$=SFR$_{UV, uncorr}$+SFR$_{FIR}$ versus SFR derived by the MPA-JHU group for the SDSS DR7. The one to one relation is represented by the black dashed line, while the thick blue line is the best-fit to the SF sample (without including the outliers, see text).  The dotted lines represent typical uncertainties (0.4 dex, see text) around the best-fitting relation. The obtained slope and dispersion for the SF sample are also shown, as well as the dispersion values for the unclassifiable galaxies. \textit{Bottom panel}: Same as in the upper panel, the colour code represents the morphological classification. The dark green line is the best-fit to the late type galaxies (Sab and Scd). Colours and symbols explained in the legend. The galaxies highlighted by a red cross in the upper panel are further discussed in the appendix and are identified by their IDs. }
\label{total_SDSS}
\end{figure*}

\begin{figure}
 \centering
\includegraphics[scale=0.55]{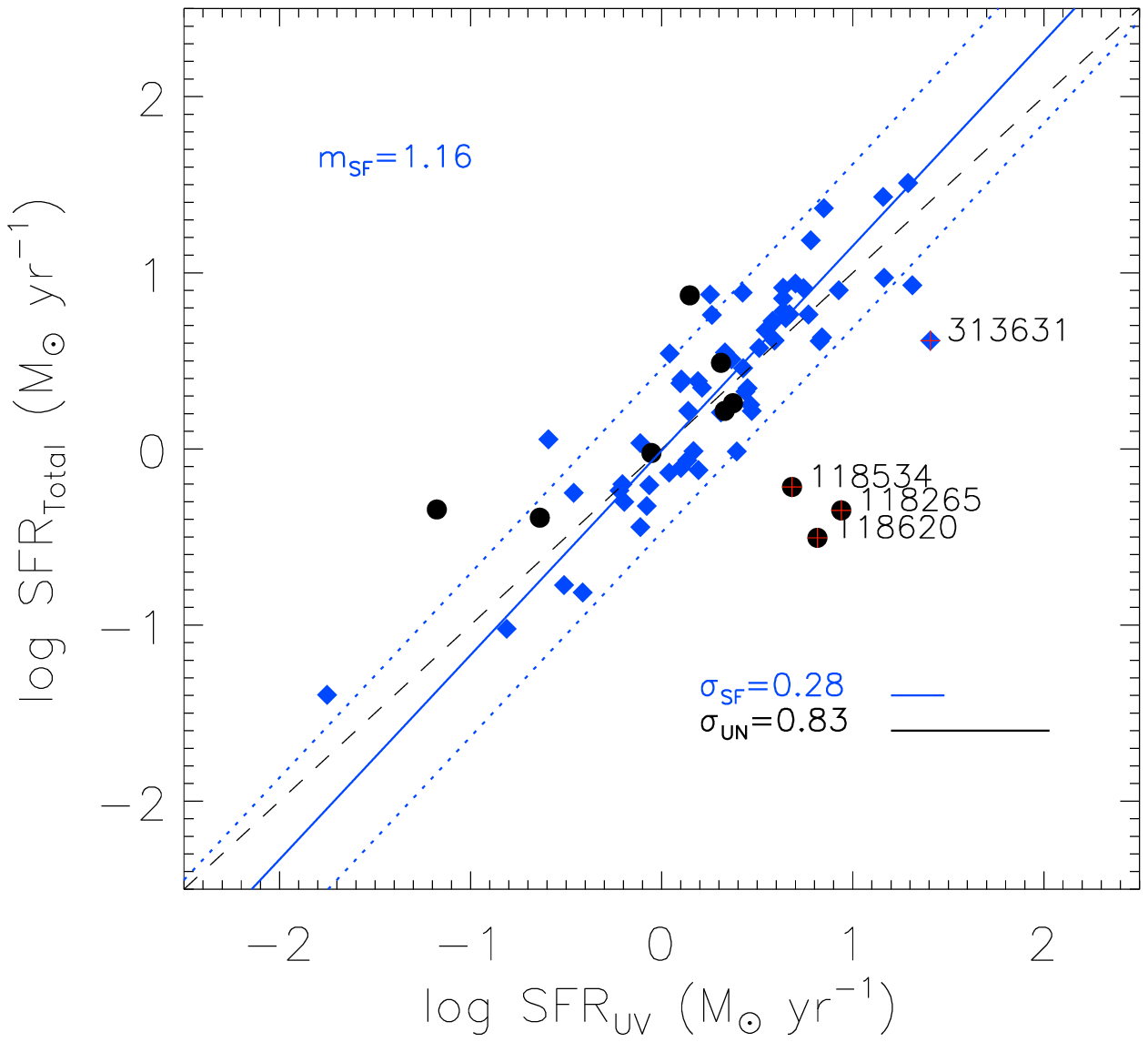}
\includegraphics[scale=0.55]{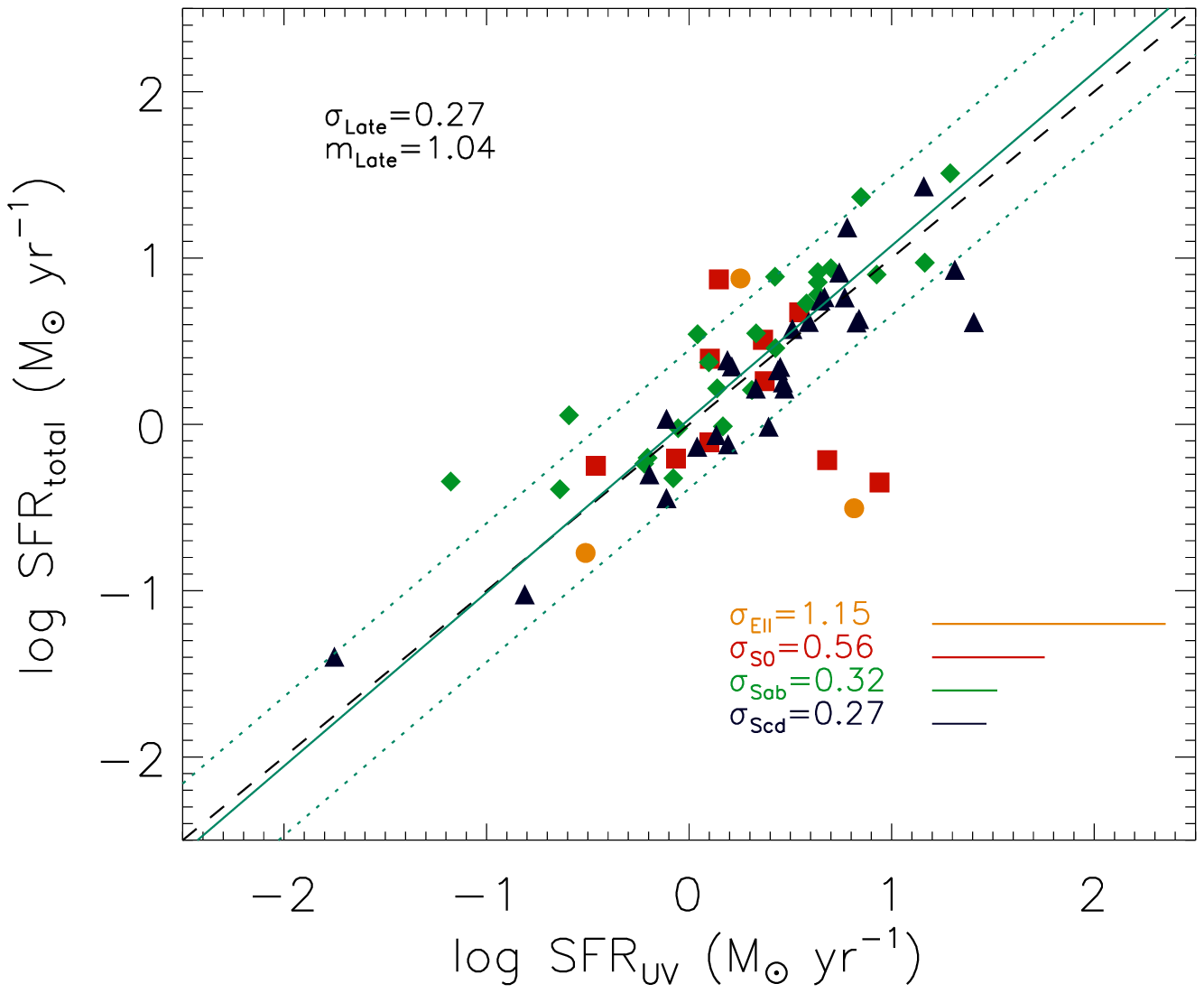}
\caption{Comparison of SFR$_{total}$ and SFR$_{UV}$. Colours and symbols as in Fig. \ref{total_SDSS}. }
\label{total_UV}
\end{figure}

\begin{figure}
 \centering
\includegraphics[scale=0.55]{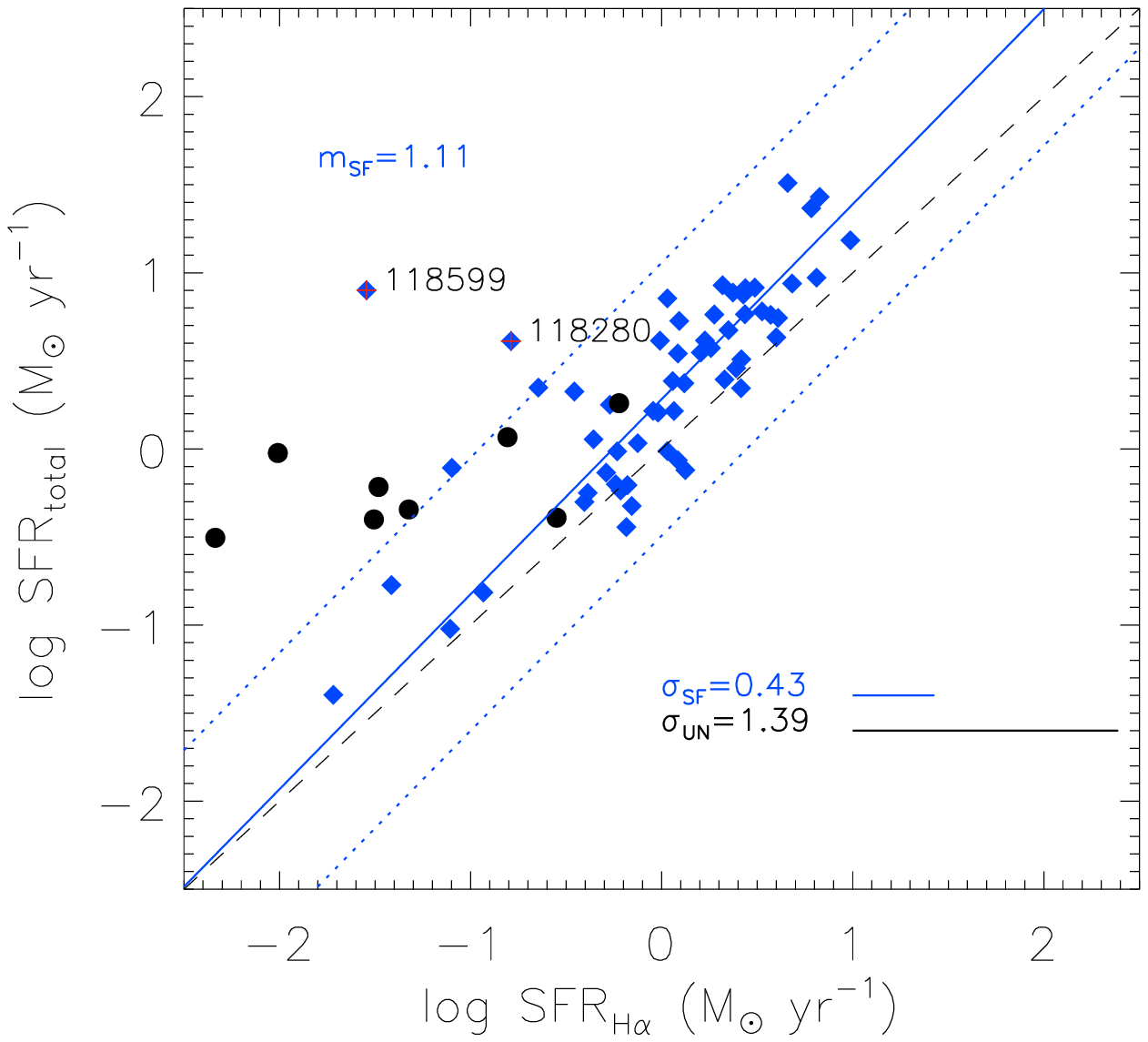}
\includegraphics[scale=0.55]{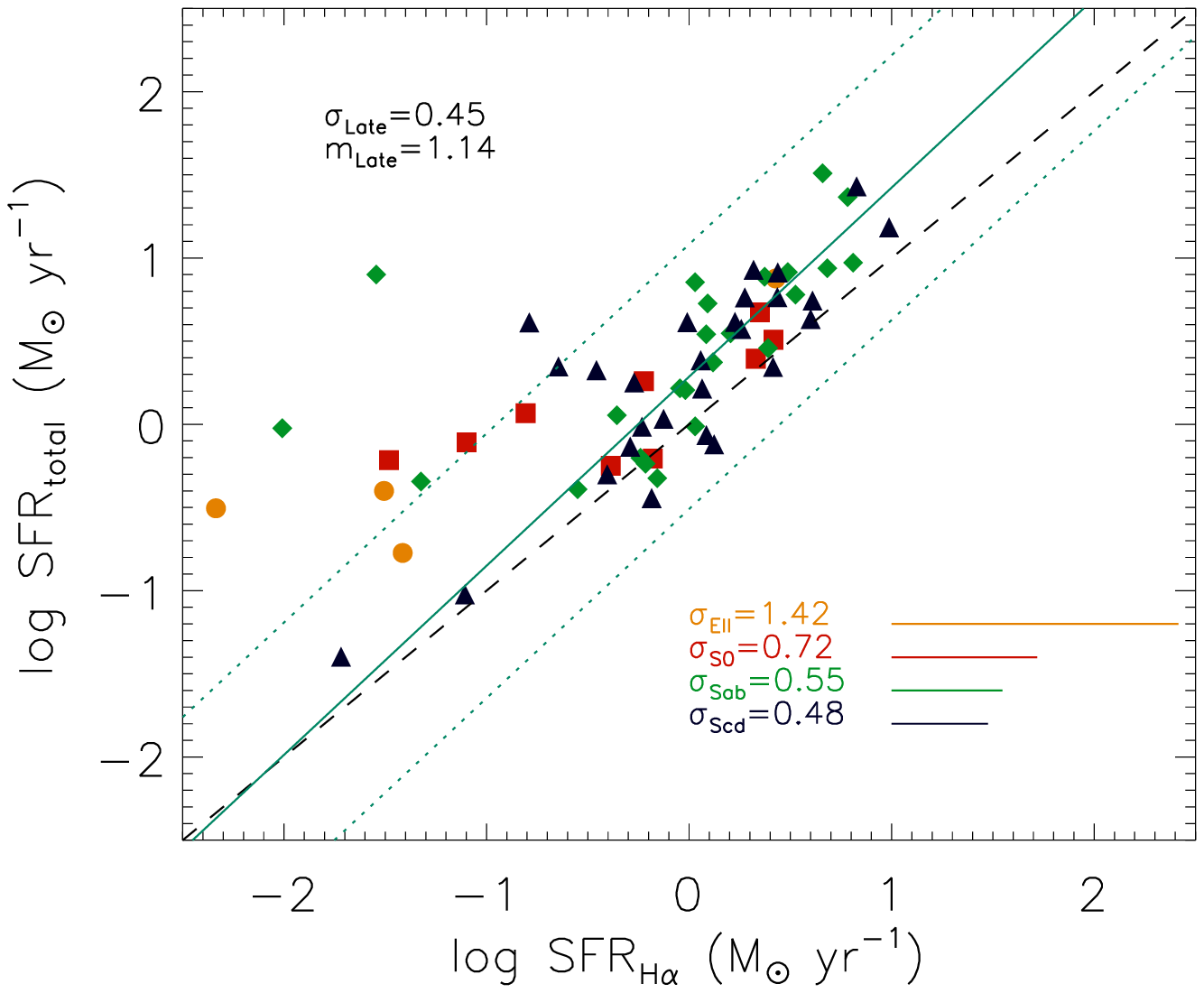}
\caption{Comparison of SFR$_{total}$ and SFR$_{H\alpha}$. Colours and symbols as in Fig. \ref{total_SDSS}.}
\label{total_Ha}
\end{figure}

 As comparing all the SFR indicators previously explained with each other would result in a too complicated analysis, for simplification we will focus on the comparison of the other SFRs indicators with the SFR$_{total}$. The inclusion of  information in two fundamental wavelengths ranges, the UV from \textit{GALEX} and the FIR from \textit{Herschel} PEP data, makes SFR$_{total}$ the more accurate SFR indicator of the above explained, as it is relatively unaffected by the dust extinction correction. Recall that the galaxies used in this comparison are 70 (58 SF and 12 unclassifiable galaxies, as we require both FIR and FUV detection) and 67 when we include the morphological information (this is why some galaxies present in the upper panels of Figs. \ref{total_SDSS},\ref{total_UV} and \ref{total_Ha} are missing in the lower panels). 

To begin with, we show in Fig. \ref{total_SDSS} the comparison between the  SFR$_{total}$ and the SFR$_{SDSS}$ (which have been converted to a Kroupa IMF, as explained in the introduction), which  includes a careful analysis of the dust extinction (see Section \ref{sect:SFR-SDSS}). In the plot we show the one to one relation, as well as the best-fit to  the SF  sample. The slope for the SF sample is m=1.05 and the dispersion for the SF and unclassifiable samples $\sigma_{SF}$=0.20, $\sigma_{un}$=0.64.
Mean SFR$_{SDSS}$ uncertainties  are $\sigma_{all}$=0.40, $\sigma_{SF}$=0.28, $\sigma_{un}$=0.63 for the whole sample of galaxies considered in this work, for the SF and for the unclassifiable sample, respectively.  This means that the dispersion of the relation is comparable with typical SFRs uncertainties.  We show this typical 0.4 dex uncertainty  as dotted lines around the best-fitting line to the SF sample to characterise the confidence region of the relation. Only 2 SF galaxies are located beyond the dotted lines and are discussed in the appendix. We did not include these galaxies in the fitting to obtain the best-fitting slope. The unclassifiable galaxies show systematically larger (by $\sim$ 0.3 dex) SFR$_{total}$ than SFR$_{SDSS}$ and most of them are located beyond the dotted lines. We recall that for these galaxies, the SFR$_{SDSS}$ is not based on the \Ha~ emission but on the D4000-SFR relation (see B04). It is difficult to assess which of the two SFR estimates is the more accurate. However, the disagreement between the two estimates is within typical SFR uncertainties, even for the unclassifiable galaxies. 

In the lower panel of  Fig.  \ref{total_SDSS}  the galaxies have been colour coded depending on their morphology. The agreement between the two SFRs for the late type galaxies is excellent (m$_{late}$=0.99) and the dispersion is small ($\sigma_{late}$=0.25). We also show typical uncertainty values around the best-fitting relation (dotted lines).The dispersion values for the Sab and Scd galaxies ($\sigma_{Sab}$=0.32, $\sigma_{Scd}$=0.20) are within typical SFR$_{SDSS}$ uncertainties. For the E an S0 galaxies the dispersion values are larger ($\sigma_{E}$=0.38, $\sigma_{S0}$=0.52) and they usually show the largest SFR$_{total}$ with respect to  SFR$_{SDSS}$. We have to keep in mind that the recipes from K98 to convert IR and UV luminosities into SFRs are derived for star-bursts or young stellar populations. It is therefore reasonable that the E and S0 galaxies, which usually have older stellar populations, do not follow the relation as tightly as the late type galaxies (Sab and Scd).

The accuracy of the \LIR~as a SFR indicator  has been widely debated and some authors say that the contribution of the older stellar population to the IR heating is not negligible (e.g. \citealt{Calzetti2012}). Fig. \ref{total_SDSS} indicates that this is not a problem for most of our sample, specifically for local main sequence disks: the \LIR~ agrees with the SFR$_{SDSS}$ and other SFR indicators (see also Figs. \ref{total_UV}, \ref{total_Ha}). The problem may be present for the unclassified objects which are below the main sequence  (Fig. \ref{SFR-mass}) but above the one-to-one relation in Fig. \ref{total_SDSS}. Unfortunately, the lack of accurate age estimates for the SDSS sample of galaxies does not allow to study this effect in detail.  Other possible reasons for the offset for the unclassified galaxies could be due to aperture correction effects, bad calibration of the D4000 relation at low SFR values or different SFH than that assumed in the K98 recipes (e.g., decaying SF instead of constant).\\

In Fig. \ref{total_UV} we compare  SFR$_{total}$ with SFR$_{UV}$. The obtained slope for the SF sample is slightly larger than 1, m$_{SF}$=1.16 but it intersects with the one to one relation near the 0,0 point, meaning that the agreement between the two indicators is excellent for the bulk of the studied galaxies. The dispersion is similar to that for the SFR$_{SDSS}$ ($\sigma_{SF}$=0.28) and smaller than typical uncertainties in the SFR derivation from the \Luv, due to the assumptions made on the metallicity, IMF or star formation history ($\sim$ 0.4 dex, \citealt{Schaerer2000}). We show this typical uncertainty in the SFR$_{UV}$ derivation as the dotted lines around the best-fitting relation for the SF sample. However, $\sim$45$ \%$ of the galaxies show  SFR$_{total}$ $<$ SFR$_{UV}$. This is a consequence of the larger values of E(B-V)$_\beta$  with respect to the E(B-V)$_{IRX}$ (see Sect. \ref{sect:dust-comparison}, SFR$_{total}$ is equivalent to correcting SFR$_{UV}$ using the Hao et al. 2011  dust extinction).  Once the typical SFR$_{UV}$ uncertainties are taken into account, only 4 galaxies have larger SFR$_{UV}$ than SFR$_{total}$ within the errors.  These outliers are discussed in the appendix (and they were not included in the fit to obtain the best-fitting slope), but we note here that 3 of them are unclassified. For the unclassifiable galaxies, the dispersion is larger than typical SFR$_{UV}$ uncertainties ($\sigma_{un}$=0.83). Besides,  they do not show a systematic deviation towards larger or smaller SFRs, as it happened in the comparison with SFR$_{SDSS}$, but are distributed both below and above the one-to-one relation.  One of the reasons for this discrepancy may arise from the FUV emission for the ETGs not being related to young stars but to some metal line blanketing the NUV flux \citep{Donas2007} (it must be noted that more than half of the unclassified galaxies show early type morphologies).  There are 3 unclassified galaxies for which SFR$_{UV}$/SFR$_{total}$ $>$ 7. The 3 of them are ETGs (2 S0, 1 E) and are outliers in the FUV-NUV vs B-V plot  (see Fig. \ref{ETG-colors}). This suggests that there is an UV excess for these galaxies which is not due to young stars.

We classify our galaxies into morphological types in the lower panel. The agreement for the two SFRs for the late type sample is also excellent (m$_{late}$=1.04, $\sigma_{late}$=0.27). The dispersion increases for earlier type galaxies, being maximal for the E sample ($\sigma_{ell}$=1.15), although the number of SF galaxies with E morphological classification is very small (only 3).

We have to keep in mind the  SFR$_{UV}$ values strongly depend on the dust extinction correction, and therefore on the prescription used to convert the UV slope into E(B-V). For example, when using the Meurer et al. 1999 prescriptions the slope for the SF sample does not drastically change (m$_{M99}$=1.08) but there is an offset $\sim$ 0.4 dex due to the zero point. Again, this is consistent with typical SFR$_{UV}$ uncertainties, but confirms the important effect that the dust extinction corrections have in the derived SFRs. Besides, there are works that suggest that the UV slope may be poorly constrained when derived from only two broad passbands filters in the UV (e.g. \citealt{Wijesinghe2011}). To summarise, the UV slope may be a problematic dust extinction estimator and should be taken with care, specially when considering ETGs.\\

In Fig. \ref{total_Ha} we do the same exercise but now we compare SFR$_{total}$ with SFR$_{H\alpha}$. We include only galaxies with S/N $>$ 1 in both \Ha~ and \Hb~emission lines to avoid huge extinction corrections uncertainties. The slope and the dispersion for the SF sample are similar to that obtained for the other SFRs (m$_{SF}$=1.11) but the dispersion is larger ($\sigma_{SF}$=0.43) and there is an offset due to the zero point, a$_{SF}$=0.28, which causes the SFR$_{H\alpha}$ to take lower values than SFR$_{total}$ for most of the galaxies (84 $\%$).   The dispersion for the unclassifiable galaxies, $\sigma_{un}$=1.39,  is by far larger than  typical SFR$_{H\alpha}$ uncertainties including assumptions in the metallicity, star formation history or IMF ($\sim$ 0.7 dex, \citealt{Schaerer2000}). We have included this typical uncertainties as dotted lines around the best-fitting relation. This could  indicate that the extinction values derived from the \Ha/\Hb~ ratio are too small, specially for galaxies with low SFRs or unclassifiable (and therefore weak \Ha~emission lines). There are no galaxies with SFR$_{H\alpha}$ larger than SFR$_{total}$, considering typical uncertainties. However, there are two SF galaxies which fall above the dotted lines (the confidence region) and are discussed in the appendix. We did not include these galaxies in the fit to obtain the best-fitting slope for the SF sample.
If we compare the SFRs for the late type galaxies, we have very similar slope and dispersion values than for the SF sample (m$_{late}$=1.14, $\sigma_{late}$=0.45). The dispersions are also larger for the ETGs, specially for the E ($\sigma_{E}$=1.42), although the number of E galaxies in the sample is small (4).

The resulting slopes and zero points from the comparison of the different SFRs are summarised in table \ref{slopes}.

\subsection{Star Formation Rate dependences}
\label{sect:SFR_dependence}

\begin{figure}
 \centering
\includegraphics[scale=0.55]{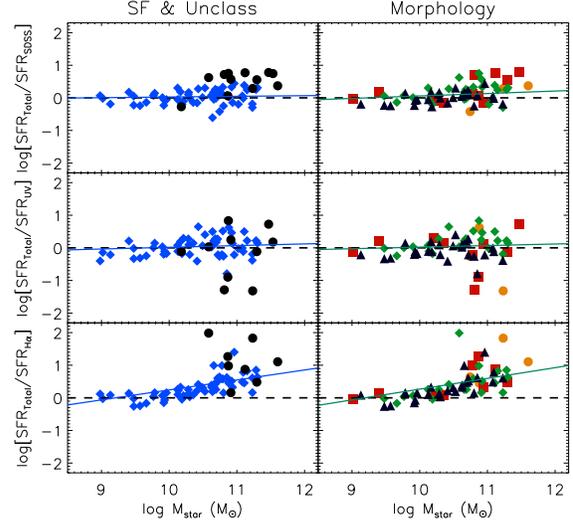}
\caption{Ratio between SFR$_{total}$ and SFR$_{SDSS}$ (\textit{upper panels}), SFR$_{UV}$ (\textit{middle panels}), SFR$_{H\alpha}$ (\textit{bottom panels}) versus the galaxy stellar mass. In the left panels galaxies are divided into SF (blue diamonds) and unclassifiable (black circles). In the right panels colours and symbols represent different morphologies: E (yellow circles), S0 (red squares), Sab (green diamonds), Scd (dark blue triangles). The thick blue and dark green lines are the best-fit to the SF and late type samples, respectively.}
\label{delta_SFR}
\end{figure}

In this section we will try to find out which are the galaxy properties that mostly affect the comparison of the SFR indicators. We first compare the ratio between  SFR$_{total}$ and the other SFR indicators (SFR$_{SDSS}$, SFR$_{UV}$ and SFR$_{H\alpha}$) with  respect to the galaxy stellar mass in  Fig. \ref{delta_SFR}. In the left panels the colours and symbols divide our galaxies into SF and unclassifiable, while in the right panels they are divided according to their morphological classification. The thick blue line in the left panels is the best fit to the SF galaxies, while the dark green line is the best fit to the late type galaxies. The behaviour of the SF and late type galaxies is very similar. We observe that the effect of the mass for the SF or late type galaxies when comparing  SFR$_{total}$ with SFR$_{SDSS}$ and  SFR$_{UV}$ is negligible. However, in the comparison with SFR$_{H\alpha}$, the mass seems to affect the result, with the largest mass galaxies showing larger SFR$_{total}$/SFR$_{
H\alpha}$ ratios. This difference becomes $\sim$ 1 dex for the more massive SF galaxies of the sample (log M $\geq$ 11 \Msun). The fact that the disagreement at high masses happens only for the comparison between SFR$_{total}$ and SFR$_{H\alpha}$ suggests that the SFR$_{H\alpha}$ values are more affected by the mass. Possible reasons for this discrepancy may be high extinction values for the most massive galaxies (so that SFR$_{H\alpha}$ is under corrected by dust) or problems related to aperture corrections (more massive galaxies are usually larger in size and therefore their \Ha~fluxes more difficult to correct the aperture effect). 

\begin{figure}
 \centering
\includegraphics[scale=0.55]{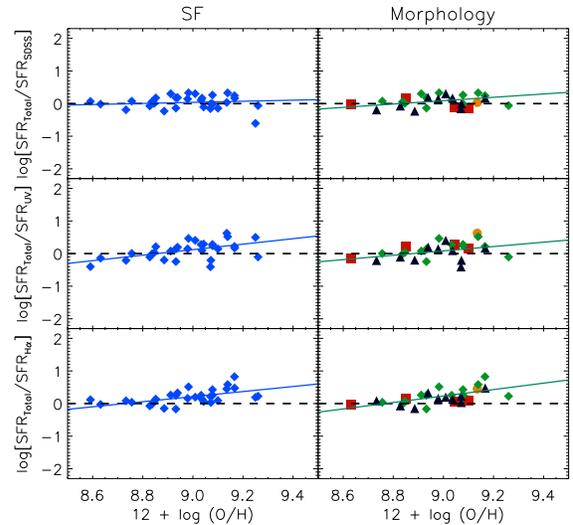}
\caption{Ratio between SFR$_{total}$ and SFR$_{SDSS}$ (\textit{upper panels}), SFR$_{UV}$ (\textit{middle panels}), SFR$_{H\alpha}$ (\textit{bottom panels}) versus the metallicity. Only 32 high S/N SF galaxies have accurate metallicities. Colours and symbols as in Fig. \ref{delta_SFR}.}
\label{delta_SFR_met}
\end{figure}

In Fig. \ref{delta_SFR_met} we plot the ratio of the different SFRs versus the metallicity of the galaxies. Unfortunately, the metallicity can only be determined for SF galaxies with high S/N emission lines, which leaves our sample with metallicity information in 32 galaxies. Again, the behaviour of the SF and late type galaxies are very similar. The metallicity does not seem to affect the comparison between  SFR$_{total}$ and SFR$_{SDSS}$. However, there is an increase of the ratio of SFR$_{total}$ and SFR$_{UV}$ or 
SFR$_{H\alpha}$ for high metallicity values. This increase reaches $\sim$ 0.7 dex in the comparison SFR$_{total}$/SFR$_{H\alpha}$ for the late type galaxies. Although this discrepancy is still within theoretical SFR$_{H\alpha}$ uncertainties (see Sect. \ref{SFR-comparison}), the effect of the metallicity in the SFR indicators has already been reported (e.g., DS2012).

\section{AGN and Composite galaxies}
\label{sect:AGN}

The recipes used in the above sections to derive SFRs are valid only for SF galaxies. The SEDs of AGNs and composite galaxies are affected by the contribution of the active nuclei over the whole wavelength range. Therefore, we study them separately in this section. 

\begin{figure}
 \centering
\includegraphics[scale=.6]{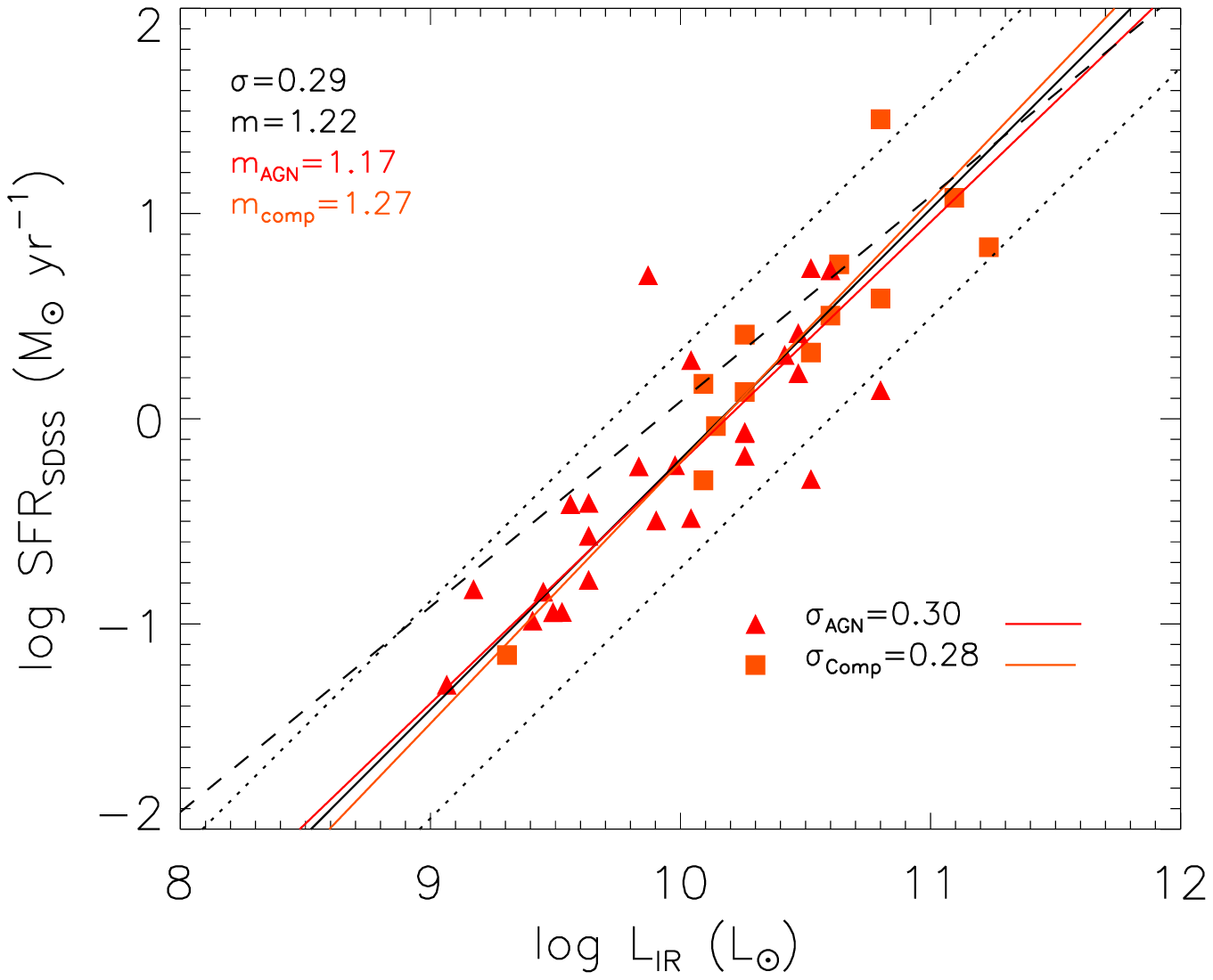}
\includegraphics[scale=.6]{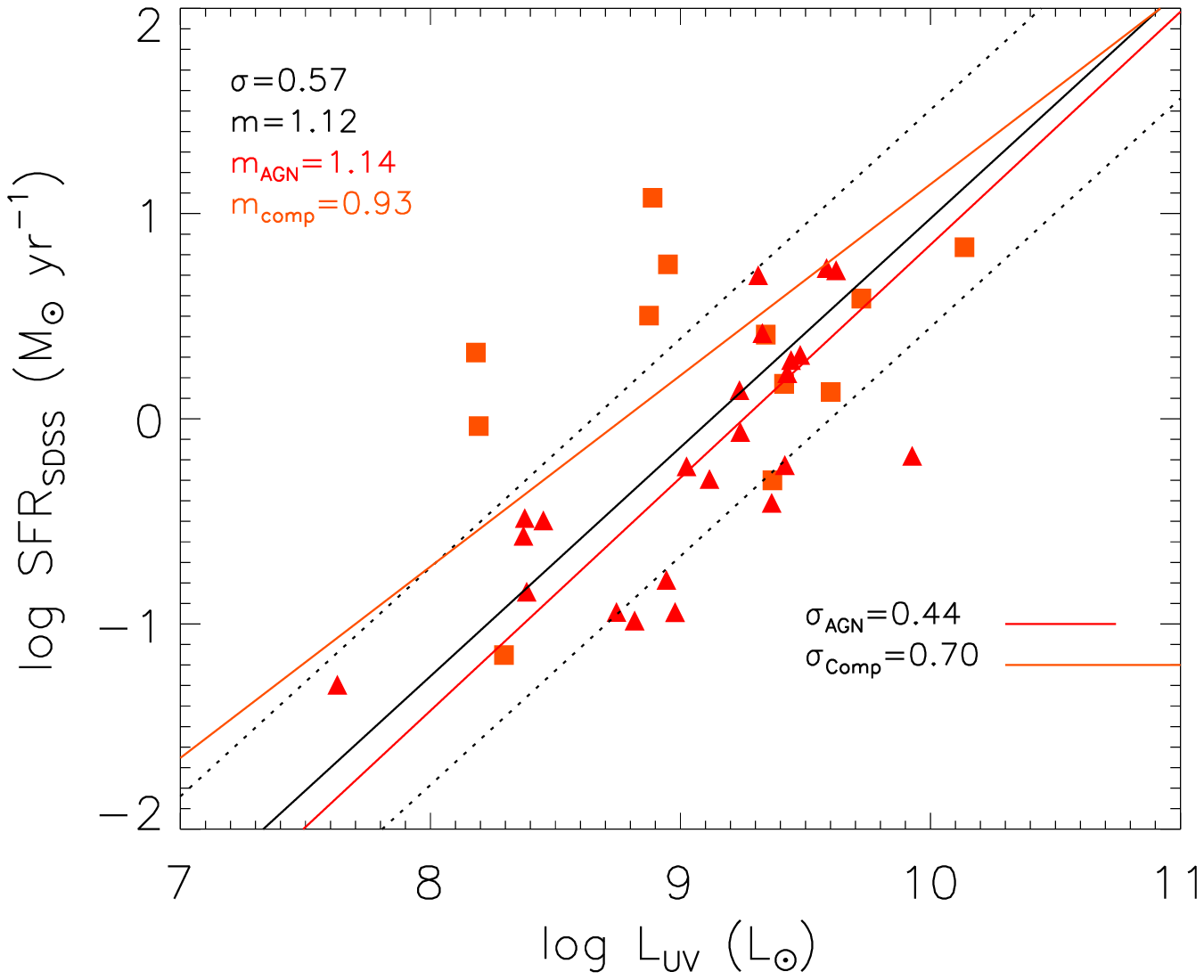}
\caption{SFR$_{SDSS}$ versus \LIR~(\textit{upper panel}) and \Luv~ uncorrected by dust extinction (\textit{bottom panel}) for AGNs (red triangles) and composite galaxies (orange squares). The black, red and orange thick line are the best fit to the AGNs plus composites, the AGNs only and the composites only, respectively. The black dotted lines represent typical $SFR_{SDSS}$ uncertainties  for active galaxies.  The dashed line is the K98 relation for SF galaxies. Also shown the best-fitting slopes and dispersions.}
\label{SDSS_AGN}
\end{figure}

Taken the SFR$_{SDSS}$ as the reference value (recall that this SFR is based in the D4000 values, see B04), we want to study if there are galaxy properties that directly correlate with the SFR for the AGN and composite samples. It has been suggested that rest-frame FIR emission is dominated by the host galaxy and it is therefore a proxy of the SF activity (e.g. \citealt{Netzer2007, Lutz2008, Hatziminaoglou2010,Mullaney2011, Santini2012, Rosario2012}). We plot in Fig. \ref{SDSS_AGN} SFR$_{SDSS}$ versus \LIR~and \Luv~ for AGNs and composite galaxies. We observe a strong correlation between the \LIR~and  SFR$_{SDSS}$ for both the AGN and composite galaxies. The derived slopes and dispersions are very similar for the two samples (m$_{AGN}$=1.17, $\sigma_{AGN}$=0.30; m$_{comp}$=1.27, $\sigma_{comp}$=0.28). The dispersion is smaller than mean SFR$_{SDSS}$ uncertainties of our composite and AGN samples ($\sigma_{AGN}$=0.56, $\sigma_{comp}$=0.49), meaning that the \LIR~is a good approximation of the SFR. The dotted lines around the best-fitting relation for both AGNs and composites represent typical SFR$_{SDSS}$ uncertainties for these kind of objects ($\sigma$ $\sim$ 0.5).  For comparison, we plot the K98 relation 
between \LIR~and SFR  for SF galaxies (Eq. \ref{eq:SFR-IR}). It can be seen that the K98 relation is almost always  above the 
predicted SFR$_{SDSS}$, although consistent within typical errors over the SFRs range studied. The larger SFR values predicted from the \LIR~ for active galaxies suggests that the contribution of the nuclear activity to the \LIR~could overestimate the SFR derivation. 

If we focus on the SFR-\Luv~ relation, now we observe that the dispersions substantially increase and they become comparable and even larger than typical SFR$_{SDSS}$ uncertainties ($\sigma_{AGN}$=0.44, $\sigma_{comp}$=0.70). Besides, the derived relations are quite different for the AGNs and composites (m$_{AGN}$=1.14, m$_{comp}$=0.93). A large fraction of the composite galaxies are located well beyond the dotted lines, representing typical SFR$_{SDSS}$ uncertainties.This suggests that the \Luv~ is not a good proxy for SFR in non-SF galaxies, as the AGN contribution to the \Luv~plays an important role. Note that the dispersion may also be caused by the fact that this \Luv~ is uncorrected for dust extinction, as the dust extinction recipes are not valid for galaxies with nuclear activity.

The results of Fig. \ref{SDSS_AGN} should be taken with caution as they  could have many interpretations, depending on type and luminosity of the AGN as well as the host galaxy properties. Decomposing the stellar and AGN contribution of the galaxy is a delicate issue which needs of careful investigation and it is beyond the scope of the present paper. Even so, recent works such as \citealt{Feltre2013} found that the AGN contribution to the \LIR~is up to 3$\%$ for non-AGN dominated galaxies and that this contribution decreases with the \LIR~coming from SF (in agreement with Fig. \ref{SDSS_AGN}). A much more detailed analysis and better statistics are needed to reach firm conclusions on the real \LIR-SFR and \Luv-SFR relations. We summarise our main best-fitting results (slopes and dispersions) in Table \ref{slopes_AGN}.

\begin{table}
  \centering
  \begin{tabular}{|lll|}\hline
    \multicolumn{1}{c}{SFR$_{SDSS}$ vs} &
     \multicolumn{1}{c}{\LIR} &
      \multicolumn{1}{c}{\Luv} 
       \\ \hline
     AGNs            & m=1.17, a=-11.94 &   m=1.14, a=-10.50 \\
     Composite       & m=1.27, a=-12.95 &   m=0.93, a=-8.17  \\
     AGN + Composite & m=1.22, a=-12.39 &   m=1.12, a=-10.18  \\
     
 \hline
  \end{tabular}
  \caption{ Best-fitting slopes (m) and zero points (a) for the AGNs and composite galaxies for the SFR-\LIR~ and SFR-\Luv~relations.}
  \label{slopes_AGN}
\end{table}

\section{Location of the SDSS FIR counterparts in the M-Z-SFR space}

There are three main properties of the galaxies to characterise their star formation history: the galaxy stellar mass, the current SFR  and the metallicity. All these three quantities are related with each other and form what is known as the Fundamental Metallicity Relation (FMR, \citealt{Mannucci2010}) or Fundamental Plane (FP, \citealt{Laralopez2010b}). In this section we confirm that the FIR counterparts of the SDSS galaxies also follow the relations derived for SDSS galaxies by \citet{LaraLopez2013} (LL2013 hereafter).

The mass and the SFR of galaxies are obviously related as the former is the integral of the latter over the lifetime of the galaxy. The SFR of a galaxy depends on its stellar mass, but also on the evolutionary stage it belongs to. There are galaxies which are still forming stars and occupy the region in the SFR-M  relation which is known as the MS, while there are galaxies which have ceased forming new stars and whose population is composed only by old stars. Besides, many studies report an evolution of the location of the MS \citep{Elbaz2007, Daddi2007, Rodighiero2010} and the characteristic mass M* of the mass function with redshift: in 
the local universe most of the massive galaxies are quiescent but if we go to z $\sim$ 2, the number density of massive 
star-forming galaxies increases by a factor of 10 (e.g., \citealt{Pozzetti2010, Ilbert2010, DS2011, Ilbert2013}).

\begin{figure}
 \centering
\includegraphics[scale=.5]{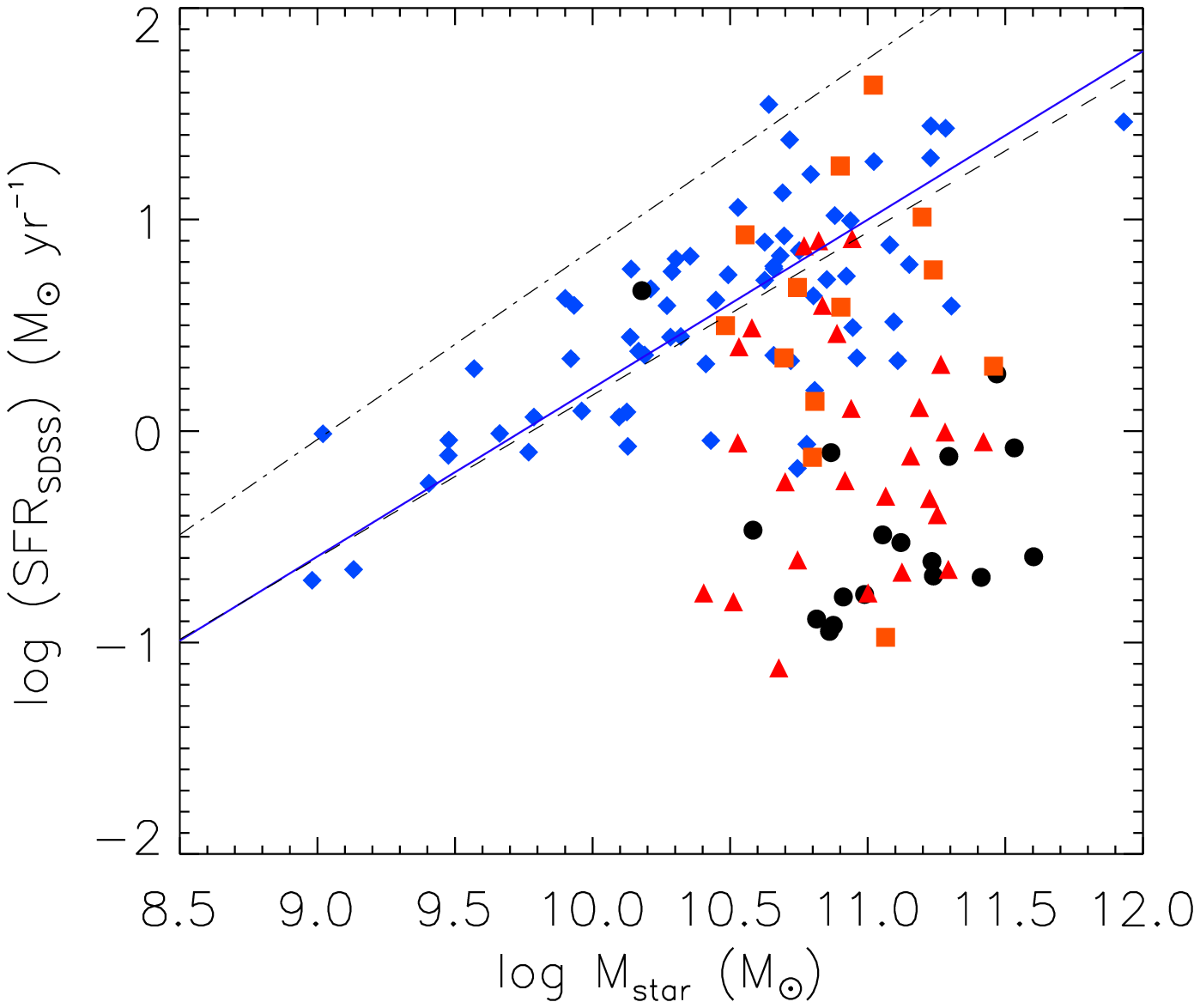}
\includegraphics[scale=.5]{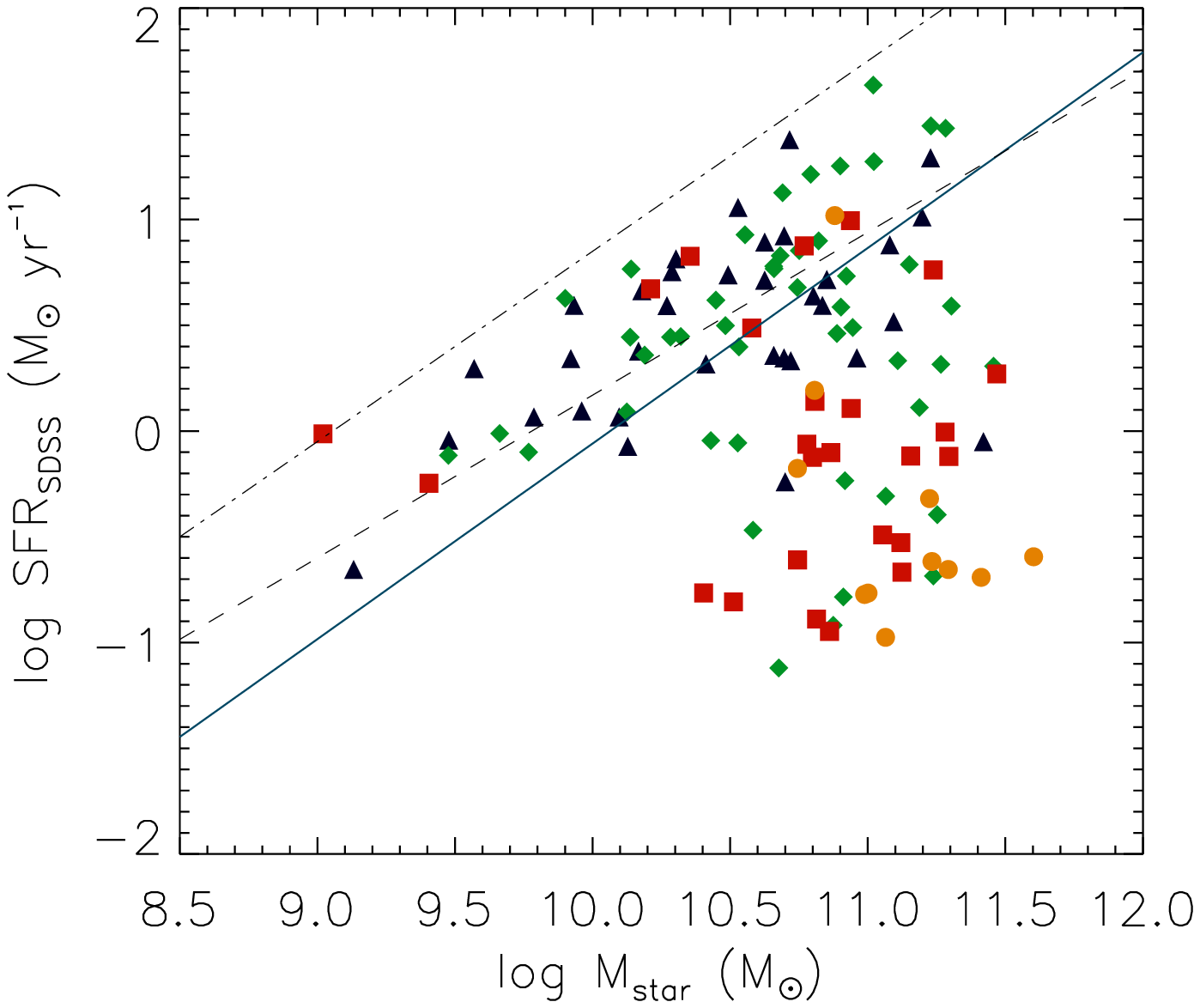}
 \caption{SFR$_{SDSS}$  versus the stellar mass for the whole sample. Colours represent different spectral (\textit{upper panel}) and morphological (\textit{bottom panel}) types. Symbols are the same of  Fig.  \ref{delta_SFR} and \ref{SDSS_AGN}. The blue and dark green thick lines are the best-fit to the SF and the late-type samples, respectively. Also shown the Main Sequence relations derived for the local SDSS sample (black dashed line, B04) and the one derived by Elbaz et al. 2007 at z $\sim$ 1 (black dotted-dashed line).}
\label{SFR-mass}
\end{figure}

In Fig. \ref{SFR-mass} we show the location of the FIR counterparts of the SDSS galaxies in the M-SFR  diagram. We use the SFR$_{SDSS}$ value; our conclusions remain unchanged when using the other SFR indicators. The SF galaxies are located along the MS and the best-fit to the SF sample  is in very good agreement with the derived value for the MS of the SDSS at z=0 from B04 (m$_{SDSS}$=0.77, a$_{SDSS}$=-7.71; m$_{FIR}$=0.80, a$_{FIR}$=-7.94; translated into Kroupa IMF). The dotted dashed-line represents the MS at z $\sim$ 1, derived by \citet{Elbaz2007}. There are almost no galaxies in that region of the M-SFR  plot, which is in agreement with the MS evolution with redshift. We can affirm that the SF FIR counterparts of the SDSS galaxies are  located in  the z=0 MS. It is also interesting the fact that the non SF sample, i.e., the AGNs, composite or unclassifiable galaxies, are  located below the MS (except for some composite galaxies). This might indicate that the presence of nuclear 
activity at low z prevents SF to take place. However, it could also be due to a selection effect in the SDSS spectral classification: moderately luminous AGN will be identified only if not overwhelmed by their host galaxy SF activity. On the other hand, some of the unclassifiable  galaxies could be some intermediate or quiescent galaxies (e.g., with FIR emission from the older stellar population) or with wrong SFRs estimates due to their weak emission lines.

In the bottom panel of Fig. \ref{SFR-mass} we show the same as in the upper panel, colours and symbols representing different morphologies. The MS slope for the late type galaxies is steeper and with a lower zero-point value (m$_{late}$=0.92, a$_{late}$=-9.48), because there is a significant number of Sab galaxies with low SFRs and high masses (i.e., low sSFR=SFR/Mass). However, there are almost no Scd galaxies with low sSFRs. On the other hand, most of the E/S0 galaxies are located below the MS. This confirms the correlation between morphology and star formation activity.

\begin{figure}
 \centering
\includegraphics[scale=.6]{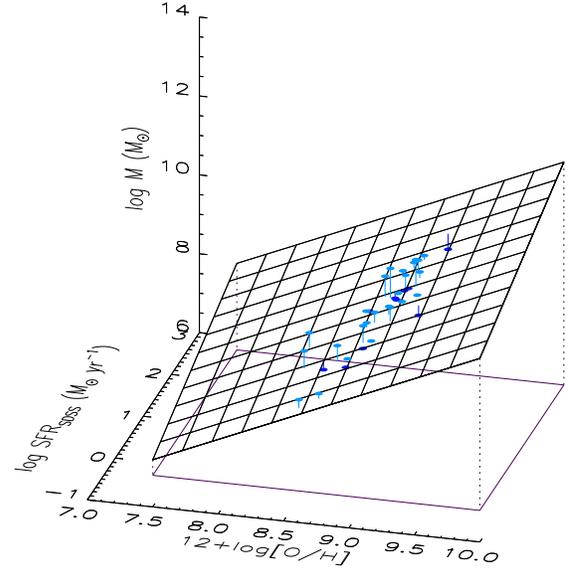}
\caption{Location of the FIR counterparts of the SDSS galaxies in the M-Z-SFR space. Only the 32 galaxies with accurate metallicities are plotted. Dark blue symbols represent galaxies below the plane of Eq. \ref{LL2013}, light blue symbols are galaxies above the fundamental plane. Blue lines highlight the distance of the galaxies to the fundamental plane.}
\label{mass-met}
\end{figure}

In a recent work by LL2013, the authors concluded that the space formed by the mass, SFR and metallicity of a galaxy could be reduced to a plane. We know that the more massive galaxies have higher metallicities. This has been suggested as the consequence of the metal enrichment of successive episodes of galaxy formation, more massive galaxies have undergone more star formation cycles and have been enriched by metals by SuperNova explosions of the most massive stars. However, there is a spread in the M-Z relation that could be explained by the SFR. At a fixed mass, galaxies with higher SFRs have lower metallicities and vice versa. The authors also derived empirical relations between the considered quantities, M($Z$, SFR), SFR($Z$, M) and Z(SFR, M), concluding that the best representation of the plane is the M=f(Z, SFR). 

\begin{equation}
 log (M/M\odot)=\alpha_z[12 + log (O/H)]+\beta_z[log(SFR)]+\gamma_z
\label{LL2013} 
\end{equation}

\noindent where $\alpha_z$=1.3824, $\beta_z$=0.5992, $\gamma_z$=-2.5729. 

In Fig.  \ref{mass-met} we  plot the 32 galaxies of our sample with accurate metallicity values (note that this are the high S/N SF galaxies) in the M-Z-SFR space together with the plane of Eq. \ref{LL2013}. To better appreciate the location of our sample, dark blue symbols represent galaxies below the plane of Eq. \ref{LL2013} ($\sim$ 30 $\%$), while light blue symbols are galaxies above the fundamental plane ($\sim$ 60 $\%$). The coloured lines are the distance of each point to the fundamental plane. Although more than half of the galaxies have larger masses than those predicted by Eq. \ref{LL2013}, the FIR counterparts of the SDSS galaxies roughly follow the prescriptions for the M-Z-SFR space derived in LL2013. Unfortunately, the scarce statistics of the sample does not allow to perform an independent fit to compare with the predictions from LL2013.


\section{Summary and conclusions}

In this work we have analysed a hundred SDSS galaxies at z $<$ 0.4 with a FIR counterpart in the PACS bands (100 and 160 $\mu$m) from the PEP survey carried out with the \textit{Herschel Space Telescope} in the COSMOS and Lockman Hole fields. From this sample, we have isolated 105 robust counterparts in FIR, with FUV emission from the GALEX data, and such galaxies constitute our main sample. We have divided this set into different spectral (SF, AGN, composites and unclassifiable galaxies) and morphological (E, S0, Sab, Scd) types.  We have made use of extensive ancillary data from the SDSS, which includes masses, SFRs, metallicities or emission lines. We have derived different SFR indicators and compared them to study their validity and limitations. We have also placed our FIR SDSS counterparts in the Fundamental plane formed by the M-Z-SFR space. Our main conclusions are:

\begin{itemize}

 \item The percentage of SF galaxies increases from  41.5$\%$ for the whole SDSS sample to 55.2$\%$ when galaxies are detected both in the FIR and FUV  bands. The unclassifiable galaxies decrease from 43.5 to 11.4$\%$. With respect to the morphological classification, the number of Scd galaxies increases from 22.5 to 31.7$\%$, while the percentage of E decreases from 20.2$\%$ to 6.9$\%$. This selection effects are expected, as FIR and FUV emission are closely related with the star-formation (which mainly occurs in SF and late type galaxies). On the other hand, the unclassifiable galaxies present weak emission lines (i.e., low star-formation) and thus no significant FIR and FUV emission.
 
 \item The distribution of redshifts and masses is not strongly affected by the FIR/FUV detection. However, the FIR counterparts of the SDSS galaxies show larger SFRs (and therefore sSFRs) and slightly larger metallicities than the whole SDSS sample.
 
 \item $L_{100}$ seems to be a very good approximation of the total \LIR, with a slope in the \LIR-$L_{100}$~ relation m=0.99 and a dispersion $\sigma$=0.07.

 \item We derive dust extinction values from two different methods: from the observed \Ha/\Hb~ ratio, E(B-V)$_{R}$, towards the emission lines,  and from the UV slope, E(B-V)$_{\beta}$, towards the continuum. We compare these extinction values with E(B-V)$_{IRX}$, from the \LIR/\Luv~ratio. E(B-V)$_{IRX}$ and E(B-V)$_{\beta}$ correlate very well, with small dispersion $\sigma$=0.06. The dispersion is larger for the  E(B-V)$_{IRX}$-E(B-V)$_{R}$ comparison, $\sigma$=0.12. We derive a conversion factor between  E(B-V)$_{UV}$ and E(B-V)$_{R}$ of 0.44, in excellent agreement with that from \citet{Calzetti2000}.
 
 \item We find a tight correlation between the E(B-V) and the \LIR, for the  three studied methods of dust attenuation (dust extinction increasing with \LIR). The correlation between E(B-V) and metallicity is also  significant (metal rich galaxies have higher dust extinctions).  The relation between the E(B-V) and the stellar mass shows a  too large dispersion  to derive any significant correlation, specially at large masses (log M $>$ 10 \Msun). 
 
 \item We have derived SFR$_{total}$ as the sum of the obscured (SFR$_{IR}$) and the unobscured (SFR$_{UV}$ without extinction correction) SFRs. The SFR$_{IR}$ represents more than 75 $\%$ of the SFR$_{total}$ for galaxies with log \LIR~$>$ 10 \Lsun~ and more than 90$\%$ for galaxies with  log \LIR~$>$ 10.7 \Lsun. However, caution must be taken when deriving the SFR from the FIR emission only for low \LIR~galaxies, as the unobscured contribution may account for $\sim$ 50$\%$  of the total SFR.
 
 \item We have compared the SFR$_{total}$ with the one derived by the MPA-JHU group, SFR$_{SDSS}$, for the SF and unclassifiable galaxies. The agreement between the two SFRs for the SF sample is excellent with a slope in the  SFR$_{total}$-SFR$_{SDSS}$ relation m=1.05 and a dispersion $\sigma$=0.20, smaller than typical SFR$_{SDSS}$ uncertainties for the SF sample ($\sim$ 0.28).
 
 \item SFR$_{total}$ and SFR$_{UV}$ or SFR$_{H\alpha}$ are also in a very good agreement for the SF sample, with slopes and dispersions $m_{UV}$=1.16, $\sigma_{UV}$=0.28, $m_{H\alpha}$=1.11, $\sigma_{H\alpha}$=0.43. The zero point in the SFR$_{total}$-SFR$_{H\alpha}$ relation is a=0.28, which causes that $\sim$ 84 $\%$ of the galaxies have SFR$_{total}$ $>$ SFR$_{H\alpha}$. There are $\sim$ 45$\%$ galaxies with SFR$_{total}$ $<$ SFR$_{UV}$, which indicates that there may be problems related to the dust extinction values derived from the UV.
 
 \item The relations obtained for the late type galaxies sample are very similar to those for the SF sample while for the unclassified and ETGs the relations show significant dispersions (larger than typical SFRs uncertainties).

 \item We have studied the dependence of the SFR comparison with the galaxy stellar mass and the metallicity.
 While the mass does not seem to affect the comparison of SFR$_{total}$ with SFR$_{SDSS}$ or SFR$_{UV}$, we find a significant difference between SFR$_{total}$ and SFR$_{H\alpha}$ for high galaxy stellar masses ($\sim$ 1 dex for log M $>$ 11 \Msun). The effect of the metallicity seems to be less important ($\sim$ 0.7 dex), but the number of galaxies with accurate metallicity values is small (32 high S/N SF galaxies) and they are mostly low metallicity galaxies (log (O/H) + 12 $<$ 9.3).
 
 \item We have studied the SFR$_{SDSS}$-\LIR~ and -\Luv~ relations for the AGN and composite galaxies. The dispersion of the SFR$_{SDSS}$-\Luv~ relation is too large ($\sigma$=0.57) to derive any recipe, but SFR seems to correlate very well with the \LIR~for both AGN and composite galaxies ($\sigma$=0.29).
 
   \item The SF sample of FIR SDSS counterparts seems to follow the MS relation obtained for the whole SDSS sample (m$_{FIR}$=0.79; m$_{SDSS}$=0.77, B04); while the AGNs, composites and unclassifiable galaxies always show lower sSFRs and are located below the MS. The best-fitting slope for the late type galaxies is larger (m$_{late}$=0.92) and shows an offset in the zero point due to the presence of late type galaxies with low sSFR. The majority of  E and S0 galaxies lie below the MS.

 \item  We have located the FIR counterparts of the SDSS galaxies in the fundamental plane formed by M-SFR-Z and confirmed that they follow the prescriptions derived by LL2013.

\end{itemize}

\section*{Acknowledgments}

The authors would like to thank the referee for the useful comments that helped to improve the clarity of this paper. This work was supported by the Spanish Ministry of Economy and Competitiveness (MINECO) under grant AYA2011-29517-C03-01. \textit{Herschel} is an ESA space observatory with science instruments
provided by European-led Principal Investigator consortia and with
important participation from NASA. The work uses Sloan Digital Sky Survey (SDSS) data. Funding for the SDSS and SDSS-II was provided by the Alfred P. Sloan Foundation, the Participating Institutions, the National Science Foundation, the U.S. Department of Energy, the National Aeronautics and Space Administration, the Japanese Monbukagakusho, the Max Planck Society, and the Higher Education Funding Council for England. The SDSS was managed by the Astrophysical Research Consortium for the Participating Institutions. 
GALEX (Galaxy Evolution Explorer) is a NASA Small Explorer, launched in 2003 April. We gratefully acknowledge NASA's support for construction, operation, and science analysis for the GALEX mission, developed in cooperation with the Centre National d'Etudes Spatiales of France and the Korean Ministry of Science and Technology. H. D. would like to thank J. Vega for technical support.

\label{lastpage}

\appendix

\section{Notes on individual sources}

In Figs. \ref{total_SDSS}, \ref{total_Ha} and \ref{total_UV} the most significant outliers have been highlighted by a red cross and identified by their ID. This ID is not the original SDSS ID, but a more convenient ID for each galaxy, included in the on-line table.  Here we analyse in deeper detail the characteristics of theses galaxies to better understand their behaviour.\\

\textit{J100245.11+014922.4:} ID 118451. This is an outlier in Fig. \ref{total_SDSS}, just in the limit of the uncertainty region. It is morphologically classified as E0 and spectrally classified as SF but  has a typical ETG spectra.\\

\textit{J100234.8+024253.2:} ID 118584: Another outlier in Fig. \ref{total_SDSS}. Spectrally classified as SF, it was not morphologically classified, visually it looks like a compact or dwarf galaxy. The spectrum suggests to host an active nuclei  and its  X-ray luminosity is typical of an AGN (L$_{x}$= 2.56$\times 10^{42}$  erg/s).\\

\textit{J100229.04+023245.8:} ID 118599: This is one of the most distant outliers in Fig. \ref{total_Ha}.  Spectrally classified as SF. It is morphologically classified as Sab, although visually could also be an S0. Nothing strange in its spectrum which could a priori explain the discrepancy between the SFRs.\\

\textit{J100116.79+021712.1:} ID 118280: Outlier in Fig. \ref{total_Ha}. Spectrally classified as SF, has an interesting spectrum with \Ha/[NII] ratio almost 1:1 and \Hb~ in absorption. Besides, the image shows a very bright structure near the bulge of the galaxy (HII region or powerful contaminating source near the galactic centre?).\\

\textit{J105150.4+573906.1:} ID 313631: Outlier in Fig. \ref{total_UV}. SF galaxy morphologically classified as Scd but which visually could be an S0. Its spectrum shows a weak starburst, \Hb~ in absorption and an intense emission at $~$410 nm. \\

\textit{J095852.79+022603.1:} ID 118534:  Outlier in Fig. \ref{total_UV}, unclassified spectral galaxy with S0 morphology. Its spectrum looks like a post-starburst (weak \Ha~emission line, \Hb~in absorption and no X-ray detection). It is a massive galaxy with low star formation, its FIR and FUV fluxes could be affected by old stellar populations.\\

\textit{J095943.91+022603.1:} ID 118265: Outlier in Fig. \ref{total_UV}, unclassified spectral galaxy with S0 morphology, very similar to the previous galaxy .\\

\textit{J100322.08+025001.1:} ID 118620: Outlier in Fig. \ref{total_UV}, unclassified spectral galaxy with E morphology. Typical ETGs spectrum  with no evidence of star formation.\\

\section{FUV-NUV versus B-V}

\begin{figure}
 \centering
\includegraphics[scale=.6]{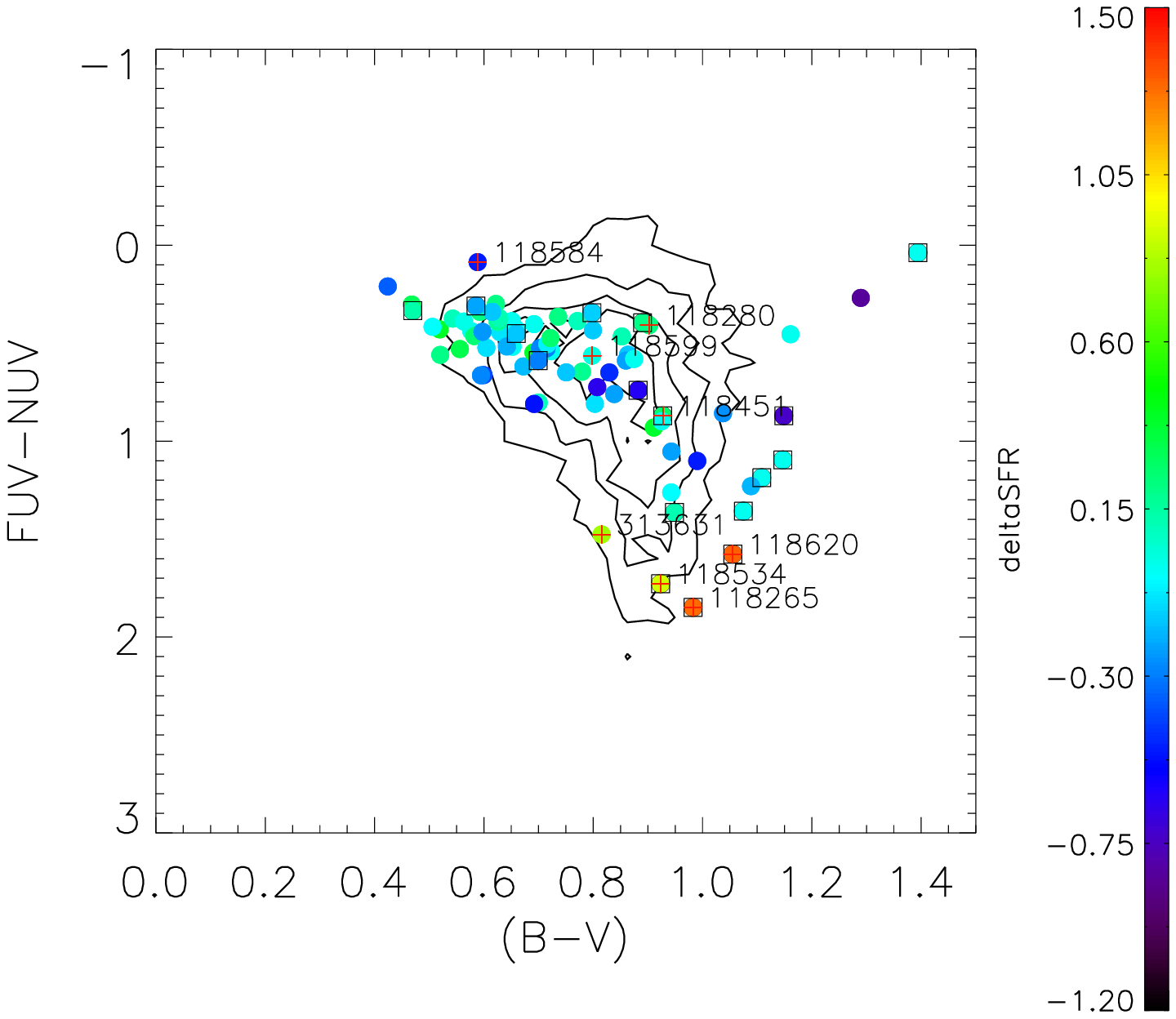}
\caption{Observed colours (not k-corrected) FUV-NUV versus B-V for the SF and unclassified  galaxies. Galaxies morphologically classified as ETGs are marked with an empty square. Galaxies have been color coded according to $\Delta$SFR=log (SFR$_{UV}$)-log(SFR$_{total})$. The contours represent the distribution of the whole sample of SDSS galaxies with UV detection from \citealt{Bianchi2011} and early-type morphology. The outliers mentioned in the appendix are highlighted by a red cross and their ID. It is clear that the galaxies for which the difference between the two SFRs is larger show the largest FUV-NUV, which points to a possible contribution to metal rich stars in the FUV.}
\label{ETG-colors}
\end{figure}

It has been mentioned in the article the possibility of FUV flux contamination for the ETGs galaxies which could affect the SFR$_{UV}$ estimation. In Fig. \ref{ETG-colors} we plot, as in Donas et al. 2007, the observed FUV-NUV versus B-V diagram (not k-corrected). The photometry is taken from SDSS DR7 and we have derived B and V magnitudes from u, g and r magnitudes using the conversions from Lupton 2005 \footnote{http://www.sdss.org/dr7/algorithms/sdssUBVRITransform.html}. The color of each galaxy represents $\Delta$(SFR)=log SFR$_{UV}$-log SFR$_{total}$. We also plot for comparison the contours of the distribution of all the galaxies from SDSS with both FUV and NUV detection from  \citealt{Bianchi2011}  and early-type classification from the catalog of Huertas-Company 2011. It can be seen that the  3 outliers with  SFR$_{UV}$ - SFR$_{total}$ $>$ 0.4  are located in the lower region of the plot, showing large FUV-NUV values. This suggests the possibility of the contribution of old metal rich galaxies in the FUV band, which affects both the dust extinction and the SFR derivation from the UV.

\section{Star Formation Rates versus \LIR~ and \Luv~ for star forming and unclassified galaxies}

Here we show the analog of Fig. \ref{SDSS_AGN} for the SF and unclassified galaxies. It is interesting to notice that the best-fitting slope for  the SF galaxies in the SFR$_{SDSS}$ versus \LIR~plot is lower than 1, m$_{SF}$=0.80. This is consistent with the results from DS2012, where the authors obtained a slope m=1.26 (1/m=0.80) in the SFR$_{FIR}$- SFR(\Ha) plot, where SFR(\Ha) was derived from the \Ha~ emission lines following the prescriptions of B04 (i.e., equivalent to SFR$_{SDSS}$).  This confirms that SFR$_{SDSS}$ is a very good proxy of the total SFR, not only the obscured part, and the importance of the \Luv~contribution to the SFR, specially at low \LIR (see Fig. \ref{SFRIR/SFRtotal}). For the unclassified galaxies the slope is larger than 1, m=1.18, and for all of them the SFR$_{SDSS}$ is lower than that predicted by K98 from the \LIR. The slopes for the \Luv~plots are m$_{SF}$=1.14 and m$_{UNC}$=1.07, meaning that the dust correction does not significantly change the slope of the \Luv~-SFR relation. Note that the dispersion for the unclassified galaxies is smaller than their dispersion when comparing SFR$_{total}$ and SFR$_{UV}$, which may indicate that the dust extinction corrections from the UV slope for these galaxies are not reliable.

\begin{figure}
 \centering
\includegraphics[scale=.6]{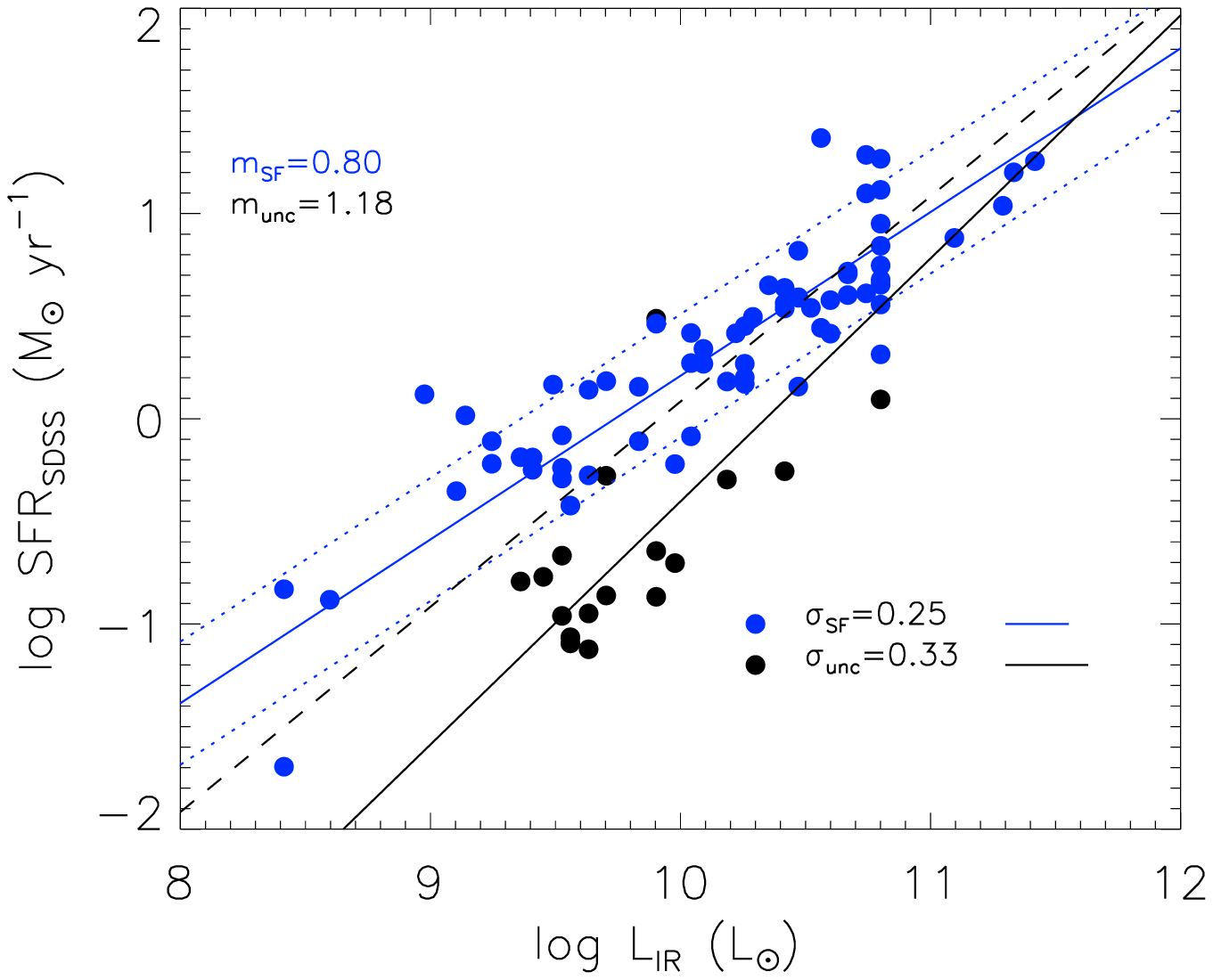}
\includegraphics[scale=.6]{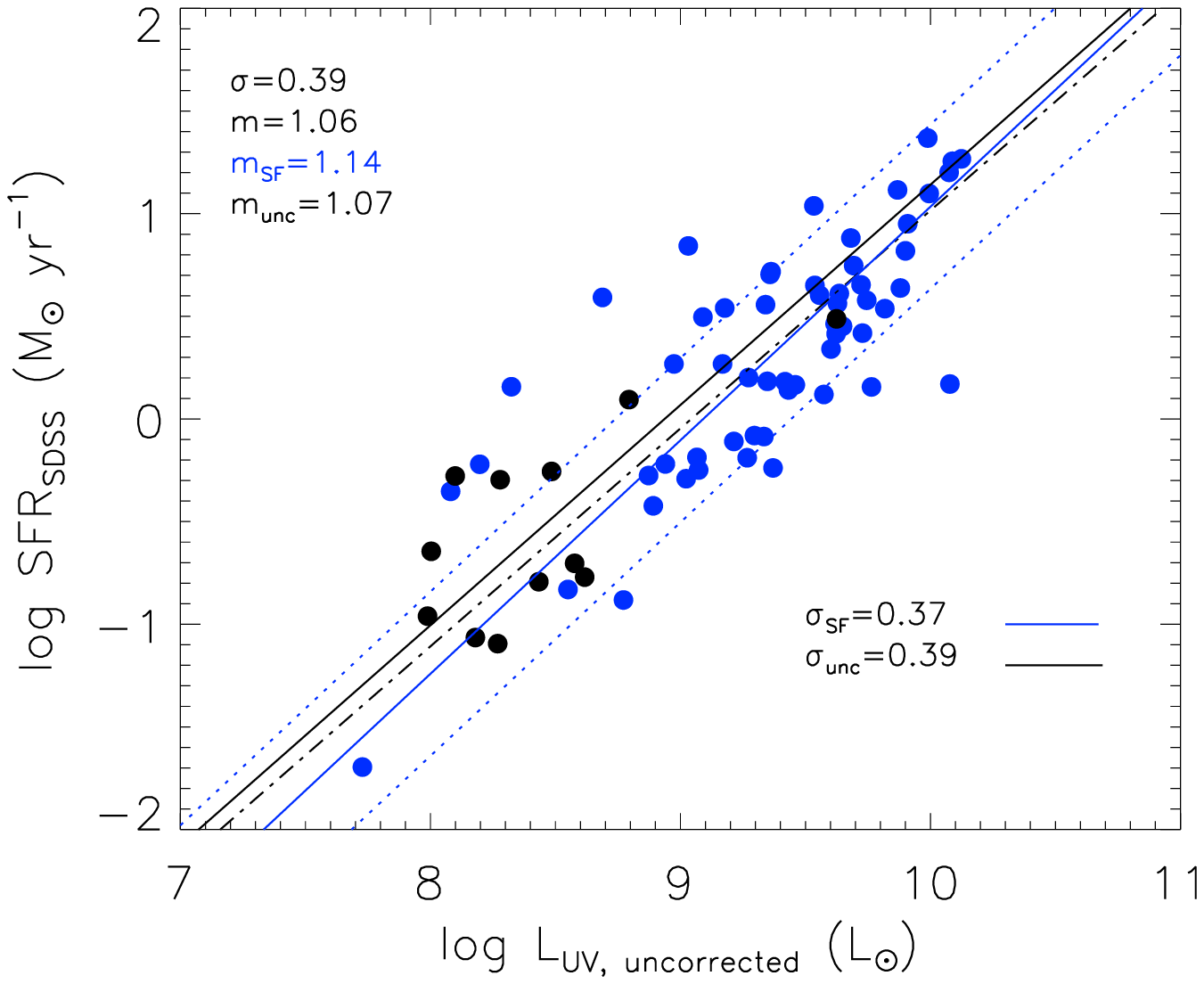}
\caption{SFR$_{SDSS}$ versus \LIR~(\textit{upper panel}) and \Luv~(\textit{bottom panel}) for SF (blue circles) and unclassified galaxies (black circles). The black and blue thick lines are the best fit to the unclassified and SF galaxies, respectively. The blue dotted lines represent typical $SFR_{SDSS}$ uncertainties  for SF galaxies.  The dashed line is the K98 relation for SF galaxies. Also shown the best-fitting slopes and dispersions obtained for each subsample.}
\label{SDSS_SF}
\end{figure}

\section{On-line data}

In order to make the results of this work accessible to the scientific community, we provide an on-line table with the main galaxy properties here analysed. In table \ref{online} we show a sample of the on-line table split in three tables to make the layout clearer.

\newpage

\bibliography{bibliografia}

\clearpage

\addtolength{\hoffset}{5.5cm}

\begin{sidewaystable}
\caption{   Main galaxy properties (sample of the on-line table data)} 
\label{online}
\centering 
\footnotesize
\begin{tabular}{*{16}{c}}\toprule
\hline
\hline
ID  & RA &  DEC  &   z & class & log M  & SFR (SDSS)         & Z                & \LIR  & Separation Herschel & pE  &  pS0     &  pSab   &  pScd   &  Separation morph & ...\\  \midrule

 Units  &     &       &     &       & [\Msun]& [\Msun yr$^ {-1}$] & [log (O/H) + 12] & [\Lsun]  &[arcsec]           &[$\%$] &   &   &   &   [arcsec]        &   \\ \midrule
 \hline
 108993  & 149.471 & 1.74141& 0.21     &  low S/N SF   & 11.23  & 1.12     &   ---     &  10.80  & 0.25  &  0.003 & 0.013 & 0.459 &   0.525   & 0.062 &  \\
 108994  & 149.391 & 1.65895 & 0.03    & high S/N  SF   & 8.99   & -0.88    &   8.59  & 8.60     & 0.54  &  ---       &  ---      &  ---      &  ---         & --- &      \\
  109395 & 149.473 & 1.97538 & 0.22    & low S/N SF  & 11.28 & 1.26     &   ---      & 11.42   & 0.11   &  0.045  & 0.091 & 0.485 &  0.379    &  1.043 &  \\
  117888  & 150.613 &1.54651  & 0.10    &  high S/N  SF   & 10.70 & 0.75    &  9.17   & 10.81   & 0.36   &  0.009  &  0.087 & 0.378 & 0.526     &  1.263 & \\
  117890   & 150.652 &1.51091  & 0.05    & high S/N  SF    &  9.48 & -0.219  &  8.49   & 9.25     & 0.54   &   0.035 & 0.114  & 0.385 & 0.467    &  1.098  & \\
   117891 & 150.511  & 1.49929 & 0.10    &  AGN  & 10.53 & 0.222   &    --- &  10.47    & 0.04    &  0.007  &  0.042 & 0.770 & 0.181    &1.319    & \\ \bottomrule
 \hline
\end{tabular}

 \vspace{2\baselineskip}
\caption{   Main galaxy properties (continuation)} 
\footnotesize
\centering 
\begin{tabular}{*{12}{c}}\toprule
\hline
\hline
ID  & NUV flux   & NUV flux error   & FUV flux   &  FUV flux error & separation GALEX & \Ha~flux                                  & \Ha~flux error   &  \Hb~flux &  \Hb~ flux error & Aperture &  ...\\  \midrule
  Units   & [$\mu$ Jy] &                  &            &                 &   [arcsec]       & [erg s$^{-1}$ cm$^{-2}$ $\times$10$^{17}$]&                  &           &                   &         &      \\ \midrule
\hline
 108993 &  27.85 &  0.29  &11.82   & 0.19 & 0.70  &   67.83     & 9.07  & 16.43  &  3.28   &  0.334  &  \\
 108994 &  69.98 &  0.38  & 47.12 &  0.34 &  1.04 &   70.87   &  1.83 &  22.35 &   1.76     &  0.992  &   \\
 109395 & 34.65  &  0.21  & 17.21 &  0.21 &  1.17 &   112.37 &  3.28 &   28.14 &   2.54     &   0.442 &  \\
 117888 & 64.42  &  0.28  & 36.44 &  0.30 &  1.99  &   287.33 & 4.97 &  66.61  &   2.89     &  0.402  &  \\ 
 117890 & 57.29 &  0.27  & 35.16  &  0.29 &  1.01  &  150.49  &  3.24 &  49.08 &   2.55     &   1.162 &\\
 117891 & 34.03 &  0.21  &  22.06 &  0.22 & 1.83   &  252.07  & 4.62 &   59.48 &   3.82     &  0.116 & \\ \bottomrule
\hline
\end{tabular}
%

\vspace{2\baselineskip}
\caption{   Main galaxy properties (continuation)} 
\footnotesize
\centering 
\begin{tabular}{*{8}{c}}\toprule
\hline
\hline
ID  &     log SFR (IR) &          log SFR(\Ha) &  log SFR(UV, corr) &          log SFR(UV, obs) &   EBV$_{IRX}$ &   EBV$_{R}$ & EBV$_{\beta}$ \\ \midrule
 Units      &      [\Msun yr$^ {-1}$] & & &    &               &             &                \\ \midrule
\hline
 108993 & 0.866  & 0.312  & 1.312   & 0.058  & 0.17 & 0.31 & 0.31   \\
 108994 & -1.337 & -0.933 & -0.413  & -0.971 & 0.02 & 0.09 &  0.14 \\
 109395 & 1.483  &  0.659 & 1.289   &  0.273 & 0.25 & 0.28 &  0.25 \\
 117888 & 0.866  &  0.438 & 0.742   & -0.079 & 0.20 & 0.35 &  0.20 \\
 117890 & -0.689 & -0.185 & -0.112  & -0.809 & 0.07 & 0.06 &  0.17 \\ 
 117891 & 0.537  &  0.029 &  0.272  & -0.342 & 0.19 & 0.34 &  0.15 \\   \bottomrule
\hline
\end{tabular}
\end{sidewaystable}

\end{document}